\documentclass[pdflatex,sn-mathphys-num]{sn-jnl}

\usepackage{graphicx}%
\usepackage{multirow}%
\usepackage{amsmath,amssymb,amsfonts}%
\usepackage{amsthm}%
\usepackage{mathrsfs}%
\usepackage[title]{appendix}%
\usepackage{xcolor}%
\usepackage{textcomp}%
\usepackage{manyfoot}%
\usepackage{booktabs}%
\usepackage{algorithm}%
\usepackage{algorithmicx}%
\usepackage{algpseudocode}%
\usepackage{listings}%

\theoremstyle{thmstyleone}%
\theoremstyle{thmstyletwo}%

\theoremstyle{thmstylethree}%

\newcommand{\kk}{\mathbf{k}}
\newcommand{\vv}{\mathbf{v}}
\newcommand{\qq}{\mathbf{q}}
\newcommand{\rr}{\mathbf{r}}
\newcommand{\Psih}{\hat{\Psi}}
\newcommand{\Psihd}{\hat{\Psi}^\dagger}

\newcommand{\ah}{\hat{a}}
\newcommand{\ahd}{\hat{a}^\dagger}

\begin{document}

\title[]{Collective excitations of driven-dissipative quantum fluids of light}


\author[1]{\fnm{Alberto} \sur{Bramati}}\email{alberto.bramati@lkb.upmc.fr}

\author*[2]{\fnm{Iacopo} \sur{Carusotto}}\email{iacopo.carusotto@ino.cnr.it}


\affil[1]{\orgdiv{Laboratoire Kastler Brossel}, \orgname{Sorbonne Universit\'e, CNRS, ENS-PSL Research University, Coll\`ege de France}, \orgaddress{\street{ 4 Place Jussieu},\city{Paris}, \postcode{75005}, \country{France}}}

\affil[2]{\orgdiv{Pitaevskii BEC Center}, \orgname{INO-CNR and Dipartimento di Fisica, Universit\`a di Trento}, \orgaddress{\street{via Sommarive 14}, \city{Trento}, \postcode{38123}, 
\country{Italy}}}

\abstract{In this work we give an overview on a couple of decades of theoretical and experimental research on the many-body physics of driven-dissipative quantum fluids of light in optical cavities.
These systems consist of a large number of photons enclosed in a --typically planar-- cavity device where spatial confinement induces a finite photon mass and the Kerr optical nonlinearity of the cavity material mediates finite photon-photon interactions. Differently from standard Bose gases of material particles like liquid Helium or ultracold atomic gases, photons are inevitably subject to radiative and/or non-radiative losses, which must be compensated by some pumping mechanism. As a result, the properties of the steady state are not imposed by a thermal equilibrium conduition, but are determined by a dynamical interplay of pumping and losses.
Among the many collective phenomena that are observed in these systems, we focus here on the collective excitations around the steady state and we characterize the impact of the driven-dissipative nature of the gas on their properties, in particular on their dispersion relation.
Different pumping configurations give very different behaviors: a specific discussion is provided for the most important configurations used in the experiments. Observable consequences of the various forms of the dispersion --gapped, gapless, sonic, band-sticking, diffusive Goldstone-- are illustrated with a special eye towards their consequences on the superfluidity properties of the fluid of light. 
We conclude the article with a presentation of a few research avenues that we personally find most exciting for the next years, namely the osbervable consequences of the concurrently superfluid and solid nature of supersolid states of light and the collective dynamics of quantum correlated states of strongly interacting photon gases in the presence of strong nonlinearities, in particular Mott insulator states.}




\maketitle

\section{Introduction}\label{sec1}

The concept of collective excitations is one of the most powerful tools to understand the physics of many-body states and of the phase transitions connecting them. It was originally investigated for weak excitations in conservative systems of material particles around thermal equilibrium, such as electron gases, magnetic materials, liquid Helium or dilute Bose-Einstein condensates, where it has provided a framework to theoretically understand crucial effects such as superfluidity and superconductivity~\cite{nozieres1999theory,pitaevskii2016bose,kittel2018introduction}. In the last decades, this concept is starting to play a central role also in our understanding of driven-dissipative systems, in particular quantum fluids of light~\cite{carusotto2013quantum} and condensates of photons or polaritons~\cite{bloch2022non}.

A cornerstone of the theory of dilute Bose-Einstein condensates of material bosonic particles around thermal equilibrium is the Bogoliubov theory of collective excitations~\cite{bogoliubov1947theory}, which predicts an analytical form
\begin{equation}
\hbar \omega^{\rm (eq)}_{\rm Bog}(k)=\sqrt{\frac{\hbar^2 k^2}{2m}\left( \frac{\hbar^2 k^2}{2m} + 2 g n\right) }
\label{eq:Bogo_standard}
\end{equation}
for their dispersion. Here $m$ is the particle mass and, at the mean-field level, the interaction energy is given by the product $g n$ of the interaction constant $g$ and the particle density $n$. 
The softness $\omega^{\rm (eq)}_{\rm Bog}(k\to 0) =0$ of this Bogoliubov  dispersion is a direct consequence of the spontaneous breaking of a continuous $U(1)$ symmetry at the condensation phase transition. 
At low-$k$, the Bogoliubov dispersion has a sonic character $\omega^{\rm (eq)}_{\rm Bog}(k)\simeq c_s k$ with a speed of sound $c_s=\sqrt{g n/m}$, which offers a simple physical interpretation of superfluidity properties of bosonic systems~\cite{pitaevskii2016bose}. At large $k$, it recovers instead a single-particle form with a Hartree energy shift, $\omega^{\rm (eq)}_{\rm Bog}(k) \simeq \hbar k^2/(2m)+gn$.

The goal of this review is to show how the concept of collective excitations can be fruitfully applied also to driven-dissipative systems where the number of particles is not conserved and the steady state does not originate from a thermal equilibrium condition, but from a dynamical interplay of pumping and losses, e.g. quantum fluids of light and condensates of photons or polaritons. These are assemblies of a large number of photons which acquire a finite mass from spatial confinement in an optical cavity and finite binary interactions from the optical nonlinearity of the cavity material~\cite{carusotto2013quantum,bloch2022non}. In this way, collective phenomena typical of many-body systems can be observed in a novel optical context. As photons have an inevitably finite lifetime as a consequence of radiative or non-radiative losses, some pumping mechanism has to be adopted to replenish the gas. While the basic building blocks of the theory remain similar to the one of standard Bose gases at equilibrium, the presence of pumping and losses is responsible for a number of new features stemming from the driven-dissipative condition. In particular, a wider variety of dispersion relations can be found depending on the specific pumping configuration adopted, with a consequently richer phenomenology. A review of  these many-body effect is the core of this work.

In Sec.\ref{sec:QFL_general} we offer a brief review of the basic concept of quantum fluids of light and of the different pumping configurations. 
In Sec.\ref{sec:coh_pump}, we focus on coherent pump configurations. As the condensate phase is in this case imprinted by the external beam, no condensation phase transition occurs and the dispersion displays different behaviors depending on the intensity and frequency of the incident beam, including sonic, gapless, gapped, and band-sticking dispersions. Experiments highlighting this physics are reviewed in Secs.\ref{subsec:coh_rest} and \ref{subsec:coh_flow} for the cases of a fluid at rest and of a flowing fluid. Additional features are discussed in Sec.~\ref{subsec:additional}.


In Sec.\ref{sec:incoh_pump}, we move to incoherent pumping and parametric oscillation configurations where the coherence of the photonic Bose gas originates from a non-equilibrium Bose-Einstein condensation phase transition. As the $U(1)$ symmetry is in this case spontaneously broken, the dispersion must contain a soft Goldstone mode whose frequency tends to zero in the long-wavelength limit in both its real and imaginary parts. In contrast to the sonic dispersion of the Goldstone mode in equilibrium systems such as liquid Helium or dilute atomic condensates, Sec.\ref{sec:diffusive_Goldstone} reviews how driven-dissipative condensates are characterized by a diffusive and non-propagating Goldstone mode. Experiments demonstrating this physics are illustrated in Sec.\ref{sec:expt_OPO}. As a key new possibility opened by driven-dissipative systems, the modifications of the collective excitations induced by an additional external field pinning the condensate phase are finally discussed in Sec.\ref{sec:gap}.

In Sec.\ref{sec:superfluidity}, we show how the dispersion of collective excitations provides a powerful framework to understand superfluid phenomena. After giving  in Sec.\ref{sec:superfl_general} a brief summary of the general concept of superfluidity as it was developed for liquid Helium and atomic gases, in Sec.\ref{sec:superfl_expt_QFL} we present the main experiments that have assessed the different aspects of superfluidity in fluids of light and we highlight those new features that directly stem from the driven-dissipative condition.

In Sec.\ref{sec:perspectives} we present two avenues of research that we find most promising for the coming years. In Sec.\ref{subsec:supersolids}, we give a short review of recent observations of supersolid behaviors in fluids of light and we sketch how collective excitations may offer insight on the concurrently superfluid and solid nature of these systems. In Sec.\ref{subsec:strongly_interacting} we leave the regime of weakly interacting fluids and we penetrate the very much uncharted land of strongly interacting quantum fluids of light where optical nonlinearities are so large that the discrete nature of the photon starts mattering and the mean-field theory ceases being accurate. A special attention is given to recent works for Mott insulating states of light and a sketch of future directions of experimental investigations are presented.

In Sec.\ref{sec:conclusions} we present our general conclusions and we give our point of view on the future developments of the field. The conceptual ideas underlying the main experimental techniques used to measure the dispersion relation of collective excitations are presented in the Appendix. 

\section{Basics of quantum fluids of light}

\label{sec:QFL_general}

In our intuitive picture, we are used to associate light to propagating electromagnetic waves or, in a corpuscular picture, to a stream of photons that travel across space at a very fast speed; refraction is explained in terms of the attraction by material media, but no attention has been paid to the possibility of interactions between photons.

The concept of {\em Quantum Fluid of Light (QFL)} defies this picture~\cite{carusotto2013quantum}. Merging ideas from condensed matter physics and optics, it deals with the collective behaviours that assemblies of photons display when they are endowed of an effective mass and sizable inter-particle interactions and are then manipulated as a standard fluid of many interacting particles.

In this first Section, we provide a brief review of the basic concepts of quantum fluids of light and on the mean-field theory that can be used to describe them in the so far experimentally most relevant regime of weak interactions. A special attention will be given to the additional terms that have to be included into the nonlinear partial differential equation describing the dynamics of the fluid to account for its driven-dissipative nature. On the other hand, we will not dwell into another family of systems commonly used for quantum fluids of light in the so-called propagating geometry: here, the dynamics is in fact conservative and more closely resembles the standard one of Bose gases of material particles. The interested reader can find an up-top-date review of these advances in~\cite{glorieux2025paraxial}.

Before proceeding, we also need to emphasize that analogous driven-dissipative equations have been derived in several other contexts, in optics and beyond. Among the many possible points of view on this physics, in this article we will follow a presentation angle inspired from the many-body theory of dilute Bose gases. From this perspective, it is natural to call the key equation that describes the photonic Bose gas a {\em generalized non-equilibrium Gross-Pitaevskii equation}. But it is important to remind that very similar equations are considered in the nonlinear optics literature: for instance, the {\em Lugiato-Lefever equation}~\cite{Lugiato:PRL1987} is still a most powerful tool for the description of optical devices based on Kerr cavity solitons and frequency combs~\cite{Lugiato:Varenna}; an overview of the theoretical and computational techniques for the study of these nonlinear equations in the optical context can be found in~\cite{oppo2025computational}.
From a more general standpoint of nonlinear dynamics, equations of these form go often under the name of complex Ginzburg-Landau equations~\cite{Aranson:RMP2002} and are of widespread use to study pattern formation phenomena~\cite{Cross:RMP1993}.

In spite of the formal similarity of the equations, different communities typically focus their attention on different aspects of the physics. 
Given the specific point of view adopted for this article and focused on the many-body aspects, we will concentrate our attention on those planar cavity devices that are typically used for photon fluids and non-equilibrium condensates, often in the strong light-matter coupling regime where photons get mixed with excitons into dressed polariton modes. However, to avoid entering into these details of the specific experimental realizations and keep the presentation as general as possible, in this work we will not make a distinction between polaritons and photons and will consider the former just as an example of dressed photons propagating in a material medium.
At the same time, readers should also keep in mind that most of the results we are going to review can be directly translated to other nonlinear optical systems, such as spatially extended laser devices and optical parametric oscillators. For all these systems, the concept of collective excitations may provide a powerful framework to unravel the different aspects of the optical dynamics at both the classical and the quantum regimes.
A main challenge for the coming years will therefore be to reinforce the connections between the different fields and exploit the concepts developed in one field to open new perspectives in the other ones.


\subsection{Conservative dynamics}

The goal of this Subsection is to introduce the theoretical framework used to describe the dynamics of the photon fluid within a planar cavity. Within a mean-field approximation, the Bose field operator describing cavity photons is replaced by a classical field equal to its expectation value, $\psi(\rr,t)=\langle \Psi(\rr,t)\rangle$ and the dynamics is described in terms of a classical partial differential equation for the evolution of $\psi$ in time. Physically, the field $\psi$ can be understood as a macroscopic wavefunction for the fully coherent photons, for instance its square modulus $|\psi|^2$ gives the two-dimensional in-plane photon density, its phase gradient gives the flow velocity, and the modulus of the spatial Fourier transform gives the momentum distribution. 

As it is discussed in textbooks~\cite{pitaevskii2016bose} and in previous dedicated reviews~\cite{carusotto2013quantum,Carusotto:CRAS2025}, this framework provides an accurate description as long as interparticle interactions are weak and the fluid is kept by the external pumping in a highly coherent state. An appetizer of the rich physics occurring away from this regime will be given in Sec.\ref{subsec:strongly_interacting}.

\subsubsection{Effective mass}

\begin{figure}
    \centering
    \includegraphics[width=0.45\textwidth]{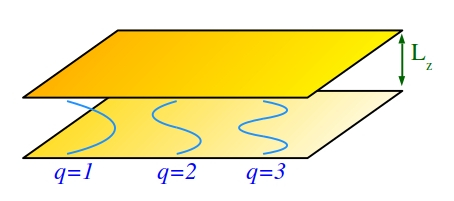}\hfill
    \includegraphics[width=0.45\textwidth]{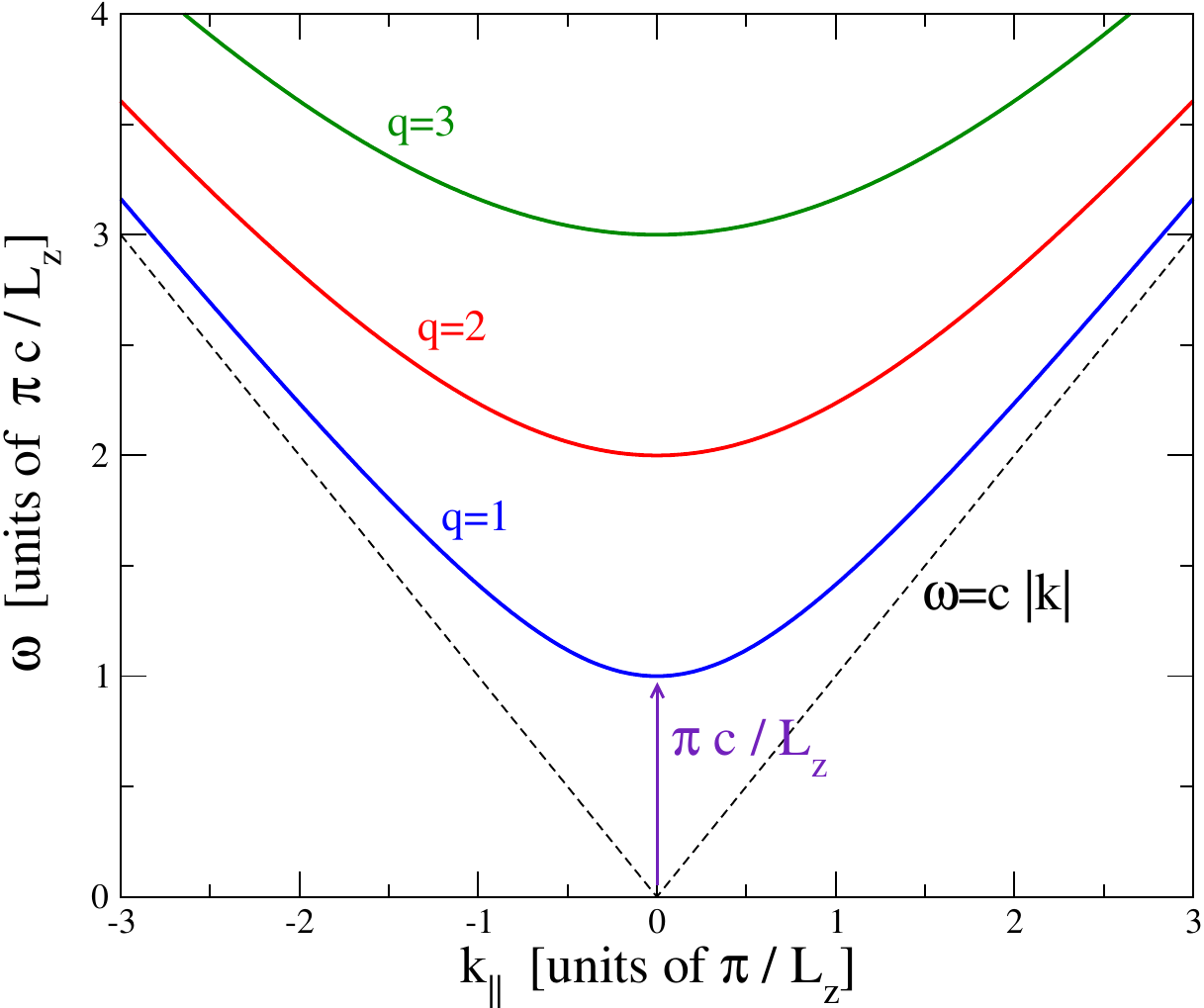}
    \caption{Light confinement in a planar microcavity (top) results in a relativistic dispersion for the in-plane motion of photons (bottom left).
    \label{fig:basics}}
\end{figure}

In vacuo, the photons that constitute light beams are massless particles that travel at a constant speed $c$. The situation is dramatically changed in spatially confined geometries, where confinement along some spatial dimension makes photons to acquire a finite effective mass for their motion along the other dimensions.

The simplest case is the one sketched in the left panel of Fig.\ref{fig:basics}, namely a cavity enclosed by a pair of plane-parallel metallic mirrors oriented along the $xy$ plane and separated by a distance $L_z$ along $z$. Boundary conditions on the electric field at the metallic mirrors force the $z$ component of the wavevector to be quantized as $k_z^{(q)}=\pi q / L_z$, the positive integer number $q>0$ corresponding to the number of field nodes along the $z$ direction. On the other hand, thanks to the translational symmetry along the cavity plane, the motion along these directions remains free and the component $\kk_\parallel$ of the wavevector along the $xy$ plane takes arbitrary values.

The dispersion of electromagnetic waves in a planar cavity filled of a material of refractive index $n$ therefore consists of a series of photonic bands with a massive relativistic-like dispersion,
\begin{equation}
\left( \hbar \omega^{(q)}(\kk_\parallel)\right)^2= \left({m^{(q)} c^2}\right)^2 + \frac{\hbar^2  c^2}{n^2} \,\kk_\parallel^2
 \label{eq:dispersion}
\end{equation}
where the finite effective mass $m^{(q)}$ grows with the quantum number $q$ as
\begin{equation}
m^{(q)} c^2 = \frac{\hbar c k_z^{(q)}}{n}=\hbar \omega^{(q)}_o\,.
 \label{eq:mass}
\end{equation}
Like in special relativity, this parameter summarizes both the {\em rest mass} related to the energy $\hbar \omega^{(q)}_o$ of a particle at rest in the zero-momentum $\hbar \kk_\parallel=0$ state and the {\em kinetic mass} related to growth of the energy as the momentum $\hbar \kk_\parallel$ is increased. In the plot of the photon dispersion shown in the right panel of Fig.\ref{fig:basics}, the rest mass corresponds to the energy gap below the dispersion of each band, while the kinetic mass is inversely proportional to the curvature at the band bottom.

In the rest of this work, we will focus on the lowest $q=1$ band and we will make a non-relativistic quadratic expansion of \eqref{eq:mass} around the band bottom,
\begin{equation}
\omega(\kk_\parallel) \simeq \omega_o + \frac{\hbar}{2m}
\,\kk_\parallel^2\,.
\label{eq:parabolic}
\end{equation}
At the level of the differential equation of motion for the photon field $\psi$, the dispersion \eqref{eq:parabolic} then leads to an evolution in the form of a Schr\"odinger equation,
\begin{equation}
    i\frac{\partial\psi}{\partial t}= \omega_o \psi -\frac{\hbar}{2m}\nabla^2_\parallel \psi\,.
    \label{eq:Schroed}
\end{equation}
For the sake of simplicity, we have neglected light polarization effects. For the sake of clarity, we have restricted our attention to metallic cavities, but analogous results hold for devices with the same translational symmetry along the  $xy$ plane, in particular dielectric microcavities~\cite{kavokin2011microcavities}.

\subsubsection{External potential}
 A straightforward way to generate an external potential for photons is suggested by the form \eqref{eq:parabolic} of the dispersion. If the explicit dependence of $\omega_o$ on the cavity parameters \eqref{eq:mass} is replaced in \eqref{eq:parabolic}, a spatial dependence of the refractive index $n(\rr_\parallel)$ and/or of the cavity thickness $L(\rr_\parallel)$ directly leads to an effective in-plane potential of the form
\begin{equation}
 V(\rr_\parallel)
 = \omega_o(\rr_\parallel)=\frac{c \pi}{n(\rr_\parallel)\, L(\rr_\parallel)}
 \label{eq:potential0}
\end{equation}
that can be used to localize the photons in suitably defined potential wells or, in a time-dependent case, to perturb the photon fluid. 
For small variations of the refractive index $n(\rr_\parallel)=\bar{n}+\delta n(\rr_\parallel)$ and/or the thickness $L(\rr_\parallel)=L_0+\delta L(\rr_\parallel)$, the potential term has the linearized form,
\begin{equation}
i\frac{\partial\psi}{\partial t}=\ldots +\omega_o\psi + \delta V(\rr_\parallel)\,\psi= \ldots+\omega_o\psi - \frac{c \pi}{\bar{n}\,\bar{L}}\,(\frac{\delta n(\rr_\parallel)}{\bar{n}}+\frac{\delta L(\rr_\parallel)}{\bar{L}})\,\psi\,.
\label{eq:potential}
\end{equation}

\subsubsection{Photon-photon interactions}


At the level of the classical electromagnetism~\cite{jackson1999classical}, the linearity of the Maxwell equations in vacuo gives a superposition principle for light fields: light beams propagate independently of each other, and no light beam can be used to modify the propagation of another beam. This cornerstone statement is no longer valid in quantum electrodynamics: here, the interconversion of photons into electron-positron pairs leads to effective photon-photon interactions. Yet, the relatively large mass of electrons and positrons ($\sim 0.5\,\textrm{MeV}$) makes the strength of this interaction for visible (or longer wavelength) photons (of energy on the eV range) to be extremely small in vacuo~\cite{heisenberg1936folgerungen}. On the other hand, a dramatic reinforcement of interactions occurs in material media, where the photon-photon interactions can be mediated by electron-hole pairs of energy in the eV range (instead of electron-positron ones in the MeV range). 

In the language of nonlinear optics~\cite{Butcher,Boyd}, photon-photon interactions are described by the nonlinear optical susceptibilities of the medium, in particular a spatially local, real-valued cubic $\chi^{(3)}$ nonlinearity corresponds to binary photon-photon contact interactions that are typically considered in the context of fluids of light. In analogy with the static potential \eqref{eq:potential}, the refractive index change induced by the nonlinearity gives a frequency-shift described by a term of the form
\begin{equation}
   i\frac{\partial\psi}{\partial t}= \ldots+g_{\rm nl} |\psi|^2 \psi 
   \label{eq:nonlinear}
\end{equation}
in the field equation. 
For the simplest case of a plane-parallel metallic cavity filled with a nonlinear medium, the coupling constant $g_{\rm nl}$ is given by
\begin{equation}
 g_{\rm nl}=-\frac{18\,\pi^2 (\hbar \omega_o)^2}{n^4\,L}\,\chi^{(3)}\,.
 \label{eq:gnl_chi3}
\end{equation}
The key approximation underlying this framework is that interactions between photons are weak enough to be treated at the mean-field level~\cite{pitaevskii2016bose}.
Combining the nonlinear \eqref{eq:nonlinear} and \eqref{eq:potential} terms with the kinetic term \eqref{eq:Schroed}, one then obtains an evolution equation in the form of the Gross-Pitaevskii equation (GPE) for the coherent photon fluid in the planar cavity,
\begin{equation}
   i\frac{\partial\psi}{\partial t}= \omega_o \psi -\frac{\hbar}{2m}\nabla^2_\parallel \psi + \delta V(\rr_\parallel)\,\psi +g_{\rm nl} |\psi|^2 \psi 
   \label{eq:photonGPE}
   \end{equation}
While nonlinear optical effects can be observed in generic optical media provided sufficient strong laser sources are used, there is an active on-going work to find materials with optical nonlinearities as large as possible. 

On one hand, this facilitates experimental investigations as it reduces the amount of optical power needed to reach a desired value of the interaction energy $g_{\rm nl}\,|\psi|^2$ in an experiments. 
On the other hand, exceptionally large nonlinearities are needed for the realization of strongly correlated quantum fluids of light such as the Mott insulating state. These states can not be captured within the mean-field approximation and will be briefly described in Sec.\ref{subsec:strongly_interacting}. 

In the last years, nonlinearities at the required single-photon level have been observed in several different platforms, including superconductor-based circuit-QED devices operating in the microwave range~\cite{carusotto2020photonic,blais_RMP2021} and optically dressed atomic gases in Rydberg-EIT configurations~\cite{peyronel2012quantum,firstenberg2013attractive}. In these extremely nonlinear media, photons behave in a  totally different way compared to the non-interacting photons of classical electrodynamics: single photons are able to trigger nonlinear effects such as steering the propagation of other beams~\cite{chang2014quantum}, and photons can be intuitively pictured as effectively impenetrable objects like billiard balls.

\subsection{Driving and dissipation}
\label{sec:theory_DD}

In contrast to fluids of material particles where the lifetime of the microscopic constituents is typically very long compared to the many-body dynamics of interest, photons in the typical cavity devices used for experiments on quantum fluids of light experience a significant decay rate, due to a combination of non-radiative losses by absorption processes in the cavity material and radiative losses through the cavity mirrors. On the one hand, radiative losses provide a direct way to observe the dynamics of the fluid of light just by looking at the emitted light. On the other hand, the interplay of losses with the pumping that is used to replenish the photon fluid is responsible for the additional driven-dissipative physics. At the mean-field level, this physics can be modeled by including additional terms into the GPE \eqref{eq:photonGPE}. The interested reader can find a formal justification of the different terms in the previous reviews~\cite{carusotto2013quantum,Carusotto:CRAS2025}.


A coherent pump driving the cavity is described by a forcing term in the equation of motion for $\psi$ in the form:
\begin{equation}
i\frac{\partial\psi}{\partial t}= \ldots + \eta\,{E}_{\rm inc}(\rr_\parallel,t)
\end{equation}
where ${E}_{\rm inc}(\rr_\parallel,t)$ is the spatio-temporal amplitude of the field incident on the planar device. The coefficient $\eta$ is proportional to the transmission amplitude through the mirror which, at a simplest level of approximation, has been taken as frequency- and wavevector independent.

At this same level of approximation, losses give linear, phase-insensitive decay terms of the form
\begin{equation}
i\frac{\partial\psi}{\partial t}= \ldots -\frac{i\gamma_{\rm loss}}{2} \psi\,. 
\end{equation}
at rate $\gamma_{\rm loss}$. 
A broadband incoherent pump can be modeled as an effective spatiotemporally local, nonlinear amplification term of the form
\begin{equation}
i\frac{\partial\psi}{\partial t}= \ldots + \frac{i\,P}{2}\,\frac{1}{(1+|\psi|^2/n_s)}\,\psi
\label{eq:P} 
\end{equation}
where the amplification rate $P$ is fixed by the strength of the incoherent pumping. For low photon densities, gain is linear and occurs at the unsaturated rate $P$. At higher photon densities, gain gets saturated with a characteristic saturation density $n_s$. This form of the incoherent pumping assumes the so-called good-cavity limit where the characteristic time-scale of the gain medium dynamics is fast compared to the cavity one. As it happened for the photon-photon interaction term \eqref{eq:nonlinear}, this mean-field description crucially depends on the assumption that a large number of photons is needed to saturate the gain; a more sophisticated fully quantum description is needed to describe, e.g., the Mott insulating state of Sec.\ref{subsec:strongly_interacting}. 

Putting all terms together, one obtains the mean-field evolution equation,  
\begin{multline}
 i\frac{\partial \psi}{\partial t}=\omega_o\,\psi -\frac{\hbar \nabla^2}{2m}\,\psi + \delta V(\rr_\parallel)\,\psi+ g_{\rm nl}\,|\psi|^2\,\psi  +\\
 +\frac{i}{2} \left[\frac{P}{1+|\psi|^2/n_s}-\gamma_{\rm loss}
 \right]\psi + \eta \, E_{\rm inc}(\rr,t)\,
 \label{eq:dd_psi}
\end{multline}
that we are going to use in the rest of the article. Depending on the specific pumping configuration under consideration, we will set the strength of the coherent and/or the incoherent pumping terms to non-zero values. For the sake of completeness, it is important to note that a direct stochastic extension of this equation including noise terms provides a simplest way to go beyond the mean-field and include the effect of thermal and quantum fluctuations~\cite{Carusotto:CRAS2025}.

\section{Coherently pumped fluids}
\label{sec:coh_pump}

The first configuration where the new possibilities opened by driving and dissipation have been explored is the coherent pumping one. While long studied in the nonlinear optics context via the so-called Lugiato-Lefever equation~\cite{Lugiato:PRL1987}, this configuration has started being theoretically investigated in the context of fluids of light in~\cite{Carusotto:PRL2004,Ciuti:PSSB2005}. The tunability of the flow speed via the incidence angle (i.e. via the in-plane wavevector) and of the speed of sound via the frequency and the intensity of the coherent drive were exploited in the first experiments demonstrating superfluid light~\cite{Amo:NPhys2009} that we will present later on in Sec.\ref{sec:superfluidity}. The full variety of dispersion relations depending on the pump parameters was experimentally mapped out in~\cite{Claude:PRL2022,Claude:PRB2023}, including gapped, sonic and band-sticking dispersions. As compared to equilibrium condensates, this variety of behaviors stems from the fact that the oscillation frequency of the field is not fixed by the interaction energy as it happens in the conservative GPE \eqref{eq:photonGPE} but is freely determined by the frequency of the incident coherent field.

\begin{figure}[htbp]
    \centering
    \includegraphics[width=0.47\textwidth]{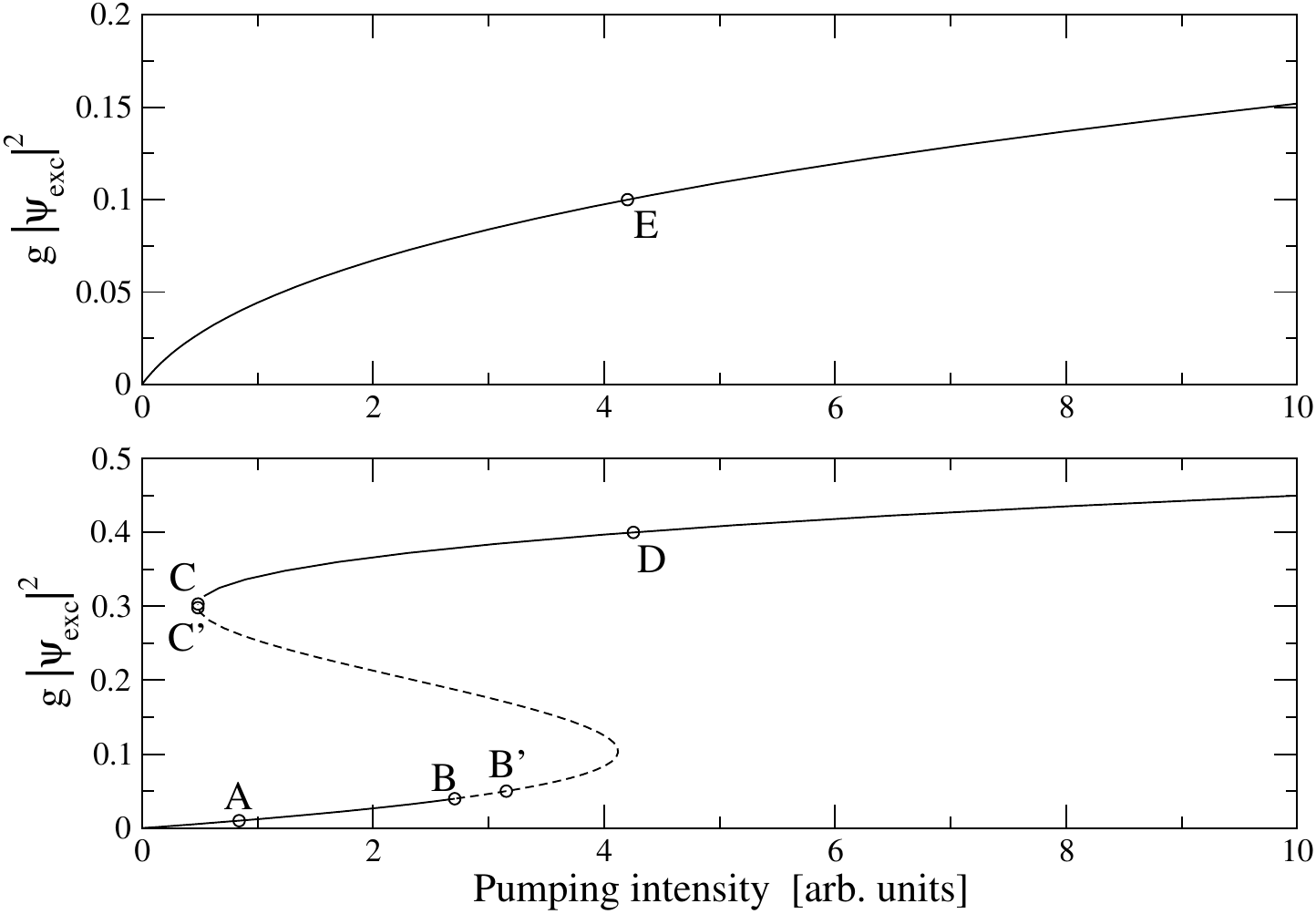}\hspace{0,05\textwidth}
    \includegraphics[width=0.47\textwidth]{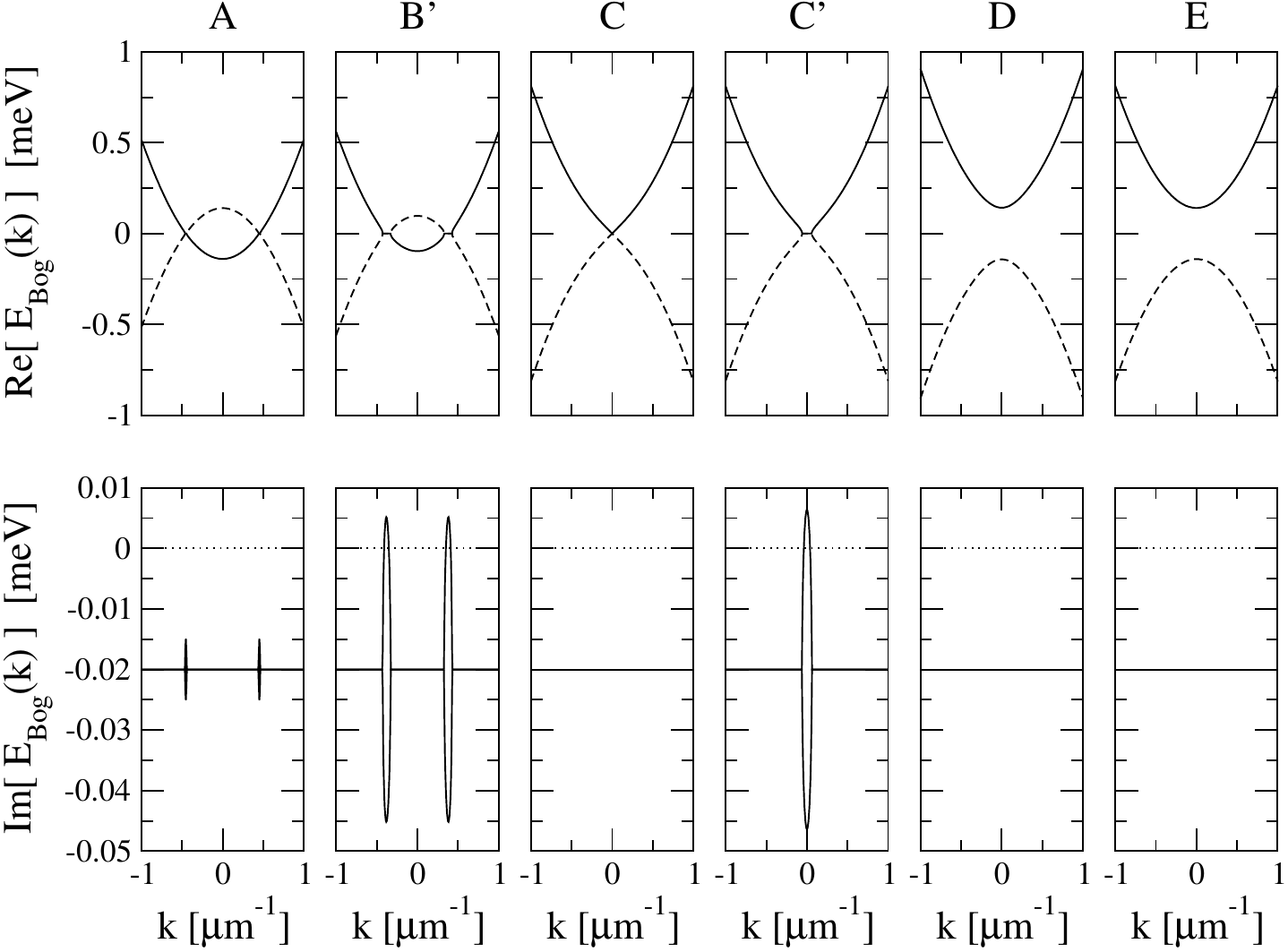}
    \caption{Left panels: plot of the equation of state of the fluid of light, expressed as fluid density as a function of pump intensity for two different incident frequencies in respectively the optical limiter regime (top) and the bistable regime (bottom). The dashed line indicates the dynamically unstable regions. Right panel: Real part (top) and imaginary part (bottom) of the excitation frequencies for the nonequilibrium Bogoliubov modes corresponding to the points indicated as A, B', C, C', D, E in the left panel. Figure adapted from~\cite{Carusotto:PRL2004}. \label{fig:coherent}}
\end{figure}

\subsection{Homogeneous fluid at rest}
\label{subsec:coh_rest}

In the simplest case of a monochromatic incident field at $\omega_{\rm inc}$ at normal incidence with a spatially homogeneous profile, we can assume that the fluid is at rest with a spatially uniform field $\psi(\rr,t)=\psi_{ss}\,e^{-i\omega_{\rm inc} t}$ and the field equation \eqref{eq:dd_psi} in the relevant case $P=\delta V=0$ gives 
\begin{equation}
\psi_{ss}=\frac{\eta E_{\rm inc}}{\omega_{\rm inc}-\omega_o-g_{\rm nl} |\psi_{ss}|^2 + i \frac{\gamma_{\rm loss}}{2}}
\label{eq:coh_ss}
\end{equation}
From this, it is immediate to derive the equation of state
\begin{equation}
\left[\left(\omega_o+g_{\rm nl}|\psi|^2-\omega_{\rm inc}\right)^2 + \frac{\gamma_{\rm loss}^2}{4} \right]\,|\psi_{ss}|^2=|\eta\,E_{\rm inc}|^2\,.
\label{eq:eos}
\end{equation}
that links in a non-linear way the density $|\psi_{ss}|^2$ to the different parameters of the incident field. For each solution of \eqref{eq:eos}, the phase of the field $\psi_{ss}$ is completely determined by the one of the incident field $E_{\rm inc}$ via \eqref{eq:coh_ss}.

The two main regimes are displayed in the left panels of Fig.\ref{fig:coherent}. Assuming for concreteness $g_{\rm nl}>0$ as in typical experiments with semiconductor devices, the top and bottom panels respectively refer to the red $\omega_{\rm inc}<\omega_o$ and blue $\omega_{\rm inc}>\omega_o$ detuning cases. In the former, the nonlinearity provides a negative feedback to the density, which grows in the sublinear way with the pumping intensity as in an {\em optical limiter} device. In the latter, the feedback is instead positive, and for sufficiently large detuning $\omega_{\rm inc}-\omega_o>\sqrt{3}\,\gamma/2$ one can observe {\em optical bistability} effects~\cite{Boyd}. This amounts to having several solutions for the density $|\psi_{ss}|^2$ for the same value of the pump intensity $|E_{\rm inc}|^2$: the upper and lower solutions are dynamically stable with respect to homogeneous $k=0$ perturbations, while the intermediate one is dynamically unstable.
Suitable temporal ramps of the pump intensity and/or frequency can be used to bring the system into the desired state. At the mean-field both lower and upper states have an infinite lifetime, but a fully quantum theory predicts that  transitions between them can be induced by quantum fluctuations~\cite{Vogel:PRA1988,casteels2017critical,rodriguez2017probing,Selvakumaran:arXiv2026}.

Linearization of the field equation \eqref{eq:dd_psi} around the stationary state \eqref{eq:coh_ss} gives the collective excitations of the fluid in a driven-dissipative generalization of the Bogoliubov theory of dilute equilibrium condensates. In the homogeneous case under consideration here, this can be formalized in the linearized field equations for the $\kk$-space field fluctuations~\cite{Carusotto:PRL2004}
\begin{multline}
 i\frac{d}{dt}
\left( \begin{array}{cc}
 \delta \psi_\kk \\
  (\delta \psi_{-\kk})^*
\end{array}
\right)= \\   
= \left(
\begin{array}{cc}
\omega_o+ \frac{\hbar k^2}{2m} + 2 g_{\rm nl} |\psi_{ss}|^2 - \omega_{\rm inc}-i \frac{\gamma_{\rm loss}}{2} & g_{\rm nl} \psi_{ss}^2 \\
- g (\psi_{ss}^{*})^2 & -\omega_o-\frac{\hbar k^2}{2m} - 2 g_{nl} |\psi_{ss}|^2 + \omega_{\rm inc}-i \frac{\gamma_{\rm loss}}{2}
\end{array} 
\right)
\left( \begin{array}{cc}
 \delta \psi_\kk \\
  (\delta \psi_{-\kk})^*
\end{array}
\right)\,.
\label{eq:linearized_coh}
\end{multline}
Straightforward algebra gives the analytic form
\begin{equation}
\omega_{\rm Bog}(\kk)= \pm \Big[\Big(\omega_o+\frac{\hbar k^2}{2m}+2 g_{\rm nl} |\psi_{ss}|^2 -\omega_{\rm inc} \Big)^2  - (g_{\rm nl} |\psi_{ss}|^2)^2\Big]^{1/2}-i\,\frac{\gamma_{\rm loss}}{2}.
\label{eq:Bogo_neq_coh}
\end{equation}
for the eigenvalues, which encode the dispersion of collective excitations as a function of wavevector $\kk$. In this equation, the $\pm$ signs correspond to the positive/negative norm branches of the Bogoliubov dispersion, also called {\em normal} and {\em ghost} branches. The finite decay rate $\gamma_{\rm loss}$ of the bare photons described by the last term in \eqref{eq:Bogo_neq_coh} directly translates into an analogous finite lifetime of the collective excitations. 
Examples of the dispersion in the most remarkable regimes are plotted in the right panels of Fig.\ref{fig:coherent}.

\begin{figure}[htbp]
    \centering
    \includegraphics[width=0.6\textwidth]{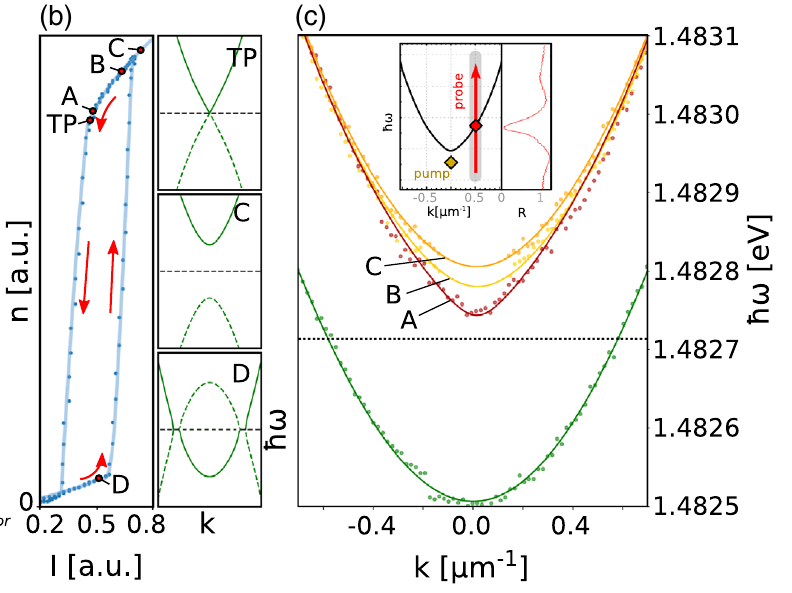}\hspace{0.05\textwidth}
    \includegraphics[width=0.95\textwidth]{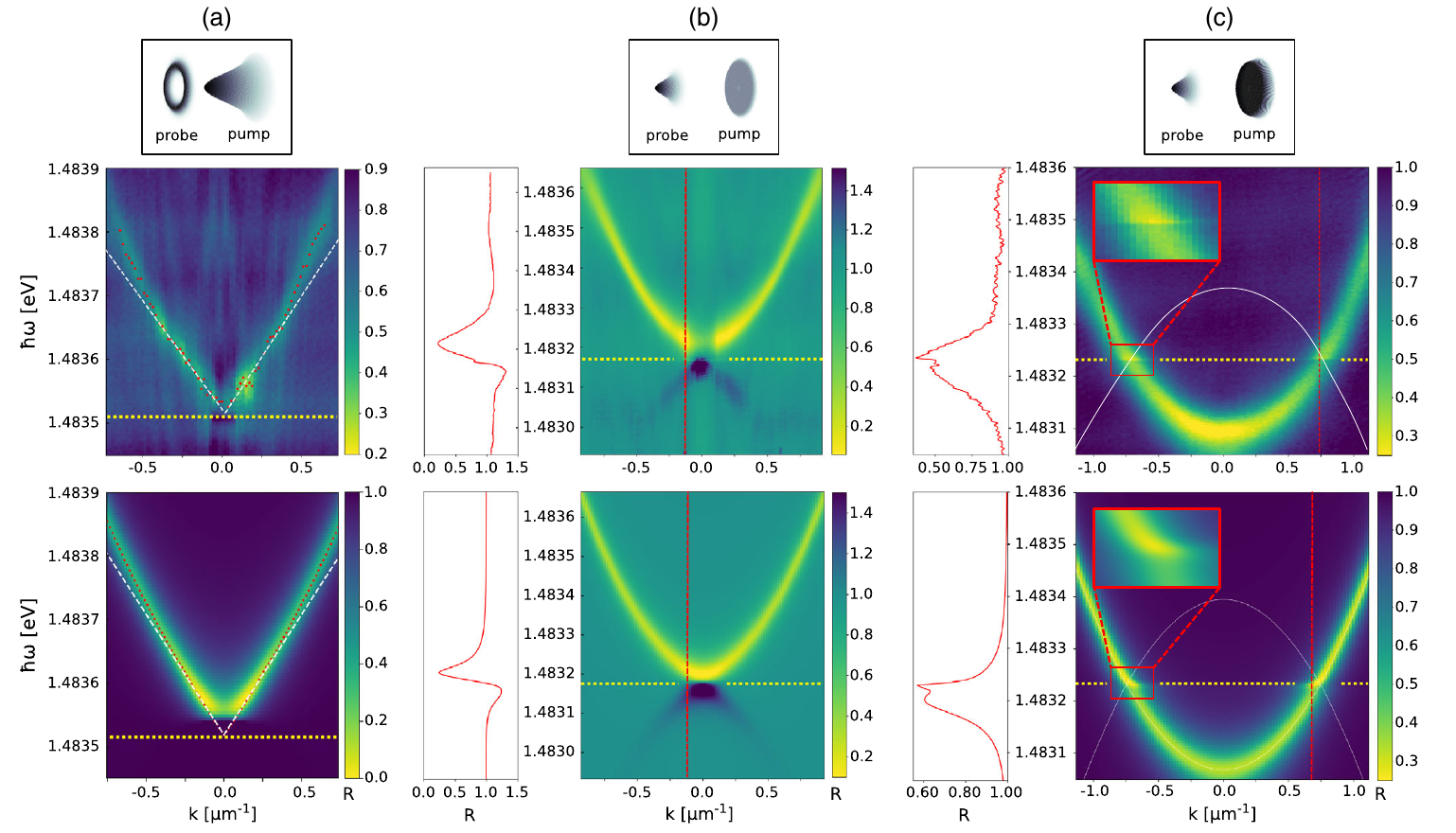}
    \caption{(Upper panel) Optical bistability in a coherently pumped fluid of light. Left: photon density as a function of the pump intensity for an blue-detuned pump frequency. Red arrows indicate the path along the hysteresis cycle during the upwards and then downwards ramp of the pump intensity. 
Right: Bogoliubov dispersion for different values of the pump intensity giving a fluid density at working points $A$, $B$, and $C$ along the bistability cycle. Points are experimental measurements and lines are theoretical fits. The black dashed
line indicates the pump energy $\hbar \omega_{\rm inc}$ and the green line is the dispersion of non-interacting polaritons in the linear regime. 
(Lower panels) Experimental measurements of the dispersion of collective excitations using a pump-and-probe technique. Within each panel, the top row shows the spatial mode profiles of the pump and of the coherent probe lasers in each configuration. 
Middle row: colorplot of the probe reflectivity as a function of the probe wavevector $k$ and frequency $\omega$ for the different pumping configurations. Bottom row: corresponding numerical simulations.
In each panel, the horizontal yellow dotted line indicates the coherent pumping frequency $\hbar\omega_{\rm inc}$. The three columns correspond to three different operating points. Left: Sonic dispersion at turning point A of the bistability loop. White dashed line is a linear dispersion fit of the speed of sound. Center: Almost sonic dispersion with a small gap for an operating point in the vicinity of the turning point A of the bistability loop. 
Right: At the low-intensity point D of the bistability loop, band-sticking is visible at $\omega=\omega_{\rm inc}$ around the crossing point 
of the normal and ghost (white curve) branches, leading to the Fano features highlighted in the inset.
Figure adapted from~\cite{Claude:PRL2022}.
\label{fig:coherent_exp}}
\end{figure}

The standard dispersion of equilibrium condensates is recovered when $\omega_{\rm inc}=\omega_o+g_{\rm nl}|\psi_{ss}|^2$, which is satisfied at point $C$ near to the turning-point of the upper branch of the bistability curve. In this case, the low-$\kk$ part of the dispersion has a linear, sonic form $\omega_{\rm Bog}(\kk)\simeq c_s k$ with the usual sound speed $c_s=\sqrt{g_{\rm nl}\,|\psi_{ss}|^2/m}$.
On the rest of the upper branch as well as in the whole optical limiting case, one has instead $\omega_{\rm inc}<\omega_o+g_{\rm nl}|\psi_{ss}|^2$ and the dispersion is gapped, as shown in the figure for points $D,E$. On the lower branch of bistability, one has instead $\omega_{\rm inc}>\omega_o+g_{\rm nl}|\psi_{ss}|^2$ and the positive and negative norm of the Bogoliubov dispersion cross at a finite $k$, leading to a band-sticking phenomenon with bubbles opening in the imaginary part of the dispersion, as visible at point B; these features are precursors of the dynamical instability that is fully developed at point B', as signaled by some modes acquiring a positive imaginary part. Finally, as typical in optical bistability, all states on the intermediate branch of the bistability loop are unstable.

A pioneering observation of the dispersion of the collective excitations on top of a coherently pumped fluid of light was reported in~\cite{stepanov2019dispersion} using a luminescence experiment of the kind reviewed in App.\ref{sec:experimental_techniques_luminescence}. A complete experimental mapping of the different regimes was carried out in~\cite{Claude:PRL2022,Claude:PRB2023} using a high-resolution angle-resolved coherent probe spectroscopy technique 
and is summarized in Fig.\ref{fig:coherent_exp}. For a pump frequency $\omega_{\rm inc}>\omega_o$, a clear bistability loop is observed when the pump intensity is scanned upwards and then downwards (upper-left panel). The collective excitations are then extracted from a measurement of the reflectivity of a weak probe beam. For each value of the probe wavevector $\kk$, the reflectivity shows a resonant feature when the probe frequency approaches a collective excitation mode (lower panels). Close to the turning point of the bistability loop, the dispersion has a sonic shape with an almost gapless and linear shape (bottom left panel). While positive-norm Bogoliubov branches typically correspond to a drop in reflectivity compensating for the transmitted light, the negative-norm branches correspond to enhanced reflectivity via a four-wave-mixing-induced amplification process: such {\em ghost branches} are visible near to the turning point and are well-resolved in flat-top fluids (bottom center panel). More complex Fano-like features are observed when the positive- and negative-norm branches cross and a band-sticking effect arises as a precursor of dynamical instability  (bottom right panel).

\begin{figure}
    \centering
    \includegraphics[width=0.8\textwidth]{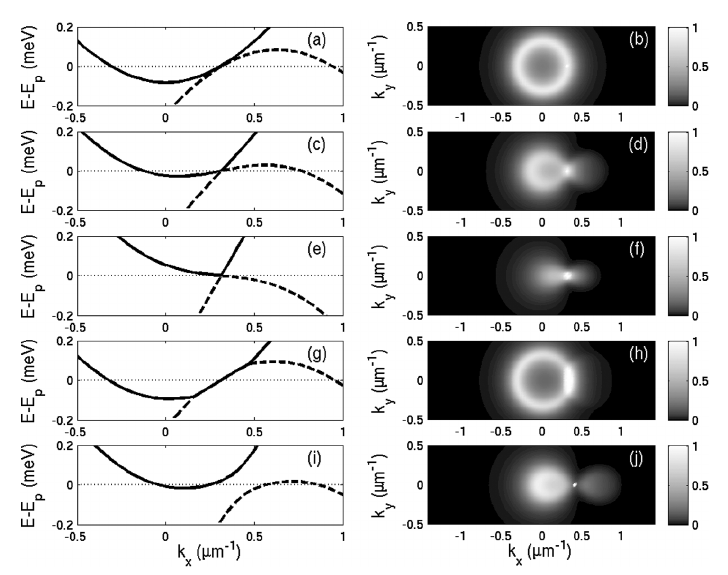}
    \caption{
    Left panels: theoretical prediction for the dispersion of collective excitations in a coherently pumped fluid of light flowing in the rightward direction. Right panels: $\kk$-space intensity distribution of the Rayleigh scattered light by a weak defect at rest.    
    The different rows correspond to different pumping configurations, as described in Figs.\ref{fig:coherent} and \ref{fig:coherent_exp}: the three first rows (a-b,c-d,e-f) correspond to the turning point of the bistability loop for growing values of the density. The fourth row (g-h) is for a band-sticking regime, precursor of dynamical instability. The fifth row (i-j) is for a gapped dispersion.
Figure adapted from~\cite{Carusotto:PRL2004}.
\label{fig:coherent_sf_th}}
\end{figure}

\begin{figure}
    \centering
    \includegraphics[width=0.6\textwidth]{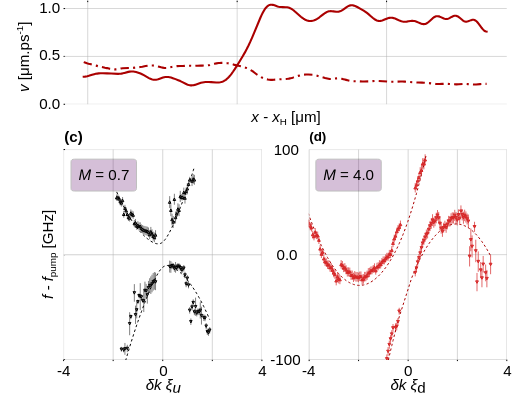}
    \caption{Dispersion of the collective excitations at different spatial positions in a non-homogeneously flowing fluid. The top panel shows the spatial profiles of the speed of sound (dot-dashed) and of the speed of flow (solid). The crossing point of the two lines gives a black-hole-like sonic horizon. The bottom panels show the dispersion in the sub-sonic $M=v/c_s=0.7$ upstream (left) and super-sonic $M=v/c_s=4$ downstream (right) regions as it was experimentally measured using the pump-and-probe technique discussed in App.\ref{sec:experimental_techniques_PP}. The wavevector $\delta k$ on the horizontal axis of the lower panels corresponds to $\qq$ in the text. Figure adapted from~\cite{Falque:PRL2025}.
\label{fig:coherent_sf_exp3}}
\end{figure}

\subsection{Collective excitations in flowing fluids}
\label{subsec:coh_flow}

Moving fluids are straightforwardly obtained by tuning the incidence angle $\theta_{\rm inc}$ of the coherent pump beam. From conservation of the in-plane wavevector, one has that $k_{\rm inc}=\frac{\omega}{c}\,\sin(\theta_{\rm inc})$. For our parabolic photon dispersion, this gives a flow speed $v=\hbar k_{\rm inc}/m$. The stationary-state equations \eqref{eq:coh_ss} and \eqref{eq:eos} keep an almost identical form except for the bare photon frequency that acquires the form $\omega_o+\hbar k_{\rm inc}^2/2m$. On the other hand, the dispersion of collective excitations 
\begin{equation}
\omega_{\rm Bog}(\delta\kk)= \vv \cdot \delta \kk \pm \Big[\Big(\omega_o+\frac{\hbar\, \delta k^2}{2m}+2 g_{\rm nl} |\psi_{ss}|^2 -\omega_{\rm inc} \Big)^2  - (g_{\rm nl} |\psi_{ss}|^2)^2\Big]^{1/2}-i\,\frac{\gamma_{\rm loss}}{2}
\label{eq:Bogo_neq_coh_v}
\end{equation}
exhibits an additional Doppler shift term proportional to the flow speed $\vv$ which physically describes the dragging of the collective excitations by the underlying fluid. Note that in this formula, the wavevector $\delta\kk$ corresponds to the wavevector of the excitation on top of the coherent fluids; in transmission/reflection experiments, this quantity corresponds to the difference between the experimental wavevector $\kk$ and the coherent fluid wavevector at $\kk_{\rm inc}$.

Examples of the dispersion of collective excitations in different cases are shown in the left panels of Fig.\ref{fig:coherent_sf_th} for a fixed value of the pump wavevector $\kk_{\rm inc}$ and, thus, of the flow speed $\vv$. Panel (g) describes the band-sticking case and panel (i) describes a gapped dispersion. 
Panels (a,c,e) illustrate the sonic case for different values of the density $|\psi_{ss}|^2$ and, thus, of the sound speed: 
changing the ratio $v/c_s$ across $1$ has a strong qualitative impact on the shape of the modes as we are now going to see.

An experimental measurement~\cite{Falque:PRL2025} of these features is summarized in Fig.\ref{fig:coherent_sf_exp3}. Using a monochromatic coherent pump beam with a structured phase profile, a complex flow geometry was generated where two almost homogeneous regions are separated by a sharp transition (upper panel).
Within each individual region, the homogeneity of the flow profile allows for a precise spectroscopic measurement of the collective excitations. The result of such spatially-resolved measurements is reported in the lower panels.
On both sides, a small but finite energy gap in the dispersion remains visible: a slight finite distance from the turning point was needed to keep the fluid configuration stable.

In the upstream region on the left, the flow has a subsonic character, the flow speed being slower than the sound speed. As such, the Doppler shift is moderate and the positive/negative-wavevector sound modes keep propagating in their standard right/left directions with positive/negative group velocities. Correspondingly, positive (negative) norm Bogoliubov modes retain a purely positive (negative) energy. 

On the other hand, in the downstream region on the right, the flow has a supersonic character, the flow speed being faster than the sound speed. Here, the Doppler shift is so large that negative-wavevector sound mode is dragged by the flow into the downstream direction and, at low-$k$, end up displaying a positive group velocity. This feature is reflected in the positive norm mode getting pulled to negative energies, with crucial consequences on the superfluid properties as we are going to discuss in Sec.\ref{sec:superfluidity}. Note that this {\em energetic} instability (i.e. the presence of negative-energy, positive-norm modes) is not necessarily associated to a {\em dynamical} instability (signaled by a positive imaginary part of the frequency of some modes). 

As it was first explored in~\cite{Solnyshkov:PRB2011,Gerace:PRB2012,Nguyen:PRL2015}, such configurations displaying a transition from an upstream subsonic flow to a downstream supersonic flow separated by a sharp transition can be seen as optical analogs of black holes in gravitational physics: in the sub-sonic region long-wavelength collective excitations are able to propagate in both directions, but in the downstream region they are doomed to be dragged away by the supersonic flow. The transition between the two regions behaves as a horizon-like no-return point. Upon quantization of the collective sound modes, the presence of both positive and negative norm modes at the same frequency is the basic ingredient required for the onset of Hawking emission processes at the horizon~\cite{recati2009bogoliubov,barcelo2011analogue,delhom2025analogue}. Recent studies of analog Hawking physics in the specific case of photon fluids can be found in~\cite{Grisins:PRA2016,Falque:PRL2025,Guerrero:PRL2025,Delhom:PRD2024}.

\subsection{Additional features}
\label{subsec:additional}

After having summarized the main general features of the collective excitations in fluids of light under coherent pumping, it is useful to conclude the Section with a brief account of a few specific additional features.

\subsubsection{Dispersion of collective excitations outside the pump spot}
All the discussion in the previous subsections focused on geometries where the fluid of light is spatially homogeneous in the region of interest. This is the case when the generation and the detection of the collective excitations are restricted to a region where a spatially uniform coherent pump is present to support the fluid of light. 

For spatially localized pumps, most of the density is concentrated under the pump spot, but a sizable density of photons is also found outside the spot as a consequence of the ballistic flow of photons outside the pump. This flow extends over a characteristic spatial decay length $v_g/\gamma_{\rm loss}$ determined by the in-plane speed $v_g$ of the photons in the $\kk$ modes that are resonantly selected by the pump frequency~\cite{carusotto2013quantum}. As in the region outside the pump spot the phase of the field is not locked to the coherent pump and is able to freely evolve, the collective excitations recover the standard sonic dispersion \eqref{eq:Bogo_standard} of equilibrium systems~\cite{Amelio:PRB2020}.  

In the presence of external potentials, the ballistic flow may display complex features, including horizons separating sub- from super-sonic flow regions, so that the gapless nature of the collective excitations can facilitate the observation of Hawking features~\cite{Nguyen:PRL2015,Grisins:PRA2016}.

\subsubsection{Effect of an incoherent reservoir}
\label{sec:coh_pump_res}
\begin{figure}
    \centering
    \includegraphics[width=0.4\textwidth]{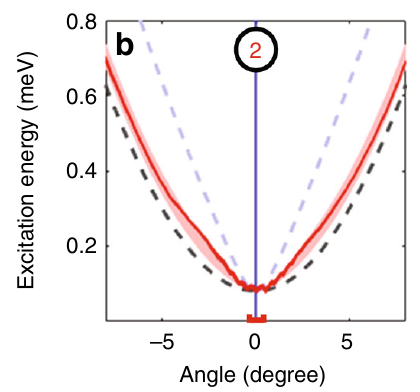}
    \caption{Experimental measurement of the effect of the incoherent reservoir on the collective excitations in a coherently pumped fluid of light. 
    The solid red curve shows the experimental measurement of the dispersion of collective excitations extracted from the luminescence spectrum as discussed in App.\ref{sec:experimental_techniques_luminescence}. The wide pink line is the theoretical fitting including the effect of the incoherent reservoir. The black dashed line shows the rigidly shifted bare photon dispersion, while the light blue dashed line is the prediction of the naive Bogoliubov dispersion \eqref{eq:Bogo_neq_coh} without including the incoherent reservoir.
    Figure adapted from~\cite{stepanov2019dispersion}.
\label{fig:exp_res}}
\end{figure}

While the theory discussed above reproduces in a qualitatively accurate way the observations, a quantitative agreement requires including the effect of the incoherent reservoir of material excitations that is generated in the cavity material by spurious absorption processes. 

At a simplest level of approximation, the effect of the reservoir can be described by combining a rate equation for the reservoir density $n_R$
\begin{equation}
\partial_t n_R= -\gamma_R n_R +\gamma_{\rm res}\,|\psi|^2\end{equation}
to the field equation for the coherently pumped fluid of light
\begin{equation}
 i\frac{\partial \psi}{\partial t}=\omega_o\,\psi -\frac{\hbar \nabla^2\psi}{2m} + g_{\rm nl}\,|\psi|^2\,\psi  + g_R n_R \psi \\ -\frac{i}{2}(\gamma_{\rm loss} + \gamma_{\rm res})\,\psi+ \eta \, E_{\rm inc}(\rr,t)\,.
 \label{eq:dd_psi_res}
\end{equation}
This latter now includes a, typically repulsive $g_R>0$, reservoir interaction term and an additional loss channel into the incoherent reservoir at rate $\gamma_{\rm res}$. Both $\gamma_{\rm res}$ and the reservoir decay rate $\gamma_R$ are typically much slower than the photon loss rate $\gamma_{\rm loss}$.

Within this model~\cite{Amelio:PRR2020}, the photons experience an overall blue shift $g_{\rm nl}\,|\psi_{ss}|^2+g_R n_R$ that combines their intrinsic nonlinearity and interactions with the reservoir and determines the potential landscape felt by the fluid in its stationary state. On the other hand, sound propagation typically occurs on time-scales much faster than $\gamma_{R, {\rm res}}$ and, thus, only experiences the the intrinsic nonlinearity. As a result, the sound speed, as it is observed right at the turning point of the bistability loop, only includes the intrinsic nonlinearity, $c_s=\sqrt{g_{\rm nl}|\psi_{ss}|^2/m}$, and is therefore significantly reduced. A direct experimental observation of the effect of the reservoir is illustrated in Fig.\ref{fig:exp_res}. Analogous precision measurements of the dispersion of collective excitations have offered detailed information on the microscopic process determining the interaction constant $g_{\rm nl}$ between photons~\cite{Richard:PRR2026}.

The effect of the reservoir is most important in the case of semiconductor microcavities in the strong light-matter coupling regime where photons are hybridized with matter excitations of excitonic nature and acquire a mixed character of exciton-polaritons~\cite{kavokin2011microcavities}. Its magnitude markedly changes from one experiment to another, with the tendency of being stronger in devices including a larger number of quantum wells in the cavity layer.

\section{Non-equilibrium Bose-Einstein condensed fluids of light}
\label{sec:incoh_pump}

In the previous Section, we have focused on the case of coherently pumped fluids, where the overall phase of the cavity field amplitude is locked to the incident field one. The situation is quite different when the dynamics obeys the $U(1)$ symmetry associated to phase rotations, i.e. it is invariant under a transformation $\Psih \to \Psih\,e^{i\varphi}$ of the order parameter with arbitrary $\varphi$. In the condensed state~\cite{bloch2022non}, a generalized Goldstone theorem then imposes the presence of a soft mode among the collective excitations, whose (complex valued) frequency tends to zero in the long-wavelength limit, $\omega_G(k\to 0)=0$. 
This Section is devoted to a review of the theory and of the first experiments investigating the collective excitations in this regime.

All along this Section, an intuitive picture on the condensation process can be obtained by keeping in mind the analogy with an easy-plane ferromagnetic transition. As the material develops a magnetization along a random direction on a plane, the spontaneously broken symmetry is a SO(2) rotational one, that is mathematically equivalent to the U(1) that is spontaneously broken in the condensation process. In agreement with the Goldstone theorem, the dispersion of the collective excitations in the ferromagnetic state is gapless and has a form
\begin{equation}
\omega(k)=c\,k^2\,.
\end{equation}
This gapless dispersion is to be contrasted to the gapped dispersion
\begin{equation}
\omega(k)=a+c\,k^2\,.
\end{equation}
of a ferromagnet subject to an external magnetic field that pins the direction of magnetization and, thus, explicitly breaks the SO(2) rotational symmetry.

\subsection{Generic theoretical model}
\label{sec:diffusive_Goldstone}

In order to understand the basic features of the theory, it is useful to first focus our discussion on a simplest model based on the field equation \eqref{eq:dd_psi}. In the context of fluids of light, models of this kind were first introduced in~\cite{Wouters:PRL2007} but they share close analogies with classical equations of nonlinear physics and pattern formation, in particular the complex Ginzburg-Landau equation (CGLE)~\cite{Cross:RMP1993,Aranson:RMP2002}.

Thanks for its generic nature, this simple model can be used to describe non-equilibrium Bose-Einstein condensation phenomena in a variety of specific platforms, from standard lasers to polariton condensates. We refer to the dedicated review~\cite{bloch2022non} for a complete account of the different physical systems where non-equilibrium BEC physics is observed and of the more sophisticated models that can be used to capture the specific features of each of them.

\subsubsection{Phase transition in the stationary state}

The dynamics of a incoherently pumped fluid of light is described by the field equation \eqref{eq:dd_psi} setting the coherent pump to zero $E_{\rm inc}=0$. 
For simplicity, we furthermore restrict our attention to spatially uniform geometries.

For a weak incoherent pump $P<\gamma_{\rm loss}$, the only stationary-state has a vanishing mean-field value $\psi=0$. The collective excitations around this trivial solution are straightforwardly obtained by linearizing \eqref{eq:dd_psi} around $\psi=0$. This gives a quadratic dispersion of the form
\begin{equation}
\omega(k)=\frac{\hbar k^2}{2m}-\frac{i}{2}(\gamma_{\rm loss}-P)\,:
\label{eq:Bogo_neq_below}
\end{equation}
For low pump intensities $P<\gamma_{\rm loss}$ the empty state is dynamically stable with a finite imaginary gap in the dispersion. As the pump intensity is increased, the pumping rate compensates more and more the losses and imaginary gap correspondingly decreases. For $P\to \gamma_{\rm loss}^-$ the $\psi=0$ solution eventually loses its dynamical stability via a Hopf bifurcation mechanism~\cite{Cross:RMP1993}, leaving space to additional finite-intensity solutions corresponding to the condensed state. 

For  $P> \gamma_{\rm loss}$, the field equation \eqref{eq:dd_psi} admits spatially-uniform stationary-state solutions in the form 
\begin{equation}
\psi(\rr,t)=\psi_{ss}\,e^{-i \omega_{ss} t} = \sqrt{n_s\,\left(\frac{P}{\gamma_{\rm loss}}-1\right)}\,e^{i \varphi }\,e^{-i \omega_{ss} t}
\label{eq:incoh_ss}
\end{equation}
where the density 
\begin{equation}
|\psi_{ss}|^2=n_s\,\left(\frac{P}{\gamma_{\rm loss}}-1\right) \label{eq:ss_dens}
\end{equation}
and the oscillation frequency 
\begin{equation}
\omega_{ss}=\omega_o + g_{\rm nl}\,|\psi_{ss}|^2
\label{eq:ss_omega}
\end{equation}
are univocally fixed by the stationarity condition for the motion equation \eqref{eq:dd_psi}. On the other hand, the phase $\varphi$ remains arbitrary and is randomly chosen at every instance of an experiment as in typical $U(1)$ symmetry-breaking phenomena.


\subsubsection{Collective excitations and diffusive Goldstone mode}

A prediction for the dispersion of collective excitations of the non-equilibrium condensate  can be straightforwardly obtained by linearizing the field equation \eqref{eq:dd_psi} around the stationary state \eqref{eq:incoh_ss}. In the spatially homogeneous geometry under consideration here, they have the form
\begin{multline}
 i\frac{d}{dt}
\left( \begin{array}{cc}
 \delta \psi_\kk \\
  (\delta \psi_{-\kk})^*
\end{array}
\right)= \\   
= \left(
\begin{array}{cc}
\frac{\hbar k^2}{2m} + g_{\rm nl} |\psi_{ss}|^2 -i \frac{P}{2n_s} \frac{|\psi_{ss}|^2}{(1+|\psi_{ss}|^2/n_s)^2}    & g_{\rm nl} \psi_{ss}^2 -i \frac{P}{2n_s} \frac{\psi_{ss}^2}{(1+|\psi_{ss}|^2/n_s)^2} \\
-g_{\rm nl} (\psi_{ss}^*)^2 -i \frac{P}{2n_s} \frac{(\psi_{ss}^{*})^2}{(1+|\psi_{ss}|^2/n_s)^2}  & -\frac{\hbar k^2}{2m} - g_{\rm nl} |\psi_{ss}|^2 -i \frac{P}{2n_s} \frac{|\psi_{ss}|^2}{(1+|\psi_{ss}|^2/n_s)^2}   
\end{array} 
\right)
\left( \begin{array}{cc}
 \delta \psi_\kk \\
  (\delta \psi_{-\kk})^*
\end{array}
\right)\,.
\label{eq:linearized_incoh}
\end{multline} 
The dispersion of collective excitations can be then obtained by diagonalization of these equations, which gives
\begin{equation}
\omega_{\rm Bog}(\kk)= -\frac{i\Gamma}{2} \pm \sqrt{[\omega^{\rm (eq)}_{\rm Bog}(\kk)]^2-\frac{\Gamma^2}{4}}
\label{eq:Bogo_neq_incoh}
\end{equation}
where 
\begin{equation}
\omega^{\rm (eq)}_{\rm Bog}(\kk)=\sqrt{\frac{\hbar k^2}{2m}\left(\frac{\hbar k^2}{2m}+2g_{\rm nl} |\psi|^2 \right)}
\end{equation}
is the standard Bogoliubov dispersion \eqref{eq:Bogo_standard} of equilibrium condensates and
\begin{equation}
\Gamma=\gamma_{\rm loss}\frac{P-\gamma_{\rm loss}}{P}
\label{eq:Gamma_eff}
\end{equation}
is an effective damping rate around the steady-state whose physical interpretation will be clear soon.

By selecting the $+$ sign, the dispersion \eqref{eq:Bogo_neq_incoh} satisfies as expected the Goldstone theorem and displays a soft Goldstone branch with $\omega_{\rm Bog}(k\to 0)=0$ in both real and imaginary parts. In contrast to the propagating sound mode of Bose-Einstein condensates at thermal equilibrium, the low-$k$ part of our Goldstone mode dispersion has a diffusive form 
\begin{equation}
    \omega_G(k)\simeq -i \alpha_2 k^2\,.
    \label{eq:diffG}
\end{equation}
with a flat and vanishing real part and a quadratically\footnote{A different power-law for the $k$-dependence of the decay rate at very low-$k$ values was predicted by more sophisticated theories~\cite{Ji:PRA2015} including thermal and quantum fluctuations around the mean-field steady-state and their nonlinear interaction within noisy Kuramoto-Sivashinsky or Kardar-Parisi-Zhang models.} growing imaginary part with a diffusion coefficient 
    $\alpha_2= {c_s^2}/{\Gamma}$. 
Physically, the dispersion \eqref{eq:diffG} describes the diffusive relaxation of slow twists of the condensate phase, i.e. of the order parameter, 
\begin{equation}
\frac{\partial \varphi}{\partial t}= \alpha_2 \nabla^2 \varphi\,.
\end{equation}
In the absence of interparticle interactions $g_{\rm nl}=0$, the diffusion coefficient $\alpha_2=0$ and the imaginary part has an even slower quartic growth $\omega_G(k)\simeq -i \alpha_4 k^4$ with a real and positive quartic diffusion coefficient $\alpha_4=\hbar^2/(4m^2\Gamma)$. While in this Section we restrict to a specific model of non-equilibrium condensation, it is important to keep in mind that the diffusive nature of the Goldstone mode is a general effect that was simultaneously predicted for several distinct models~\cite{Wouters:PRB2006,Szymanska:PRL2006,Wouters:PRA2007,Wouters:PRL2007}.

In the usual Mexican hat picture of phase transitions, the Goldstone mode corresponds to the dynamics along the rim, while the dynamics in the radial direction corresponding to the condensate density gives the Higgs mode~\cite{SachdevBook}. In our model, this gapped Higgs-like density mode is described by selecting the $-$ sign in \eqref{eq:Bogo_neq_incoh}:  differently from the equilibrium case, its gap is here in its imaginary part $\omega_A(k\to 0)=-i\Gamma$ and is associated to an overdamped relaxation dynamics of the condensate density at the rate $\Gamma$ defined in \eqref{eq:Gamma_eff}. This latter vanishes as the transition point is approached $P\to \gamma^+_{\rm loss}$ in the typical critical slowing-down effect of second-order phase transitions. On the other hand, at large $k$, the dispersion \eqref{eq:Bogo_neq_incoh} recovers the standard equilibrium Bogoliubov dispersion $\omega_{\rm Bog}^{\rm (eq)}(\kk)$ plus an overall damping $\Gamma$. 
The two behaviours are separated by exceptional points at the $k$ value for which $\omega^{\rm (eq)}_{\rm Bog}(k)=\Gamma/2$: here the argument of the square-root in \eqref{eq:Bogo_neq_incoh} is zero, giving a singularity in the dispersion. 

Besides the rich features of condensation in spatially finite geometries~\cite{Richard:PRL2005,Wouters:PRB2008}, a growing interest is being devoted to condensates in the chiral edge modes of a topological system, the so-called topological lasers~\cite{Bahari:Science2017,Harari:Science2018,Bandres:Science2018,Price:JPhysPhot2022}. Calculations of the collective excitation modes in these systems were reported in~\cite{amelio:PRX2020,Zapletal:Optica2020,Loirette:PRA2021}: in spatially homogeneous geometries, the main new feature beyond \eqref{eq:Bogo_neq_incoh} is a Doppler tilting of the dispersion analogous to \eqref{eq:Bogo_neq_coh_v}. The physics gets more intriguing in spatially-finite geometries where the Hermitian and non-Hermitian parts of the Bogoliubov matrix have a non-trivial interplay~\cite{Amelio:PRA2022} and connections can be drawn to the physics of non-Hermitian topological systems such as, e.g., the Hatano-Nelson model~\cite{Brunelli:SciPostPhys2023}.

As a final point, it is important to stress that our discussion based on \eqref{eq:dd_psi} has assumed to be in the so-called good-cavity limit where the internal dynamics of the incoherent pump process is much faster that the one of the in-cavity field $\psi$ and can be therefore be adiabatically eliminated. In the language of Sec.\ref{sec:coh_pump_res}, this condition corresponds to having $\gamma_R\gg \gamma_{\rm loss}$.

When this condition is broken, a much richer physics can be observed, where the dynamics of the cavity field dynamics interplays with the one of the amplifying bath. As a result, the collective excitations show complex dispersions and, in some case, may also develop secondary instabilities and correspondingly noisy emission spectra~\cite{Baboux:Optica2018}.


\begin{figure}[tbp]
    \centering \includegraphics[width=0.8\columnwidth,angle=0,clip]{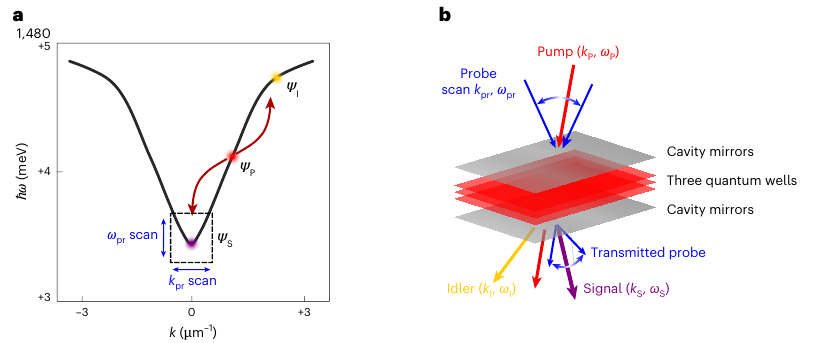}\vspace{0.5cm}
    \centering \includegraphics[width=0.7\columnwidth,angle=0,clip]{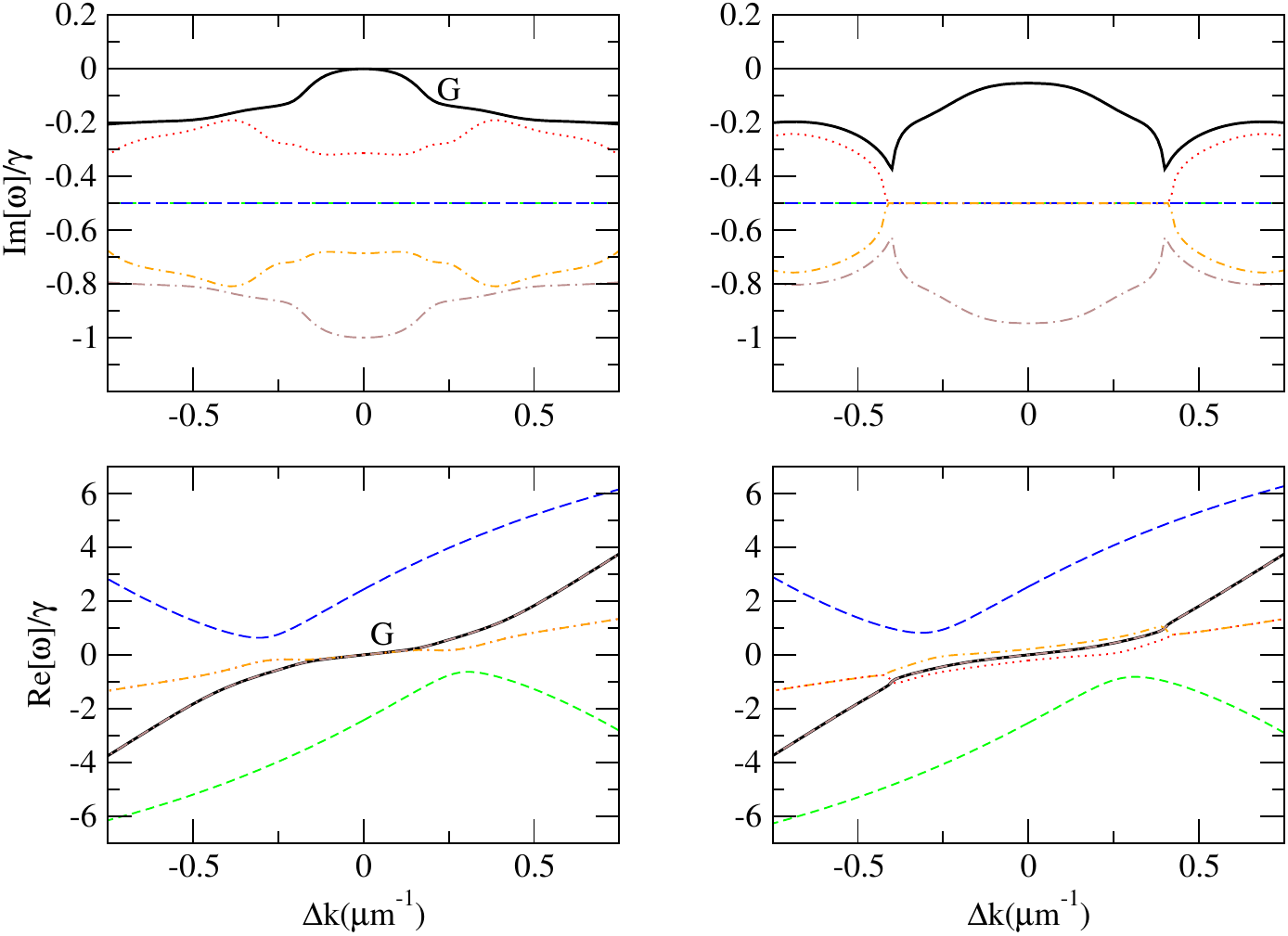}
    \caption{ Top row of panels: sketch of the polariton dispersion at linear regime and of the scattering process underling optical parametric oscillation (left); of the semiconductor planar cavity device and the optical beams involved in the process. A pair of photons from the externally pumped $p$ mode are converted into a pair of signal $s$ and idler $i$ photons that form coherent output beams. An additional probe beam tuned next to the signal mode is used to measure the collective excitation spectrum. Figure from~\cite{claude2025observation}.
    Lower group of panels: imaginary (upper panels) and real (lower panels) part of the numerically calculated dispersion of the collective excitation modes on top of a non-equilibrium condensate in a spatially extended optical parametric oscillator configuration. The wavevector $\Delta k$ is measured from the condensate one, corresponding here to the signal wavevector $\kk_s$. 
    The left panels refer to the case in which the signal/idler phase rotation symmetry is spontaneously broken and a Goldstone mode $G$ is present (heavy black line).  The right panels refer to the case in which a weak phase-locking beam is applied, so that the $U(1)$ symmetry is explicitely broken and a gap opens in the imaginary part of the dispersion relation.
    Figure from~\cite{Wouters:PRA2007}. 
\label{fig:Goldstone_OPO_th}}
\end{figure}

\begin{figure}[htbp]
    \centering
    \includegraphics[width=0.95\textwidth]{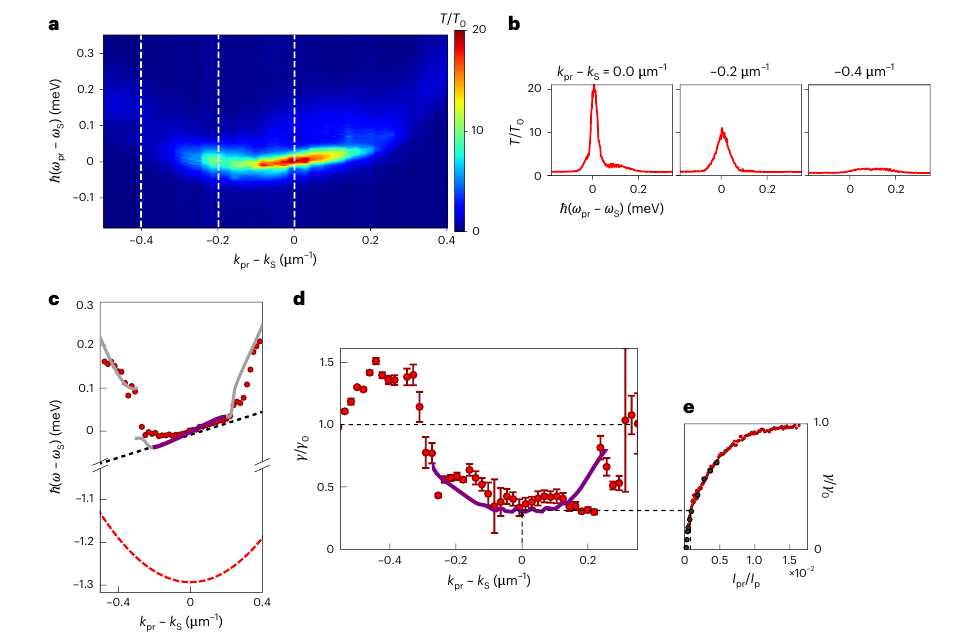}
    \caption{Experimental observation of the diffusive Goldstone mode of a non-equilibrium condensate. Panel (a): colorplot of the probe transmission as a function of the wavevector and frequency of the probe beam measured from the condensate ones. Cuts along the white dashed lines at different wavevectors are shown in panel (b). Real (c) and imaginary (d) parts of the dispersion of the collective excitations extracted by fitting the frequency-cuts of the colorplot (a) at different $k$ values. Panel (e) illustrates the physical origin of the finite linewidth as an effect of the finite probe intensity. Figure adapted from~\cite{claude2025observation}.    
\label{fig:Goldstone_OPO_exp}}
\end{figure}

\subsection{The Goldstone mode of parametric oscillators: theory and experiments}
\label{sec:expt_OPO}

While it is simplest derived for the field dynamics of \eqref{eq:dd_psi}, the diffusive nature of the Goldstone mode of non-equilibrium phase transitions is a general result that goes far beyond this specific model of non-equilibrium condensation. Actually, its first prediction was put forward for a different example of non-equilibrium phase transition in a spatially extended optical parametric oscillator based on exciton-polaritons~\cite{Wouters:PRB2006,Wouters:PRA2007} and, almost simultaneously, found also in fermionic models~\cite{Szymanska:PRL2006}.

As it is sketched in the top panel of Fig.\ref{fig:Goldstone_OPO_th}, the field dynamics of optical parametric oscillators involves three field components $\Psi_{s,p,i}$ centered around the pump, signal, idler wavevectors $\kk_{p,s,i}$, on which the field $\Psi$ can be expanded 
\begin{equation}
\psi(\rr,t)= \Psi_{s}\,e^{i\kk_s\rr} e^{-i\Omega_s t} + \Psi_{p}\,e^{i\kk_p\rr} e^{-i\Omega_p t}+ \Psi_{i}\,e^{i\kk_i\rr} e^{-i\Omega_i t}\,.
\end{equation}
The pump mode $p$ is coherently driven by the external laser field of amplitude $E_{\rm inc}$ that fixes its phase and its frequency $\Omega_p=\omega_{\rm inc}$. The amplification mechanism for the signal $s$ and idler $i$ modes is provided by resonant parametric scattering processes that convert two pump polaritons into a pair of polaritons in respectively the signal and idler modes: the specific polariton dispersion is instrumental to be able to simultaneously satisfy the momentum $\kk_s+\kk_i=2\kk_p$ and energy $\Omega_s+\Omega_i=2\Omega_p$ conservation conditions when the pump is tuned at the so-called magic angle. An additional weak probe $pr$ beam is then used to probe the collective excitations of the system.
 
An explicit expression for the complete equations of motion of the slowly-varying field components $\Psi_{s,p,i}(\rr,t)$ can be found in~\cite{Wouters:PRA2007,claude2025observation}.
At the steady-state, the fields satisfy the system of equations,
\begin{eqnarray}
\Omega_p \Psi_p &=&%
\left[ \omega_p-i\frac{\gamma_{\rm loss}}{2}\right] \Psi_p+
g_{\rm nl}\,\left[\left( \vert \Psi_p\vert ^{2}+2\vert \Psi_s\vert ^{2}+2\vert \Psi_i\vert
^{2}\right) \Psi_p+2\Psi_p^{\ast
}\Psi_s\Psi_i\right]+\eta E_{\rm inc} \label{eq:OPO_p}
\\
\Omega_s \Psi_s &=&
\left[ \omega_s-i\frac{\gamma_{\rm loss}}{2}\right] \Psi_s+
g_{\rm nl}\,\left[\left( 2\vert
\Psi_p\vert ^{2}+\vert \Psi_s\vert ^{2}+2\vert \Psi_i\vert^{2}\right) \Psi_s+\Psi_p^{2}\Psi_i^{\ast }\right] 
\label{eq:OPO_s}\\
\Omega_i \Psi_i&=&
\left[ \omega_i-i\frac{\gamma_{\rm loss}}{2}\right] \Psi_i+
g_{\rm nl}\,\left[\left( 
2\vert \Psi_p\vert ^{2}+2\vert \Psi_s\vert ^{2}+\vert
\Psi_i\vert ^{2}\right) \Psi_i+\Psi_p^{2}\Psi_s^{\ast }\right],
\label{eq:OPO_i}
\end{eqnarray}
where $\omega_{p,s,i}=\omega_o+ \hbar k_{p,s,i}^2/2m$ are the bare frequencies of the photon modes at the pump, signal, idler wavevectors and $\Omega_{p,s,i}$ are the frequencies of the running optical parametric oscillator, typically close but not exactly equal to the bare frequencies $\omega_{p,s,i}$. While the coherent pump forces the pump mode to oscillate at $\Omega_p=\omega_{\rm inc}$, the signal and idler ones are dynamically chosen by the steady-state equations.
As a key feature, these equations are invariant under  a $U(1)$  symmetry
\begin{equation}
\Psi_{s,i} \to \Psi_{s,i} e^{\pm i \varphi}\,; \;\;\; \Psi_{p} \to \Psi_{p}
\label{eq:psi_OPO}
\end{equation}
where the signal and idler beams acquire equal and opposite phases and the pump one stays fixed, locked to the incident beam amplitude $E_{\rm inc}$. Under this symmetry, the field configuration \eqref{eq:psi_OPO} experiences a global spatial shift. Like in the simpler model discussed above, by the Goldstone theorem, such a continuous $U(1)$ symmetry implies the presence of a soft Goldstone mode in the dispersion of collective excitations.

The collective excitation modes of the running optical parametric oscillator are then obtained by linearizing the field equations for $\Psi_{p,s,i}(\rr,t)$ around the steady-state, allowing for a slow variation in space and time. As compared to the $2\times 2$ matrix involved in \eqref{eq:linearized_incoh}, this now requires finding the eigenvalues of a $6\times 6$ matrix. An example of calculation is illustrated in the bottom left panels of Fig.\ref{fig:Goldstone_OPO_th}. The Goldstone mode is highlighted in black: as the condensation process occurs at finite $\kk$, a finite drift velocity is visible on top of the diffusive behaviour. In addition to the gapped branch corresponding to intensity fluctuations, the other collective excitation modes result from a complex interplay of sound modes in the different $p,s,i$ field modes involved in the process~\cite{Berceanu:PRB2015}.

After a number of pioneering steps~\cite{assmann2011polariton,nakayama2017observation,ballarini_observation_2009,ballarini_directional_2020}, a comprehensive experimental investigation of this physics was recently reported in~\cite{claude2025observation} and the main observations are summarized in Fig.\ref{fig:Goldstone_OPO_exp}. The dispersion is extracted from a frequency- and angle-selective measurement of the cavity transmission [panel (a)] carried out in close analogy to the one for the coherent pump case discussed in Sec.\ref{sec:coh_pump}. The real and imaginary components of the frequency of the main peak are extracted with a Lorentzian fitting of the cuts along the frequency direction [some examples of such spectra are shown in panel (b)]. 
The real part [panel (c)] displays a clear diffusive plateau tilted according to the drift velocity. The imaginary part [panel (d)] displays a marked narrowing around the condensate wavevector by a significant factor of at least $3$. 

The fact that the linewidth does not decrease to zero can be explained in terms of the finite light intensity used in the probe beam [panel (e)] that has the side effect of pinning the condensate phase and, thus, opening a gap in the dispersion as we are going to see shortly. In spite of this limitation, the transmitted intensity in the central region exceeds by a large factor the bare transmission of the cold cavity, signaling an enhanced response via parametric amplification processes.

\subsection{Opening a gap by explicitly breaking the symmetry}
\label{sec:gap}

While the previous Subsections were devoted to the physics of a spontaneous breaking of the $U(1)$ phase symmetry,
in this Subsection we are going to discuss the consequences of the explicit symmetry breaking when an external field pinning the condensate phase is applied to an equilibrium or a non-equilibrium condensation phase transition. In this latter case, the phase-pinning mechanism shares interesting analogies with the injection locking mechanism widely studied for generic oscillators~\cite{Adler1946ASO,Liu:JLT2020,Huard:PRApp2019}.

\subsubsection{Equilibrium condensate}

The idea of applying an external field to a quantum Bose field has been long studied as a way to better understand the Bose-Einstein condensation phase transition and the underlying $U(1)$ spontaneous symmetry breaking mechanism~\cite{Gunton:PR1968}. If a term of the form
\begin{equation}
H_\eta=-\int \, d\rr\, \left[\hbar \eta^* \Psih(\rr) + \hbar \eta \Psi^\dagger(\rr)\right]
\end{equation}
is added to the Hamiltonian of an equilibrium condensate, energy minimization pins the condensate wavefunction $\psi$ to have the same phase as $\eta$. 

While most works have focused on the case of an infinitesimally small $\eta$, with the only purpose of explicitly breaking the symmetry in the thermodynamic limit, it is interesting to see the consequence of a finite $\eta$ on the dispersion of collective excitations. Even though this is a somehow academic problem in the atomic case --the phase symmetry is in fact intrinsically connected to particle-number conservation-- its solution can help to understand the general features of the gap opening effect.

\begin{figure}
    \centering \includegraphics[width=0.95\columnwidth,angle=0,clip]{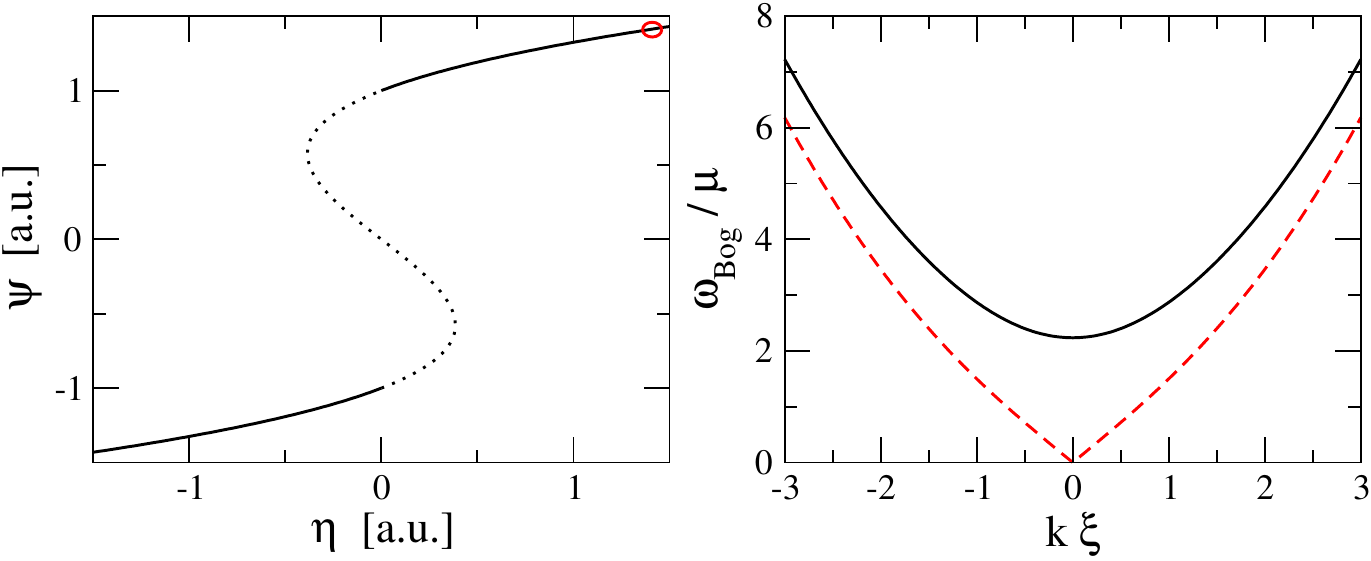}
    \caption{Left panel: plot of the generalized equation of state \eqref{eq:state_eq_h} for the condensate order parameter $\psi$ as a function of the external field $\eta$, both in arbitrary units. The solid lines indicate the stable ground state solution. The dashed line indicates the unstable regions.
    Right panel: dispersion of the collective excitations in the absence (red dashed) and in the presence (solid black) of the external field. The operating point is indicated by the red circle in the left panel. Wavevector is normalized to the healing length $\xi=\sqrt{\hbar^2/(m \mu)}$.
\label{fig:Bogo_eta}}
\end{figure}

Including the $\eta$ term, the generalized Gross-Pitaevskii energy functional reads
\begin{equation}
E_{GP}(\psi) = \int d\rr\,\left[ \frac{\hbar^2}{2m}|\nabla \psi|^2+\frac{\hbar g_{\rm nl}}{2}\,|\psi|^4 - \mu |\psi|^2 - \hbar \eta^* \psi -\hbar \eta \psi^*\right]
\label{eq:GPE_funct_eta}
\end{equation}
where $\mu$ is the chemical potential. The ground state is found by minimization of \eqref{eq:GPE_funct_eta}, which gives a cubic equation that, assuming a real $\eta$, reads
\begin{equation}
\hbar \eta = \hbar g_{\rm nl} |\psi_g|^2\psi_g - \mu \psi_g\,.
\label{eq:state_eq_h}
\end{equation}
This equation plays the role of a generalized equation of state for the Bose condensate in the presence of the external field $\eta$. It directly imposes that $\psi_g$ is real and the solution for a positive $\mu$, plotted in the left panel of Fig.\ref{fig:Bogo_eta}, closely resemble the hysteresis loop of the magnetization of a ferromagnet~\cite{Huang}: the $\eta$ field plays the role of the external magnetic field $B$ and the Bose field plays the role of the magnetization $M$. As in ferromagnets, the energy minimum condition requires that the order parameter $\psi_g$ has the same sign as the applied field $\eta$ and restricts the acceptable solutions\footnote{It is interesting to note that the central, negative slope branch of the hysteresis loop in Fig.\ref{fig:Bogo_eta} corresponds to an energy maximum along the real-$\psi$ line. On the other hand, the external solution for which $\psi$ and $\eta$ have opposite signs is an energy minimum along the real-$\psi$ line, but an energy maximum when $\psi$ is varied along the phase direction. A similar behavior can be observed in ferromagnets where the order parameter can freely rotate along the easy-plane, but not in Ising-like ferromagnets where the magnetization is restricted to a given direction.} to the corresponding quadrants of the $\eta-\psi_g$ plane of the Figure. 

Linearization of the generalized GPE corresponding to the energy functional \eqref{eq:GPE_funct_eta} gives equation of motion of a similar form as the standard Bogoliubov theory
\begin{equation}
 i\frac{d}{dt}
\left( \begin{array}{cc}
 \delta \psi_\kk \\
  (\delta \psi_{-\kk})^*
\end{array}
\right)= \left(
\begin{array}{cc}
\frac{\hbar k^2}{2m} + 2 g_{\rm nl} |\psi_{g}|^2 - \mu  & g_{\rm nl} \psi_{g}^2 \\
- g_{\rm nl} (\psi_{g}^{*})^2 & -\frac{\hbar k^2}{2m} - 2 g_{nl} |\psi_{g}|^2 + \mu
\end{array} 
\right)
\left( \begin{array}{cc}
 \delta \psi_\kk \\
  (\delta \psi_{-\kk})^*
\end{array}
\right)\,.
\label{eq:linearized_coh_incoh}
\end{equation}
Taking into account the modified equation of state \eqref{eq:state_eq_h}, the dispersion around the ground state solution $\psi_{g}$
\begin{equation}
\omega_{\rm Bog}(\kk)= \pm \Big[\Big(\frac{\hbar k^2}{2m}+g_{\rm nl} |\psi_{g}|^2 + \frac{\hbar \eta}{\psi_g}
\Big)^2  - (g_{\rm nl} |\psi_{g}|^2)^2\Big]^{1/2}
\label{eq:Bogo_neq_coh_eta}
\end{equation}
recovers the usual gapless Bogoliubov dispersion only in the absence of external field $\eta\to 0$ while is gapped for any finite $\eta$. An example of such a gapped dispersion is shown in the right panel of Fig.\ref{fig:Bogo_eta}. On the other hand, the dispersion of the collective excitations around the physically unacceptable solutions indicated by the dotted line in Fig.\ref{fig:Bogo_eta} display a dynamical instability at low wavevectors. 

To conclude our discussion, it is interesting to draw a connection between these behaviors and the physics of the driven-dissipative fluid under a coherent pump discussed above: in that case, the dispersion \eqref{eq:Bogo_neq_coh_eta} can be stable and gapped or dynamically unstable depending on the positive/negative sign of the effective detuning $g_{\rm nl}\,|\psi_{ss}|^2-\omega_{\rm inc}$; here, the same role is played by the alignment/anti-alignment with the external field, $(\hbar \eta)/{\psi_g}=\hbar g_{\rm nl}\,|\psi_g|^2-\mu$.

\subsubsection{Driven-dissipative condensates}

A first theoretical investigation of the gap opening phenomenon for the driven-dissipative case of non-equilibrium condensates in the presence of an explicit breaking of the $U(1)$ symmetry was reported in~\cite{Wouters:PRA2007} and is illustrated in the bottom-right panels of Fig.\ref{fig:Goldstone_OPO_th}. In the formalism of Sec.\ref{sec:expt_OPO}, this would correspond to an additional driving term in the equation \eqref{eq:OPO_s} for the signal amplitude. As expected, the presence of an coherent external field locking the signal phase is responsible for the opening of a gap in the dispersion. In contrast to the equilibrium case of Fig.\ref{fig:Bogo_eta}, however, the gap is here in the imaginary part of the dispersion and leads to a broadening of the spectral features corresponding to the Goldstone mode~\cite{Wouters:PRA2007}.

\begin{figure}
    \centering
    \includegraphics[width=0.95\textwidth]{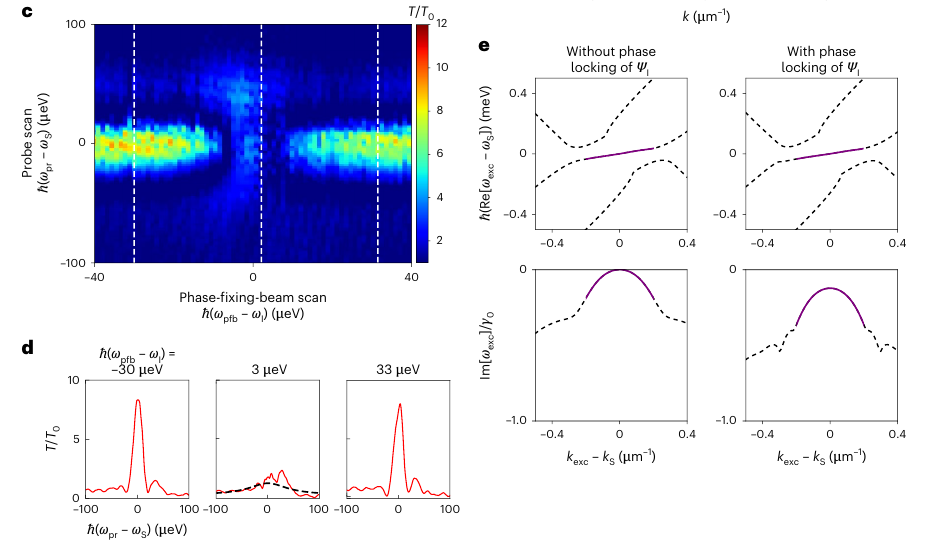}
    \caption{Experimental observation of the opening of a gap in the dispersion of collective excitations of a non-equilibrium condensate when a phase-locking beam is applied to the condensate. The upper-left panel shows a colorplot of the cavity transmission as a function of the probe frequency (vertical axis) for a series of values of the phase-locking beam frequency $\omega_{\rm pfb}$ (horizontal axis). A few vertical cuts at different values of $\omega_{\rm pfb}$ are shown in the bottom panel. 
    Figure adapted from~\cite{claude2025observation}.    
\label{fig:Goldstone_OPO_exp2}}
\end{figure}

An experimental observation of this behavior was reported in~\cite{claude2025observation} and is summarized in Fig.~\ref{fig:Goldstone_OPO_exp2}. In the simplest case of a phase-locking beam close to resonance, the Goldstone mode is replaced by a much broader feature with a linewidth set by the width of the imaginary gap. This behavior is visible in the spectrum shown in the central subpanel of Fig.~\ref{fig:Goldstone_OPO_exp2}(d) and agrees with the dispersion shown in the theoretical calculation in Fig.\ref{fig:Goldstone_OPO_th}. 
When the phase-locking beam is instead far from resonance, it is not able to efficiently lock the phase of the condensate. In this case, the strong and narrow transmission peak corresponding to the Goldstone mode is again clearly visible, basically unaffected by the phase-locking-beam. 

\begin{figure}
    \centering \includegraphics[width=0.6\columnwidth,angle=0,clip]{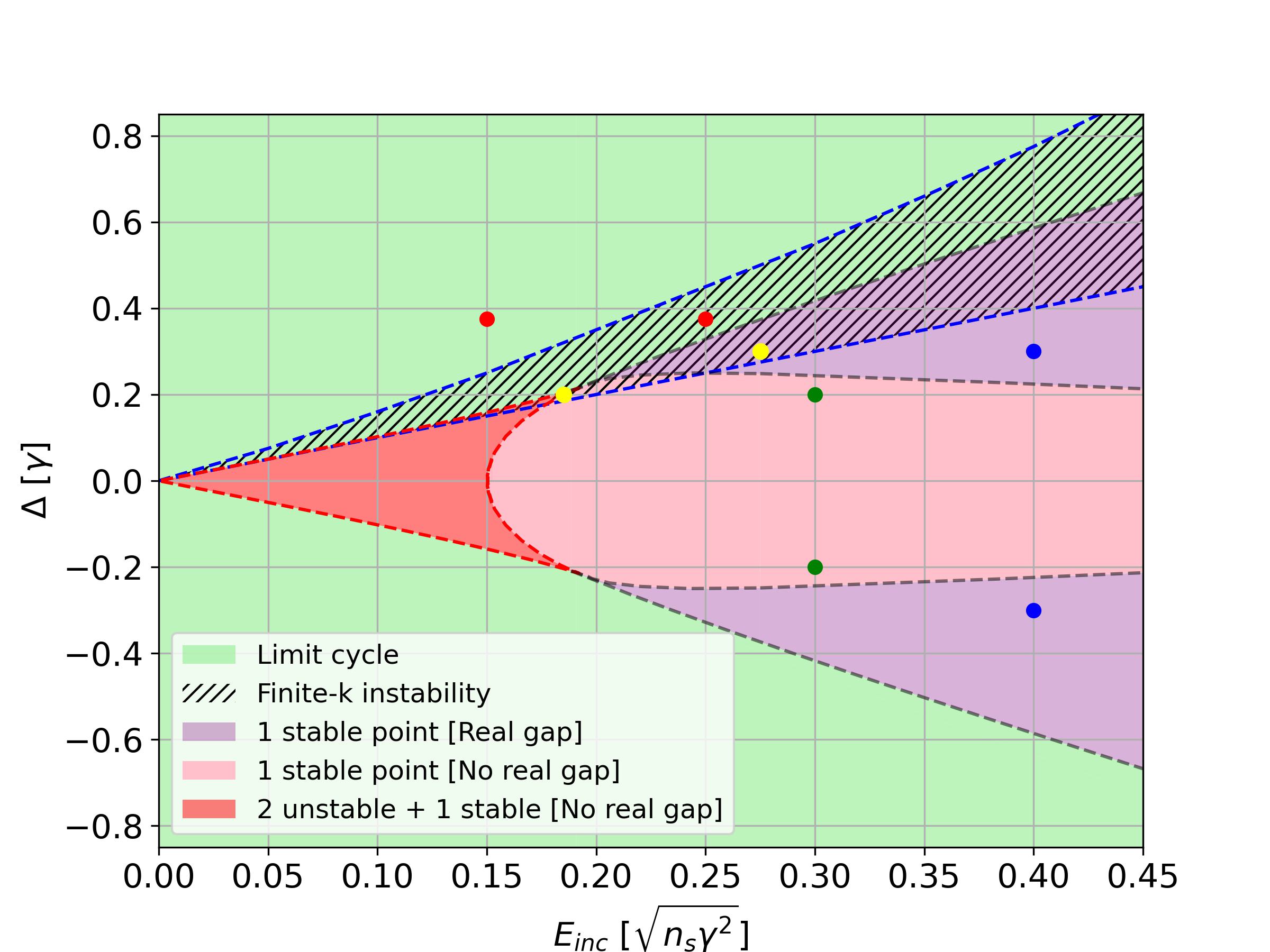}
    \caption{Phase diagram of the phase-locking mechanism for a non-equilibrium condensate described by the field dynamics \eqref{eq:dd_psi} as a function of the amplitude and the detuning of the phase-locking beam. The incoherent pump is $P/\gamma_{\rm loss}=2$ and interactions vanish $g_{\rm nl}=0$. The red/pink/purple (green) regions indicates phase-locking (no phase-locking). The gap is purely imaginary in the red and pink regions and acquires a non-zero real part in the purple one.
    Figure from~\cite{Stazzu:PRA2026}.
\label{fig:gapped_Gold_th}}
\end{figure}

A detailed theoretical study of this physics was carried out in~\cite{Stazzu:PRA2026} using the model equation \eqref{eq:dd_psi} in the most general case where both a coherent $E_{\rm inc}$ and an incoherent $P$ pumps are simultaneously present. 
In spite of its simplified nature as compared to the full optical parametric oscillator of~\cite{Wouters:PRA2007}, this model already displays an extremely rich phenomenology and is able to capture much of the experimental observations.

An example of phase diagram is shown in Fig.\ref{fig:gapped_Gold_th} as a function of the detuning $\Delta=\omega_{\rm inc}-\omega_o$ and the ampplitude $E_{\rm inc}$ of the phase-locking beam: the inner red/pink/purple regions indicate efficient locking of the condensate phase and, thus, a disruption of the Goldstone mode. On the other hand, the external green regions indicate absence of locking and, thus, the presence of a narrow Goldstone mode. In agreement with the experiment, phase locking is favoured by a strong amplitude and a small detuning of the phase-locking beam from the natural frequency of the condensate  \eqref{eq:ss_omega}.

As an additional feature that is still awaiting experimental observation, the theoretical calculations also predict that the gap can now appear either in the imaginary part or in both the real and the imaginary parts of the dispersion. The latter situation typically occurs when the coherent pump frequency is significantly detuned (purple region in the figure) from the natural oscillation frequency of the condensate \eqref{eq:ss_omega}.

\section{Superfluidity}
\label{sec:superfluidity}

In combination with non-equilibrium phase transitions associated to condensation and lasing, superfluidity has been another long-lasting thread in the development of the field of quantum fluids of light. In this Section, we are going to first review the basic concept of superfluidity as it was developed in the context of superfluids of material particles. Then, capitalizing on the discussion of the collective excitations presented in the previous Sections,  we will highlight the new features introduced by the driven-dissipative nature of the photon fluid.

\subsection{General concepts}
\label{sec:superfl_general}


Since the pioneering observation of frictionless flow in liquid Helium by Kapitsa~\cite{Kapitsa:Nature1938} and Allen and Misener~\cite{Allen:Nature1938}, several definitions have been used to put the qualitative idea of superfluidity into a conceptually coherent framework. In the next subsections we follow the footsteps of~\cite{Leggett:RMP1999} to briefly outline each definition and give a critical assessment of its pros and cons.

\subsubsection{Landau criterion}
\label{sec:superfl_Landau}

A simplest formulation of the concept of superfluidity is based on energetic arguments to determine whether a fluid flowing at velocity $\vv$ through a pipe at rest can experience friction as a consequence of the generation of excitations in the fluid.

Within a perturbative framework, the kinetic energy of the flow can be dissipated only if the spectrum of the collective excitations in the moving fluid contains modes that display a negative energy in the laboratory frame. By Doppler-transforming the dispersion $\omega_\kk$ in the comoving frame of the fluid to the lab frame, 
one obtains a condition
\begin{equation}
\omega'_\kk=\omega_\kk+\vv \cdot \kk <0
\label{eq:Landau}
\end{equation}
for such negative energy modes to be available. In this, case an imperfection of the pipe can make the macroscopic flow to relax by generating excitations into these modes. 

From this condition, it is immediate to deduce that no excitations can be generated if the speed of flow is below the so-called Landau critical velocity for superfluidity defined as
\begin{equation}
    v_c=\textrm{min}_\kk\frac{\omega_\kk}{k}\,:
\end{equation}
for sufficiently slow velocities $v<v_c$, no excitation is energetically allowed, so the fluid flows undisturbed through the pipe in an effectively frictionless way. If instead $v>v_c$, collective excitations can be generated by any imperfection of the pipe and a finite friction appears.  

\begin{figure}
    \centering
\includegraphics[width=0.6\textwidth]{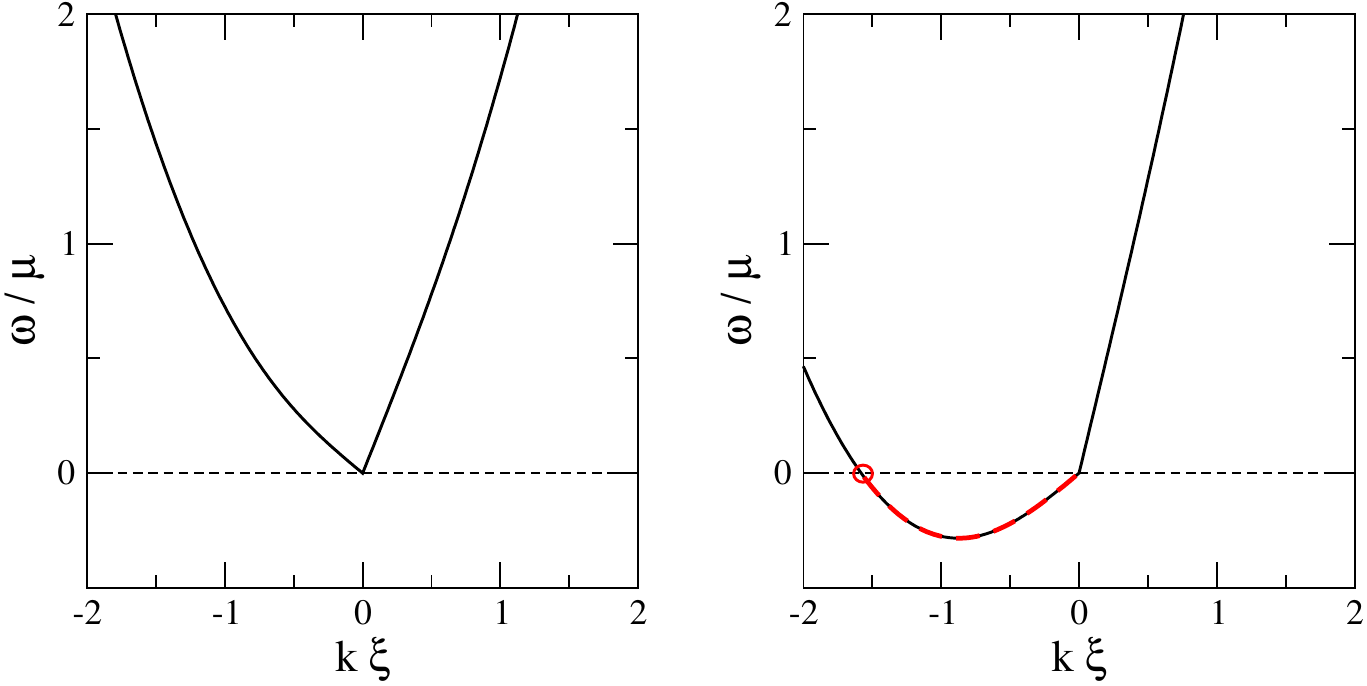}
    \caption{Dispersion of collective excitations in a weakly interacting Bose-Einstein condensate in motion at $v<v_c$ (left) and $v>v_c$ (right). In the left panel no mode satisfies the excitation condition \eqref{eq:Landau} and the flow is frictionless and superfluid. In the right panel, a range of modes --indicated by the red dashed line-- can be excited and the flow experiences friction. The red dot indicates the mode that can be elastically excited by a rigid pipe or an impurity.
    \label{fig:Landau}}
\end{figure}

The energetic inequality \eqref{eq:Landau} is graphically illustrated in Fig.\ref{fig:Landau} for the specific case of the Bogoliubov dispersion \eqref{eq:Bogo_standard} of a weakly interacting Bose-Einstein condensate. In this case, the Landau critical velocity $v_c$ coincides with the speed of sound $c_s=\sqrt{\mu/m}$. The left and right panels correspond to the two cases of a frictionless superfluid flow at $v<v_c$ and of a dissipative flow at $v>v_c$, respectively: in the former case, all $\kk$ modes have a positive frequency $\omega'_\kk>0$ in the laboratory frame, while in the latter case there is a window of negative frequency modes which can give rise to friction. 

If one focuses on configurations where the pipe is a rigid container with no internal dynamics and can not absorb any energy, the inequality sign in \eqref{eq:Landau} must be replaced by an equality sign so to restrict our attention to elastic processes. In plots of the dispersion relation like the one in Fig.\ref{fig:Landau}, these elastic processes are associated to zero-crossing points. 

Modulo a Galilean boost, this same condition also describes the excitations that are elastically generated in the fluid by a spatially-localized impurity traveling through the fluid at a constant speed $-\vv$. The close connection between this formulation of superfluidity and the classical problems of Cherenkov emission by moving charges and the wake generated on the surface of a fluid is highlighted in~\cite{Carusotto:Ducks2013}. In this same work, a detailed analysis of the connection between the locus of zero-crossing points in $\kk$-space and the spatial shape of the emission pattern in real-space is presented for generic geometries beyond the one dimensional case of Fig.\ref{fig:Landau}.

A direct experimental illustration of this physics was reported in~\cite{Onofrio:PRL2000}: the fluid is a Bose-Einstein condensate of ultracold atoms and the moving impurity is created by means of an externally-imposed localized repulsive potential generated by a focused laser beam and scanned across the condensate. Superfluidity was then detected by looking at density modulation generated by the moving impurity into the fluid. Evidence of superfluidity consisted in the disappearance of the density modulation in the fluid when the motion is below a critical velocity.

A crucial assumption of this formulation of the Landau criterion is that the interaction potential between the pipe/impurity and the fluid is perturbative, so that the dispersion of the collective excitations is not modified. When this assumption is no longer satisfied, energy can be dissipated also in the form of large, nonlinear excitations in the fluid such as solitons or vortices, As it was originally studied in~\cite{Frisch:PRL1992,Hakim:PRE1997,Pavloff:PRA2002} and experimentally observed in an atomic gas~\cite{Engels:PRL2007,Neely:PRL2010}, these processes have a lower critical velocity and can occur also below the Landau critical speed $v_c$.

While physically intuitive and easy to visualize, the perfectly frictionless flow predicted by the Landau criterion only holds at zero temperature. At finite temperatures, the Landau criterion does not provide information on the friction arising from the scattering  onto the impurity of collective excitations already present in the fluid. In general, this definition of superfluidity is not able to provide a quantitative definition of the superfluid and normal fractions. As a further difficulty, its direct application would predicts no superfluidity $v_c=0$ if the fluid displays a spatially periodic modulation of the density and the collective excitations are organized in Bloch bands of Bogoliubov modes~\cite{Price:PRA2023}.

\subsubsection{Response to rotations} 
\label{sec:superfl_bucket}

Another definition of superfluidity is based on the rotational response of a fluid contained in a rotating bucket. The idea is that for sufficiently small rotation speeds, the normal fraction equilibrates with the container walls and rigidly follows their motion, while the superfluid fraction remains at rest in the frame of the fixed stars. This results in a reduced value of the moment of inertia of the fluid which can be measured, e.g., via a low-frequency torsion pendulum experiment~\cite{Andronikashvili:RMP1966}. In contrast to the Landau criterion presented in the previous Section, this procedure provides an operative definition of the normal and superfluid fractions of the fluid and it is valid independently of the nature and the temperature of the system.

\begin{figure}[htbp]
    \centering
   \begin{minipage}[c]{0.49\textwidth}
   \includegraphics[width=0.99\textwidth]{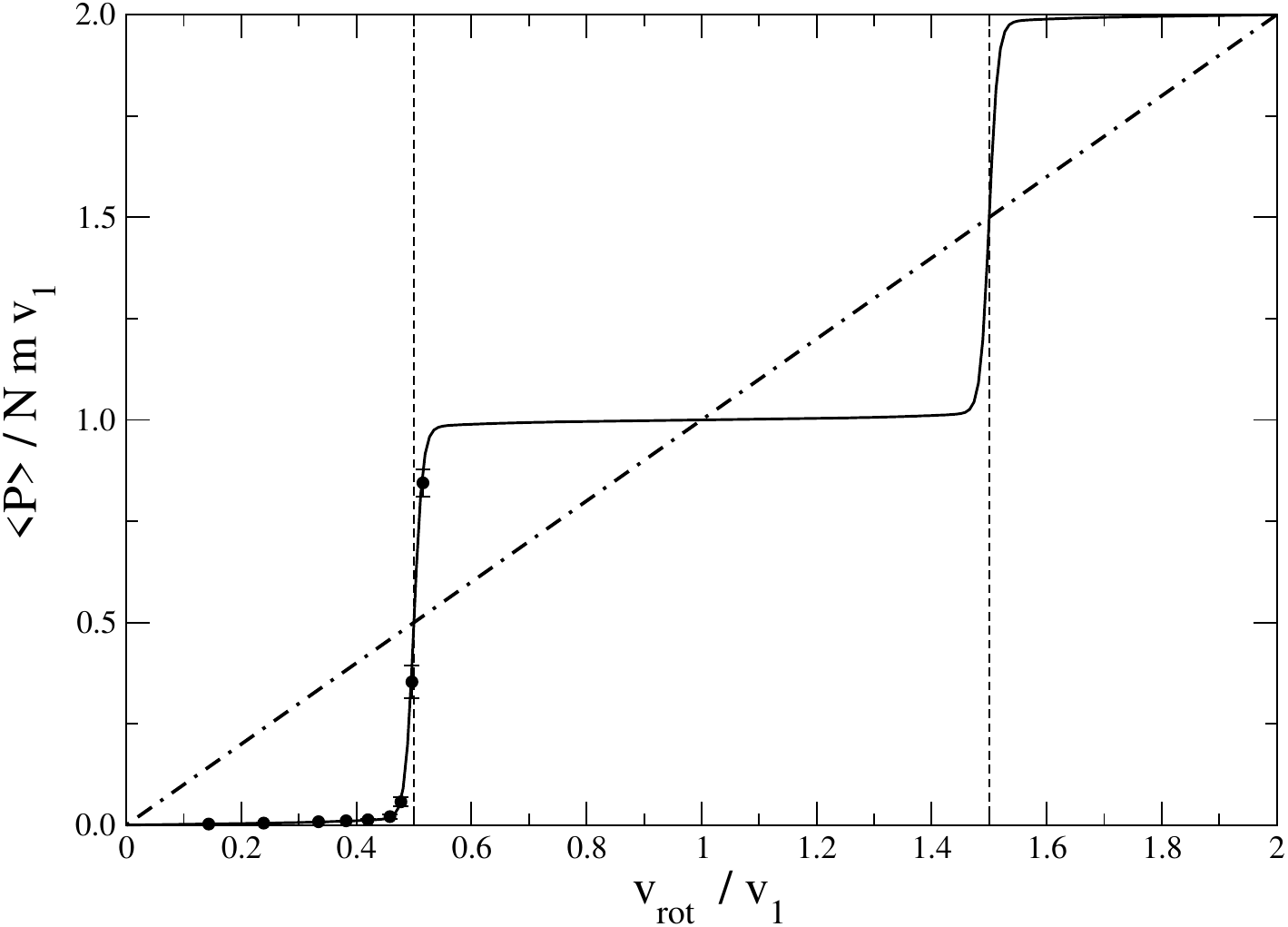} 
   \end{minipage}
    \begin{minipage}[c]{0.49\textwidth}
        \includegraphics[width=0.98\textwidth,trim=0.3cm 0.5cm 0.5cm 1.5cm,clip]{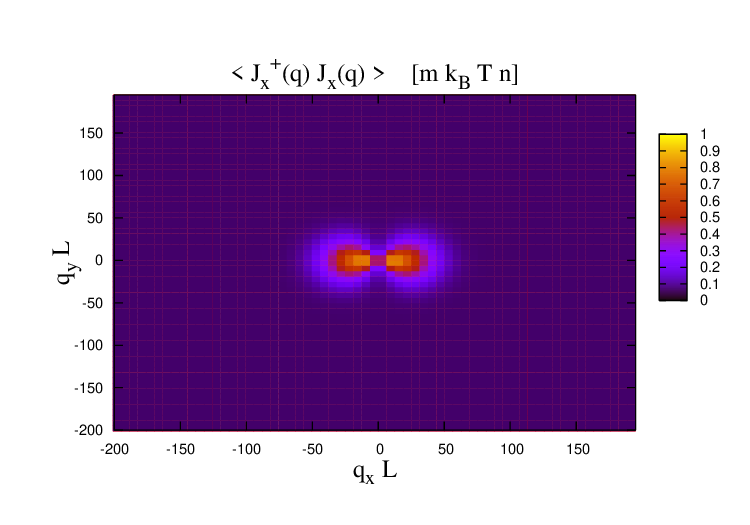}\\
        \includegraphics[width=0.98\textwidth,trim=0.3cm 0.5cm 0.5cm 1.5cm,clip]{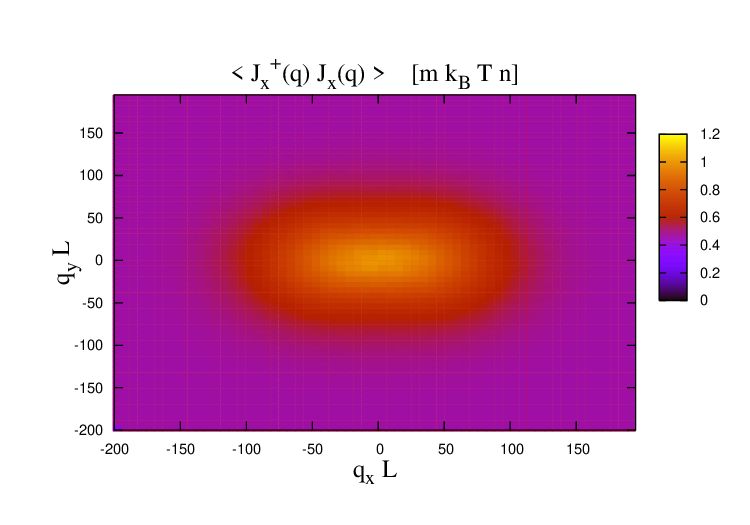}
    \end{minipage}   
    \caption{Numerical calculations of the superfluidity properties of dilute Bose gases at equilibrium. Left panel: 1D Bose gas of $N=42$ atoms in periodic boundary conditions at a low but finite temperature $ $. The plot shows the total momentum $\langle P_x \rangle$ as a function of the rotation speed $v_{\rm rot}$, compared to the rigid body value (dash-dotted line). Solid line: many-valley Bogoliubov theory. Points: Quantum Monte Carlo calculation. Figure adapted from~\cite{Carusotto:CRAS2004}. Right panels: colorplots of the zero-frequency limit  ${\chi}_{xx}(\kk,\omega\to 0)$ of the $xx$ component of the current-current susceptibility matrix defined in \eqref{eq:chi_J}. The top and bottom panels respectively refer to low and high temperature cases, $T/T_{\rm deg}=0.05, 0.3$ (where $k_B T_{\rm deg}=2\pi \hbar^2 n_{2d}/m$). The susceptibility matrix is extracted from a semiclassical calculation of the correlation function of current-current fluctuations and is converted into the susceptibility via the fluctuation-dissipation theorem. Figure adapted from~\cite{Carusotto:IHP2007}.
\label{fig:multivalley_sf}}
\end{figure}

\paragraph{Low speeds: superfluid and normal fractions}

At a simplest level, this idea can be formalized by considering a superfluid enclosed in a box with periodic boundary conditions and looking at its thermodynamical equilibrium state at temperature $T$ in a frame moving at a small velocity $\vv$. This can be described by a usual thermal equilibrium state under the modified Hamiltonian 
\begin{equation}
    H=H_0-\mathbf{P} \cdot \vv
    \label{eq:P-v}
\end{equation} 
where $\mathbf{P}$ is the total momentum of the fluid.
The normal fraction is then defined as the momentum response 
\begin{equation}
    f_n=\lim_{v_x\to 0} \frac{\langle P_x \rangle }{m N v_x}
\label{eq:f_n_v}
\end{equation}
where the average is taken in a thermodynamical equilibrium state under the Hamiltonian \eqref{eq:P-v} with a velocity $\vv$, assumed to be directed along the $x$ direction. The denominator corresponds to the momentum of a rigid-body motion. If the fluid rigidly follows the moving frame, the normal fraction is complete $f_n=1$. If the fluid is superfluid, it stays at rest and the normal fraction vanishes $f_n=0$.

An example of explicit calculation for a one-dimensional weakly interacting BEC under periodic boundary conditions was reported in~\cite{Carusotto:CRAS2004} and is shown in Fig.\ref{fig:multivalley_sf}: at velocities, the average momentum is well below the rigid body value, which indicates a large superfluid fraction. The residual normal fraction is a finite temperature effect. In experiments, geometries with periodic boundary conditions can be realized in liquid Helium or in atomic BECs using ring-shaped containers or traps: thermodynamical equilibrium in the moving frame can be obtained by introducing a suitable roughness of the container that moves at the desired speed $\vv$.

This definition of superfluidity can be equivalently formulated in terms of the response of a spatially extended two- or three-dimensional fluid to a rotation~\cite{pitaevskii2016bose}. The thermodynamical equilibrium state in a rotating frame  at an angular velocity $\mathbf{\Omega}$ can be described in terms of the Hamiltonian 
\begin{equation}
    H=H_0-\mathbf{L} \cdot \mathbf{\Omega}
    \label{eq:L-Omega}
\end{equation} 
in this case, the definition of normal/superfluid fraction involves the average value of the angular momentum compared to the rigid body value,
\begin{equation}
f_n=\lim_{\Omega \to 0} \frac{\langle L_z \rangle}{I_{\rm rig} \Omega}
\label{eq:f_n_I}
 \end{equation}
where rotation is assumed to occur around the $z$ axis and $I_{\rm rig}$ is the moment of inertia of the fluid under a rigid-body assumption: once again, superfluidity is visible as a suppressed response to rotation.

This idea was put into practice in the pioneering experiment by Hess and Fairbank in liquid Helium~\cite{Hess:PRL1967}: when a normal fluid, rigidly rotating at low angular speed is cooled across the transition to the superfluid state, the superfluid fraction stops its motion and gets to a sudden rest, making the total angular momentum of the fluid to correspondingly drop and transferring the extra angular momentum to the container. As we will see in Sec.\ref{subsec:supersolids}, a drop in the moment of inertia as described by \eqref{eq:f_n_I} was the signature underlying the (later retracted) claim of supersolidity in solid Helium.

\paragraph{Linear response to a transverse vector potential}

This physics can be reformulated at a more abstract level by introducing a coupling of the fluid to a generic spatially-dependent vector potential
\begin{equation}
    H=H_0-\int\!d\rr\;\mathbf{j}(\rr)\cdot \mathbf{A}(\rr,t)
    \label{eq:H+A}
\end{equation}
and measuring the current response quantified by the expectation value of the (canonical)-current-density operator
\begin{equation}
\mathbf{j}(\rr)=\frac{\hbar}{2im}\left[\Psihd(\rr) \nabla\Psi(\rr) - \textrm{h.c.}\right]
\label{eq:current_can}
\end{equation}
in the equilibrium state under the perturbed Hamiltonian \eqref{eq:H+A}.

Assuming the fluid to be homogeneous in the region of interest, one can move to Fourier space. Here, the components at wavevector $\kk$ and frequency $\omega$ of the current-density are related to the corresponding component of the vector potential by
\begin{equation}
    \mathbf{j}(\kk,\omega)=\overleftrightarrow{\chi}(\kk,\omega)\,\mathbf{A}(\kk,\omega)\,.
    \label{eq:chi_J}
\end{equation}
In a rotationally symmetric system, the current-current susceptibility matrix $\overleftrightarrow{\chi}(\kk,\omega)$ can be split into its longitudinal and transverse components $\chi_{L,T}(\kk,\omega)$ with respect to the wavevector $\kk$. 
While the low-frequency limit of the longitudinal component is fixed by gauge-invariance and particle-number conservation arguments related to the $f$-sum rule to $\chi_{L}(\kk,\omega\to 0)=n/m$~\cite{Dalfovo:PRB1992}, the transverse part provides a quantitative definition of the normal part  
\begin{equation}
    f_n=\frac{m}{n}\lim_{k\to 0}\lim_{\omega\to 0}\chi_T(\kk,\omega)
    \label{eq:f_n_AT}
\end{equation}
that 
can be used to extract the superfluid fraction of generic systems and at any temperature. 

Examples of numerical calculations are shown in the right panels of Fig.\ref{fig:multivalley_sf}. At low temperature, the response in the transverse direction (along the vertical $k_x=0$ axis in the figure) is much smaller than the one in the longitudinal direction (along the horizontal $k_y=0$ axis), signaling a superfluid behaviour. At high temperature, the responses in the two directions have the same long-wavelength $\kk\to 0$ limit signaling the absence of superfluid behavior.
As a further advantage of this scheme, a measurement of the response of an atomic BEC to a spatially localized synthetic vector potential gives the possibility to reconstruct the spatial profile of the normal and superfluid components of spatially inhomogeneous clouds, as theoretically explored in~\cite{Carusotto:PRA2011}.

This abstract formulation is equivalent to the ones in \eqref{eq:P-v} based on a global motion\footnote{In spite of the formal similarity, it is important to note a crucial difference between the global perturbation \eqref{eq:P-v} and a longitudinal vector potential.
Via a gauge transform, a slowly-varying, long-wavelength longitudinal vector potential $\mathbf{A}_L(\rr,t)=i\kk e^{i(\kk \rr-\omega t)}\,\bar{A} + \textrm{c.c.}$ can be recast in the form of a scalar potential $V(\rr,t)=-\omega\,e^{i(\kk \rr-\omega t)}\,\bar{A}$ acting on the density. No such gauge transform is instead available for the global perturbation \eqref{eq:P-v}, which is intrinsically related to the current response and can be understood in terms of a magnetic field piercing the hole of the torus encoding the periodic boundary conditions.} and to the one based on the  reduced moment of inertia. In particular, the Hamiltonian  \eqref{eq:L-Omega}
describing equilibrium in a rotating frame at an angular velocity $\mathbf{\Omega}$ can be straightforwardly reformulated in terms of the coupling \eqref{eq:H+A} to a vector potential $\mathbf{A}=m\mathbf{\Omega}\times \rr$ of purely transverse nature. Note that in this case the canonical current operator \eqref{eq:current_can} describes the current in the lab frame~\cite{CCT:CdF}, so that the superfluid state indeed corresponds to an absence of rotational flow.

An implementation of the Hamiltonian \eqref{eq:P-v} using a synthetic gauge field for ultracold atoms was proposed in~\cite{Cooper:PRL2010}: in this case, the mechanical current operator is obtained from the canonical one via
\begin{equation}
    \mathbf{j}_{\rm mech}(\rr)=\mathbf{j}(\rr)-\frac{n(\rr)}{m}\,\mathbf{A}(\rr)
\end{equation}
where $n(\rr)$ is the density profile of the fluid. Interestingly, for a normal state the two terms compensate so the mechanical current $\mathbf{j}_{\rm mech}(\rr)$ eventually vanishes. On the other hand, in the superfluid state the canonical current $\mathbf{j}(\rr)$ vanishes and the state is characterized by a non-vanishing mechanical current opposing the applied synthetic vector potential.

This difference between mechanical and canonical current is clearer in the case of charged fluids and has to be taken into account when assessing the superconductivity properties of materials.
Here, the physical current is determined by the mechanical current operator 
\begin{multline}
\mathbf{j}_e(\rr)=q \, \mathbf{j}(\rr) - \frac{q^2}{mc} n(\rr) \, \mathbf{A}_{em}(\rr,t)=\\=
\frac{q}{m}\left[\frac{\hbar}{2i}[\Psihd(\rr) \nabla\Psi(\rr) - \textrm{h.c.}] -  \frac{q}{c} \,\Psihd(\rr) \Psi(\rr)\,\mathbf{A}_{em}(\rr,t)\right]\,:
\label{eq:J_e}
\end{multline}
subtracting the e.m. vector potential $\mathbf{A}_{em}(\rr,t)$ term from the canonical current-density response predicted by \eqref{eq:f_n_AT} recovers the perfect diamagnetic response of a superconductor underlying the Meissner effect~\cite{Tinkham}. 

In the simplest geometry of mesoscopic rings, a magnetic field threaded through the ring gives a vector potential winding along the ring in a form analogous to \eqref{eq:P-v}: as a consequence of the vector potential term in \eqref{eq:J_e}, a superconducting state shows a diamagnetic current in response to the magnetic field, while in a normal metallic state the two terms of  \eqref{eq:J_e} cancel each other and the current eventually vanishes. An even more intriguing mechanism is at play in mesoscopic ring configurations where a current is anticipated to arise from the Berry phase acquired by the spin-polarized electrons in textured magnetic fields~\cite{Loss:PRL1990}.


\paragraph{Higher speeds: macroscopic currents}
The discussion carried out so far is based on a linear response theory which, by definition, assumes that the applied perturbation is weak, that is that the rotation occurs at a slow angular speed. 
At higher rotation velocities, the physics gets richer and the equilibrium state of superfluids can display macroscopic currents. 

This phenomenology is easiest understood in the idealized ring geometry with length $L_x$ and periodic boundary conditions described by \eqref{eq:P-v}. For the sake of simplicity, we restrict our discussion here to the simplest case of a pure condensate for which the analysis is most straightforward. A more complete treatment for the case of a  weakly interacting one-dimensional BEC at finite temperature can be found in~\cite{Carusotto:CRAS2004}.

At low but finite speeds $v_x$, the zero-momentum state remains the many-body ground state. Via the definition \eqref{eq:f_n_v}, this gives a zero normal fraction and thus a perfect superfluid. 
When the speeds exceeds a threshold value $v_x>v_1/2$ with
\begin{equation}
    v_1=\frac{\hbar}{m}\left(\frac{2\pi}{L_x}\right)\,,
\end{equation}
the zero-momentum state stops being the lowest energy state and the one at $k_1=2\pi/L_x$ starts having a lower energy. As such, the many-body ground state consists of all particles in the $k_1$ state and the fluid displays a macroscopic current at speed $v_1$. While for $0<v_x<v_1/2$ superfluidity makes the flow to be slower than the rigid body, for $v_1/2<v_x<v_1$ the flow turns our to be faster than the externally imposed $v_x$. This physics is again illustrated in the left panel of Fig.\ref{fig:multivalley_sf}: the finite temperature of the numerical calculation is responsible for a smoothening of the jump around $v_x\sim v_1/2$ and for a finite normal fraction at low rotation speeds.

The behavior then repeats for even higher rotation speeds.
For instance, for $v_x > 3v_1/2$, the state at $k_2=2k_1$ gets energetically favored and the macroscopic current becomes correspondingly larger. All together, the dependence of the macroscopic current as a function of $v_x$ displays a series of plateaux corresponding to quantized values of the circulation, separated by sudden jumps. In larger dimensions, a similar phenomenology underlies the sequential entrance of quantized vortices as the rotation speed is increased, with their eventual stabilization into an Abrikosov vortex lattice~\cite{Fetter:JPhys2001,Abo:Science2001,Peretti:SciAdv2023}.

\subsubsection{Metastability of supercurrents}
The previous discussion was focused on a system that is at thermal equilibrium in a moving frame or in the presence of the vector potential. A related but conceptually different effect is the so-called metastability of supercurrents, namely the extremely long lifetime of a macroscopic current when a superfluid is initially prepared in a rotating state and the container is suddenly put to a rest. Evidence of this effect is available for liquid Helium systems~\cite{Hall:ProcRoySoc1957,Vinen:Nature1958,Davis:PRL1991} as well as for atomic condensates~\cite{Ryu:PRL2007,Ramanathan:PRL2011,Moulder:PRA2012}.

This dynamical effect must not be confused with the equilibrium Hess-Fairbank effect discussed above and, in contrast to the response to a slow rotation, does not provide a quantitative definition of the superfluid fractions. Nevertheless, the metastability of supercurrents is often used as a qualitative smoking gun of a superfluid state and is also being exploited in gyroscope applications~\cite{Packard:PRB1992,Avenel:PRL1997}.

In conservative fluids, the physical origin of this  metastability effect can be traced in the very large value of the energetic barrier separating states with different values of the quantized macroscopic current. Such a transition involves in fact creating a node in the condensate wavefunction, which has a very large cost in terms of interaction energy, and is favored by the presence of weak links along the flow~\cite{Avenel:PRL1985}.

\subsection{Superfluid light}
\label{sec:superfl_expt_QFL}

In conservative fluids of material particles, all these definitions are in basic agreement and superfluidity is a quite univocal property~\cite{Leggett:RMP1999}. The situation is more subtle in driven-dissipative fluids where the different criteria may give apparently contradictory conclusions on the superfluidity of a given system.

\begin{figure}[htbp]
    \centering
    \includegraphics[width=0.8\textwidth]{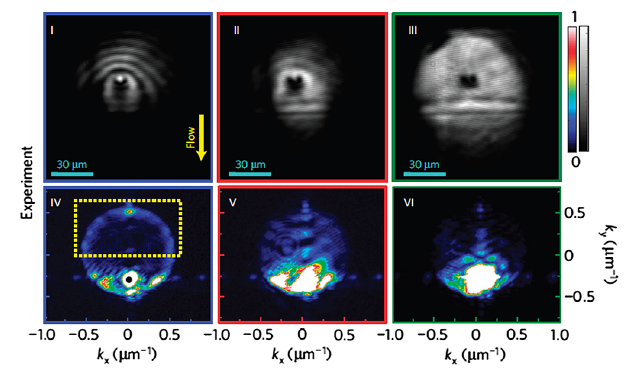}
    \includegraphics[width=0.5\textwidth]{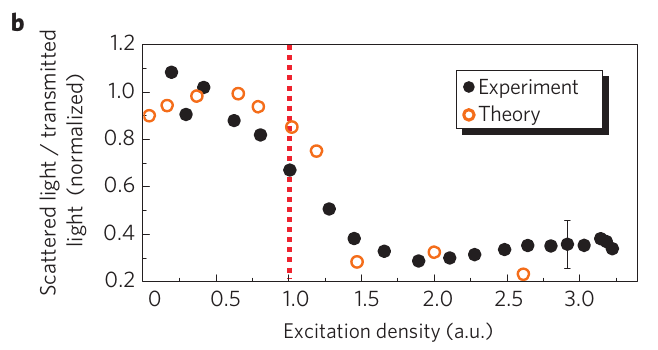} \\
    \includegraphics[width=0.4\textwidth]{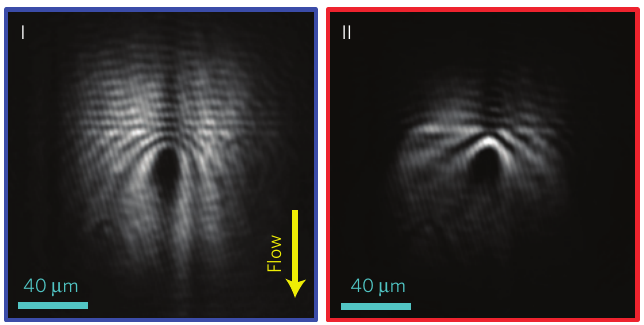}
    \caption{Experimental real-spaceand $\kk$-space images of a photon fluid flowing against a static defect (top and bottom rows of top panel). The coherent pump has a finite incidence angle giving a finite momentum in the downwards direction on the figure. The pump intensity grows from left to right, from an almost non-interacting fluid (left) to a superfluid regime (right). Central panel: plot of the relative scattered intensity as a function of the excitation density, displaying a sudden drop at the onset of the superfluid behavior. The scattered intensity is integrated in $\kk$-space within the yellow rectangle in the top panel. Bottom panels: experimental real-space images of a photon fluid hitting a static defect at a higher speed for low (left) and high (right) densities.    
\label{fig:coherent_sf_exp}}
\end{figure}

\subsubsection{Landau criterion}
Right from early days of research on quantum fluids of light, the Landau criterion presented in Sec.\ref{sec:superfl_Landau} appeared to provide a most direct strategy to investigate superfluidity properties~\cite{Carusotto:PRL2004}. As a simplest geometry, a spatially homogeneous fluid flowing at a finite speed and hitting a single, stationary and spatially-localized defect at rest was considered: for a sufficiently weak interaction potential, the dominant scattering occurs in the elastic channel and the Landau criterion predicts that the perturbation is concentrated in the zero-frequency modes that can be resonantly excited by the defect. Such zero-crossing modes are indicated in the sketch of Fig.\ref{fig:Landau} by a red dot. 

\paragraph{Superfluidity in coherently pumped fluids}

The first configuration that was theoretically and experimentally considered is the one under a coherent pump discussed in Sec.\ref{sec:coh_pump}.
Theoretical calculations~\cite{Carusotto:PRL2004} of the $\kk$-space features of superfluidity in this configuration are displayed in Fig.\ref{fig:coherent_sf_th}. The right (b,d,h,j) panels summarize the $\kk$-distribution of the light that is scattered by the defect: a sizable scattering is visible at the $\kk$ points that correspond to the locus of zero-crossing of the corresponding dispersion shown in the left (a,c,g,i) panels, with a sizable broadening due to the photon linewidth. On the other hand, a suppressed scattering is visible in the case of panel (f), for which the dispersion does not show any zero-crossing point: in this case, the scattered light is much weaker and is concentrated around the pump wavevector with a weak non-resonant diffuse component. 

While the pioneering work in~\cite{Amo:Nature2009} was confronted to the severe complications of a parametric configuration~\cite{Berceanu:PRB2015}, a detailed experimental confirmation of the theoretical predictions was reported in~\cite{Amo:NPhys2009} and is summarized in Fig.\ref{fig:coherent_sf_exp}. In this experiment, the static defect consisted in a naturally occurring imperfection in the planar cavity device and the fluid was coherently pumped into a finite-velocity state by means of a pump at a finite angle, as discussed in Sec.\ref{subsec:coh_flow}.
The central row of panels displays the $\kk$-space distribution of scattered light: for increasing values of the density (left to right), the ring-shaped distribution of scattered light --the so-called resonant Rayleigh scattering ring-- collapses towards the pump wavevector. 

The real-space manifestation of this physics is illustrated in the top panels of the same figure. At low density (left panel), the interference between light on the $\kk$-space scattering ring and the unperturbed light gives rise to a system of fringes. 
At high intensities (right panel), the absence of scattering is visible in the flat density profile of the fluid: except for a marked spatially-localized depletion at the defect location, no fringes are visible away from it. This is a clear qualitative evidence that the fluid is able to flow around the defect as a superfluid without being perturbed by it.

The details of the wake pattern generated by the defect are better visible in the experimental images for a larger flow speed (bottom panel). As we discussed in Sec.\ref{subsec:coh_flow}, a higher flow speed is straightforwardly obtained by increasing the incidence angle of the pump beam. 
The left subpanel is for an almost non-interacting, low-density fluid, where the wake pattern consists of fringes with an almost parabolic shape. These can be physically understood as resulting from the interference of the plane wave of the coherent fluid with the cylindrical wave of the scattered photons~\cite{Carusotto:Ducks2013}. 

The right subpanel is for a stronger pump tuned in the vicinity of the turning point of the bistability loop where the collective excitations have a sonic dispersion. As the flow speed exceeds the sound speed, superfluidity is broken and a conical pattern appears in the wake that closely resembles the Cherenkov cone of a superluminally moving charge or the Mach cone of a supersonic jet: the aperture $\varphi$ of the cone is geometrically determined by the ratio of the flow and sound speeds as $\sin\varphi=c_s/v$. 
A comprehensive discussion of the variety of wake patterns that can be observed for different shapes of the dispersion relation can be found in~\cite{Carusotto:Ducks2013}.

While these real-space and $\kk$-space patterns offer a visual evidence of the onset of superfluidity, a more quantitative information can be obtained from the dependence of the total scattered intensity on the flow speed. As it is displayed in the central panel of Fig.\ref{fig:coherent_sf_exp}, the onset of superfluidity corresponds to a marked drop in the scattered intensity for growing pump power. Still, extracting from these data a quantitative estimation of the superfluid fraction does not seem to be possible. In particular, it is clear from the figure how some residual scattering is always present. Besides experimental imperfections, finite-size and finite temperature effects, an intrinsic contribution to the scattering unavoidably follows from the finite linewidth of the collective excitation modes which makes non-resonant scattering processes possible even in the absence of zero-crossing points. 

Going beyond the perturbative regime of a weak defect for which the Landau criterion was originally formulated, follow-up experiments have addressed the more rich phenomena that take place in the presence of a large and impenetrable defect, such as the hydrodynamic nucleation of solitons~\cite{Amo:2011Science} or vortices~\cite{Nardin:2011NatPhys,Sanvitto:NPhot2011}. 


\begin{figure}
    \centering
    \parbox[c]{0.28\textwidth}{\includegraphics[width=0.27\textwidth,clip]{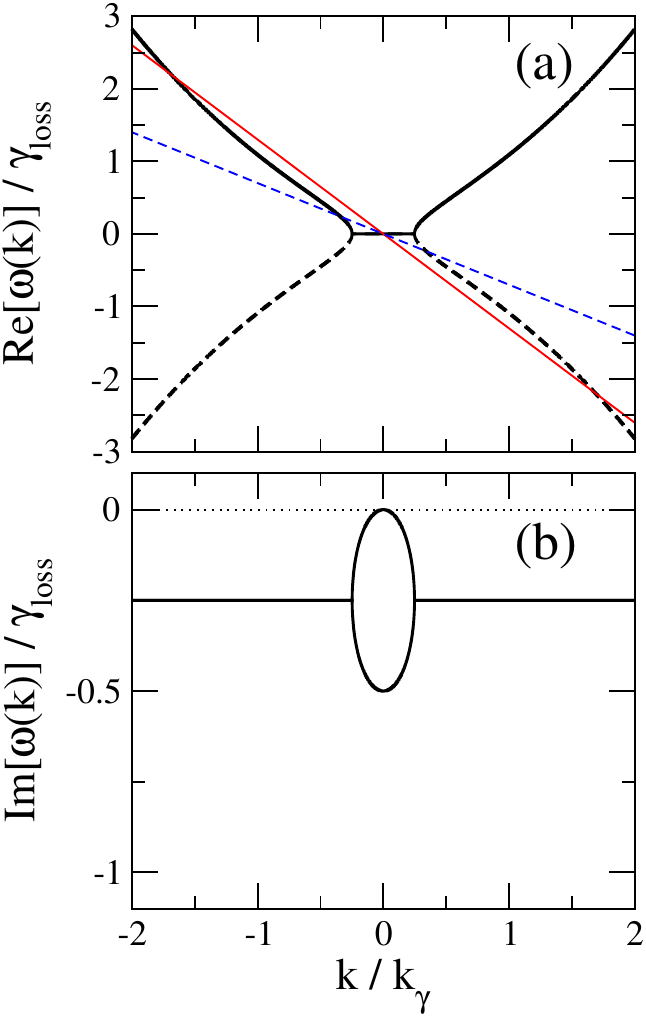}}
\parbox[c]{0.71\textwidth}{\includegraphics[width=0.7\textwidth,clip]{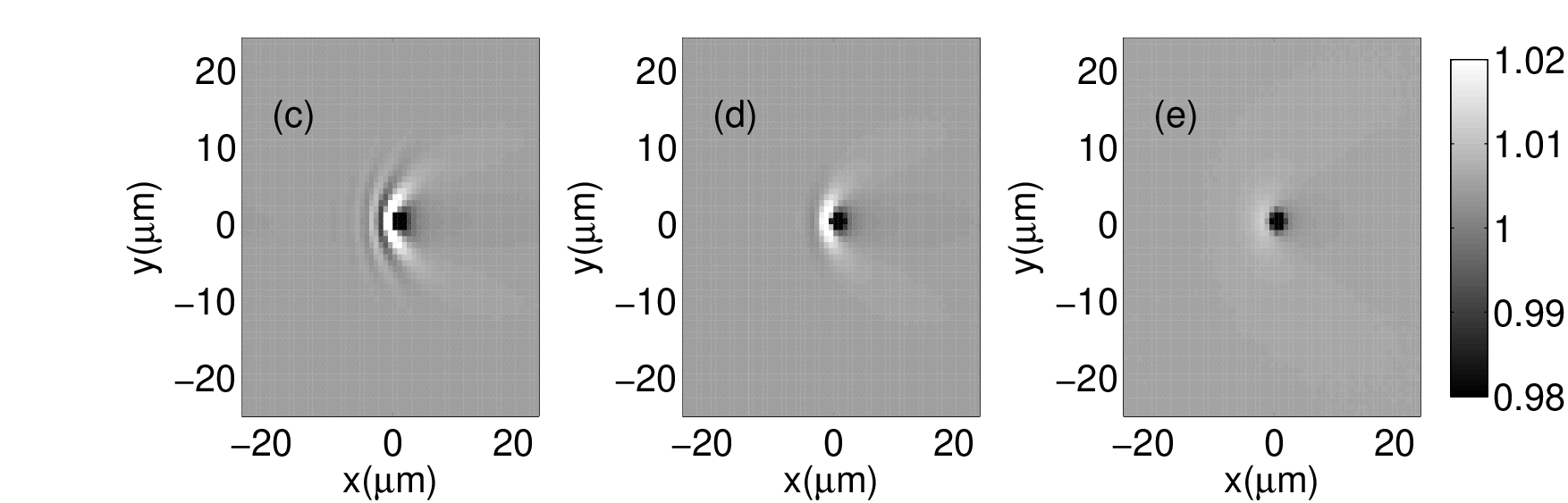}\\
    \includegraphics[width=0.7\textwidth,clip]{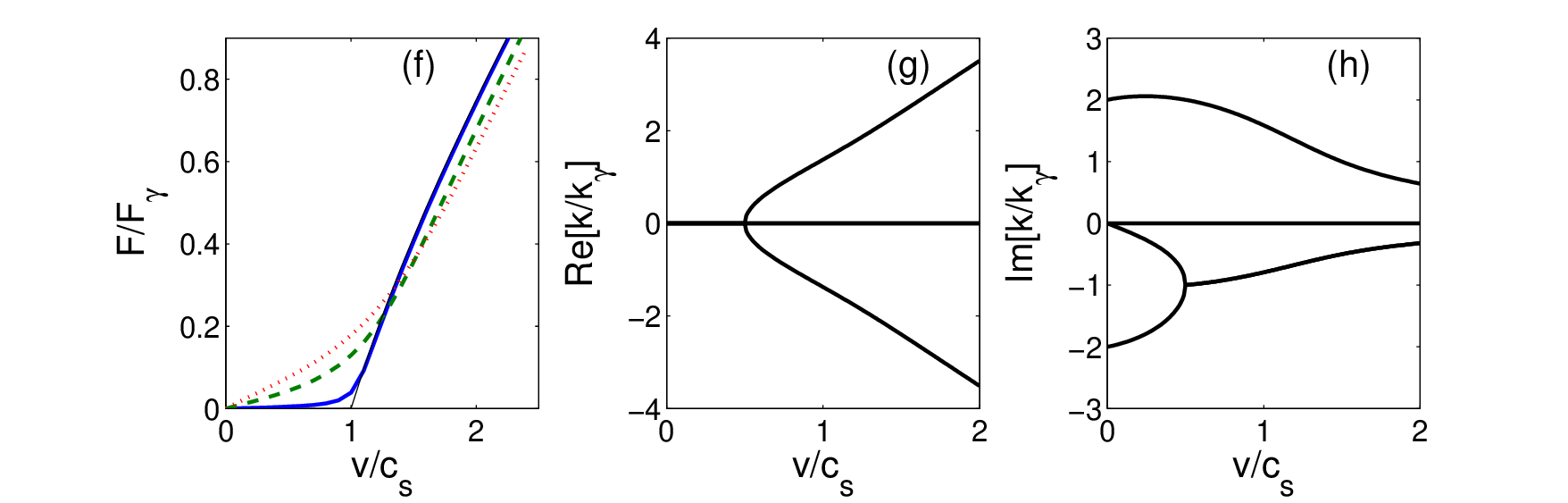}}
    \caption{ Left panels: plot of the real (a) and imaginary (b) parts of the dispersion of the collective excitations of a non-equilibrium condensate under incoherent pumping at $P/\gamma_{\rm loss}=2$ with $\gamma_{\rm loss}/g_{\rm nl}\,|\psi|^2=1$. For small $k$ values, the dispersion displays a diffusive plateau with a zero real part and a quadratically growing imaginary part. The oblique straight lines correspond to the Landau-Cherenkov $\omega=k v$ condition for a particle moving moving at $v/c_s=0.7$ (blue dashed) or $v/c_s=1.3$ (red solid) in the negative $x$-direction. Wavevector is measured in units of $k_\gamma$ such that $\hbar k_\gamma^2/2m = \gamma_{\rm loss}$. Right (c-h) panels: generalized Landau criterion for non-equilibrium condensates. Panels (c-e) show the density perturbation induced in a fluid moving in the positive $x$ direction by an impurity at rest. The different panels refer to different values of the condensate velocity $v/c_s =1.5, 1, 0.4$ across the (equilibrium) speed of sound $c_s=\sqrt{g_{\rm nl}\,|\psi|^2/m}$. Panel (f) shows the (normalized) force exerted by the fluid on the defect as a function of the condensate velocity $v/c_s$ for different values of the non-equilibrium parameter $\gamma_{\rm loss}/g_{\rm nl}\,|\psi|^2=0,\,0.1,\,1,\,2$. For the smallest $\gamma_{\rm loss}$ values, the sudden onset of friction that is visible in the vicinity of the critical speed $v/c_s= 1$  is a non-equilibrium counterpart of the Landau critical velocity of equilibrium superfluids~\cite{pitaevskii2016bose,Carusotto:Ducks2013}.
    Panels (g-h) show a cut of the real and the imaginary part of the wavevector of the collective excitations emitted in the negative $x$ direction as a function of the speed for an intermediate loss case with $\gamma_{\rm loss}/g_{\rm nl}\,|\psi|^2=1$. The sudden onset of friction in (f) can be related to the singular point that is visible in (g-h) slightly below $v/c_s= 1$. Panels (c-h) are adapted from~\cite{Wouters:PRL2010}.  The whole figure is adapted from~\cite{Carusotto:CRAS2025}.\label{fig:incoh_mean-field}}
\end{figure}

\paragraph{Superfluidity in driven-dissipative condensates}


The situation is even more subtle in $U(1)$-spontaneous-symmetry-breaking non-equilibrium condensates where diffusive Goldstone modes are present. In spite of the pioneering development reported in~\cite{Amo:Nature2009}, a complete experimental study of this configuration is still lacking. The main conceptual difficulties of superfluidity of non-equilibrium condensates are summarized in Fig.\ref{fig:incoh_mean-field}. 

In the left (a-b) panels, we plot the dispersion of collective excitations in the frame of the fluid: the oblique straight lines describe the modes that can be resonantly excited by a defect moving at speed $v$ in the negative $x$ direction. As the key feature, the intersection with the real part $\textrm{Re}[\omega(\kk)]$ of the dispersion [panel (a)] is non-empty for any value of the defect speed $v$. As such, one might expect that the critical speed vanishes and no superfluidity effect in the sense of the Landau criterion can be ever observed. 

In contrast to this naive prediction, the numerical solutions of the driven-dissipative field equation under incoherent pumping shown in panels (c-e) for different defect speeds predict a clear transition between markedly different behaviors~\cite{Wouters:PRL2010}: a fast moving defect creates the usual pattern of parabolic fringes [panel (c)], while a slowly moving defect only induces a weak and spatially localized perturbation in the fluid [panel (e)].
The critical speed for the transition between the two regimes can be localized by looking at the dependence of the drag force experienced by the defect as a function of its speed. Following~\cite{Astrakharchik:PRA2004}, the drag force can be estimated from the numerics as 
\begin{equation}
F=-\int\!d^2\rr\,n(\rr)\,\nabla_\rr V_{\rm def}(\rr)\,,
\end{equation}
where $V_{\rm def}(\rr)$ is  the interaction potential between the defect and the fluid and  $n(\rr)$  is the perturbed density profileof the fluid. 
The result is plotted in panel (f): a clear threshold is visible around the sound speed of the corresponding conservative fluid, $v\sim c_s$ with $c_s=\sqrt{g n/m}$. The weaker the loss rate compared to the interaction energy in the fluid $\gamma/gn$, the sharper the threshold.

This unexpected result can be physically understood introducing a generalized form of the Landau criterion that keeps track of the complex-valued nature of the dispersion of the collective excitations in the driven-dissipative fluid. Working in the reference frame of the defect, 
the system is overall stationary in time and the emitted perturbation has a zero frequency $\omega'=0$. Assuming that the defect breaking translational invariance is a spatially point-like one, modes at any wavevector can be excited. In particular, it is useful to characterize the modes that satisfy the $\omega'=0$ emission condition for generic complex values $\tilde{\kk}$ of the wavevector, \begin{equation}
\omega'=\omega(\tilde{\kk})-\tilde{\kk}\cdot \vv=0\,.
\end{equation}
Restricting for simplicity to $k_y=0$ modes propagating in the direction $x$ parallel to the speed $\vv$, panels (g-h) show the real and imaginary parts of the wavevector $k_x$ of such modes as a function of the velocity $v$: a singularity is clearly visible that separates a low-speed regime of a purely imaginary $k_x$ from a high-speed regime featuring complex-valued $k_x$ with a finite real part. As a finite real part of the wavevector is typically associated --for instance via the Poynting vector of electromagnetic waves in media-- to the transport of energy, it is natural to associate this singularity to the onset of friction. The negative imaginary part of $k_x$ for these modes indicates that they propagate in the negative-$x$ direction and correspond to the density modulation created by the defect at $x<0$ positions in front of its motion. This connection of the mathematical singularity and the physical onset of friction is confirmed by the reasonable match between the location of the singularity in panels (g-h) and the threshold in the drag force visible in panel (f). A similar analysis in terms of complex wavevector modes was carried out for the coherent pumping case in~\cite{Amelio:PRR2020}.

These arguments allow to overcome the conceptual difficulties inherent to the application of the Landau criterion to driven-dissipative fluids and provide a satisfactory qualitative understanding of the superfluidity properties in response to a moving defect. Still, this theory is hardly useful for a quantitative determination of the superfluid vs. normal fractions of the fluid. For this purpose, it is therefore important to explore how the other definitions of superfluidity apply to our context of driven-dissipative fluids of light.

\subsubsection{Response to rotations} 
The very fast value of the characteristic time- and velocity-scales of fluids of light hinders a direct realization of a rotating bucket experiment by mechanically rotating the cavity device. In the last years, alternative approaches have been successfully implemented, where time-dependent rotating shapes of the external potential and of the gain profile are generated via the interference of incoherent pump beams with different angular momentum and different frequency. In this way, the nucleation of quantized vortices in response to a sufficiently fast rotation has been demonstrated~\cite{Gnusov:SciAdv2023,DelValle:NanoLett2023}. 

A conceptually different strategy is to replace the mechanical motion with a so-called synthetic magnetic field for light: a suitable engineering of the photonic devices allows to introduce an effective vector potential term \eqref{eq:H+A} into the dynamics of the photon 
field~\cite{ozawaRMP2019topological}. Following the discussion in Sec.\ref{sec:superfl_bucket}, detecting the response of the polariton fluid to the vector potential may shine new light on its superfluidity properties, as theoretically pioneered in~\cite{Keeling:PRL2011,Juggins:NatComm2018}. As we are going to mention later on in Sec.\ref{subsec:supersolids}, this technique appears as a most promising tool to establish mechanical superfluidity properties of supersolid states of photonic matter.

\subsubsection{Metastable supercurrents}
\label{sec:superfl_metastable}
Over the years, a number of experiments has investigated metastability of supercurrents in driven-dissipative condensed fluids. The spontaneous appearance of stationary quantized vortices in condensates under the effect of disorder was highlighted in~\cite{Lagoudakis:NatPhys2008}. 

A controlled generation of supercurrent states can be obtained in fluids of light by seeding the condensation process with a short pulse of external coherent field with a suitable spatial profile, which forces condensation to occur in the desired state.
In~\cite{Sanvitto:NatPhys2010}, this idea was implemented in a parametric oscillation configuration in a planar geometry: a short laser pulse in a Laguerre-Gauss form injected into the signal mode at the time of condensation was used to trigger condensation into a vortex state. Well after the seed photons have decayed, the signal-idler condensate maintains the vortex phase profile which can be tracked over macroscopically long times, showing metastability of the corresponding supercurrent state until the vortex exits the cloud.

A further and comprehensive evidence of the metastability of supercurrents in ring traps was recently reported in~\cite{Yao:Optica2025}. In this geometry, vortices can not disappear by drifting out the cloud, so the supercurrent can survive for indefinitely long times as confirmed by the experiment. Interestingly, in this experiment the supercurrent was not imprinted at the time of condensation but was mechanically induced at a later time.

On top of the energetic barrier mentioned earlier for conservative fluids, decay of the supercurrent in driven-dissipative condensates is further hindered by the gain and loss dynamics that forces the photon density to stay constant~\cite{Wouters:PRL2010}. In laser physics, a similar mechanism underlies the robustness of laser emission against mode jumps. 

From an applied perspective, it is worth mentioning that the response of a light field to rotations is at the heart of a widespread concept of gyroscope device, the so-called ring laser gyroscope~\cite{Macek:APL1963}. 
When the set-up is set into overall rotation, the Sagnac effect induces a differential shift in the frequency of counter-propagating modes around the ring cavity. The instantaneous angular velocity can be directly read out from the beat note of the two modes. 

Such devices are routinely applied in aerospace technologies as well as for studies of fundamental physics~\cite{Giovinetti:FQST2024}. Their operation crucially relies on the absence of mode jumps in response to rotation, on the simultaneous laser operation in a pair of counter-propagating modes and on the fact that the phases of the two modes are not locked. While all these features are normally discussed in the language of laser physics, we anticipate that new insight can be obtained by using the concepts of superfluid light that we have reviewed in this Section.

\section{New perspectives}
\label{sec:perspectives}

In the final part of this article, we are now going to see how the concepts developed so far are a powerful workhorse to push forward the study of quantum fluids of light in new directions.

\subsection{Supersolids}
\label{subsec:supersolids}

In the last decades, an intense debate has taken place within the quantum condensed matter community about the possibility of observing supersolid states of matter that combine frictionless flow (super) with spatial order (solid). This research direction was initiated by theoretical works~\cite{Andreev:JETP1969,Chester:PRA1970,Leggett:PRL1970} and exploded with the claim of experimental observation of a supersolid state in a sample of solid $^4$He cooled to very low temperatures~\cite{Kim:Science2004,Kim:Nature2004}. While solidity of the sample can be assessed by Bragg diffraction or from its shear modulus, superfluidity was probed by implementing its definition in terms of the response to rotation. Along the lines of \eqref{eq:f_n_I}, the experimental evidence supporting the claim of superfluidity was a sudden drop in the moment of inertia of the Helium sample embedded in a torsional oscillator device. Soon after, these claims of superfluidity were questioned by follow-up experiments that suggested alternative explanations  of the observations, most likely as an effect of the temperature dependence of the shear modulus of the solid~\cite{Day:Nature2007,Syshchenko:PRL2010,Balibar:Nature2010,Kim:PRL2012,Hallock:PhysToday2015}. To date, there is no convincing evidence of a supersolid state in Helium systems.

While the quest for supersolid states of liquid Helium is still facing serious obstacles,  a new avenue for realizing supersolid states of matter was identified in ultracold atomic gases~\cite{Boninsegni:RMP2012}. In suitable configurations a Bose-Einstein condensate can in fact spontaneously develop a spatially periodic modulation of the density under the effect of dipolar interactions~\cite{Recati:NatRevPhys2023} or spin-orbit coupling~\cite{Li:Nature2017} or interaction with cavity modes~\cite{leonard2017supersolid}, while maintaining a robust long-range order in the phase of the macroscopic matter field. Given the concurrent breaking of the condensate phase symmetry and of the translational one, these novel states of matter can be seen as examples of supersolid behavior. As a key difference to solid Helium experiments, atomic ones start with a pre-formed superfluid order that eventually develops a spatial order. The fact that each site of the spatial lattice contains a large number of atoms makes the superfluid order robust to quantum and thermal phase fluctuations. 

This field of research has now reached full maturity:  many of the crucial features of a supersolid have finally received a clear experimental verification, including superfluid flow~\cite{Biagioni:Nature2024}, quantized vortices~\cite{Casotti:Nature2024}, and, most importantly, the multiple branches of collective excitation associated to the different spontaneously broken symmetries, namely the superfluid sound and the phonon modes in the spatial crystal~\cite{Guo:Nature2019,Natale:PRL2019,Tanzi:Nature2019,Chisholm:Science2026}. 
Among the next steps, we may mention a direct~\cite{Tanzi:Science2021,Norcia:PRL2022} measurement of the response of the supersolid to a synthetic magnetic field to highlight the reduced moment of inertia and a measurement of the transverse phonon mode to extract the shear modulus of a supersolid in $d\geq 2$~\cite{Senarath:PRA2025}.

As a major new avenue of research, recent atomic experiments have pushed forward the investigation of supersolid features in driven-dissipative conditions~\cite{Liebster:PRX2025}. Here, the atomic gas is kept in a non-equilibrium state by a periodic modulation of the atom-atom interaction strength: as a result of parametric scattering mechanisms from the condensate into finite-momentum modes, a two-dimensional spatial density modulation spontaneously appeared for a sufficiently strong modulations. A subsequent work by the same group has characterized the sonic vs. diffusive nature of the three phase, longitudinal and transverse collective excitation modes of the two-dimensional driven-dissipative supersolid~\cite{Liebster:NatPhys2025}.

In parallel to these advances in ultracold atomic gases, the last few years have witnessed a rapid development in the study of supersolid states in driven-dissipative fluids of light. This Section is devoted to a brief account of recent accomplishments and to a sketch of the most interesting open problems that the new systems allow to investigate in the next future.

\subsubsection{First generation of experiments}

The idea of secondary instabilities leading to the successive breaking of several symmetries is well known in the theory of pattern formation in nonlinear dynamical systems~\cite{Cross:RMP1993}. 
In spite of the intense work devoted to spatiotemporal patterns in different models of nonlinear optics~\cite{Lugiato:Varenna}, only a few realizations of a concurrent breaking of phase and translational symmetries in the optical context are however present in the literature, e.g. in the context of Turing patterns in laser devices~\cite{Bao:PRR2020}. On top of this, only little attention has been so far given to the symmetries that are spontaneously broken at the phase transitions and none --to the best of our knowledge-- to the collective excitation modes in the different regimes. This gap is presently being filled by the surge of experimental activity on supersolid states of the fluid of light.

Experimental investigations of supersolid states of polariton condensates were recently reported in a photonic crystal waveguide device~\cite{trypogeorgos2025emerging} and in a microcavity filled with a nematic liquid crystal layer~\cite{Muszynski:arXiv2025}. In both cases, supersolidity was assessed from the long-range coherence of the field and from the onset of a periodic spatial modulation of the density profile. On the other hand, work is in progress in the direction of investigating the mechanical superfluidity properties.

Even though these two independent and almost simultaneous experiments are closely related from the point of view of the broken symmetries, the microscopic mechanisms leading to the supersolid state are very different in the two cases and require a separate discussion.

\paragraph{Cascaded parametric scattering}
The mechanism underlying supersolidity in~\cite{trypogeorgos2025emerging} is based on a parametric oscillation process driven by the polariton condensate that spontaneously appears in the cavity under incoherent pump and leading to the macroscopic occupation of a pair of side-band states. As such, the origin of supersolidity can be intuitively understood in terms of a secondary instability mechanism on top of a laser oscillation.

\begin{figure}
    \centering
    \includegraphics[width=0.65\textwidth]{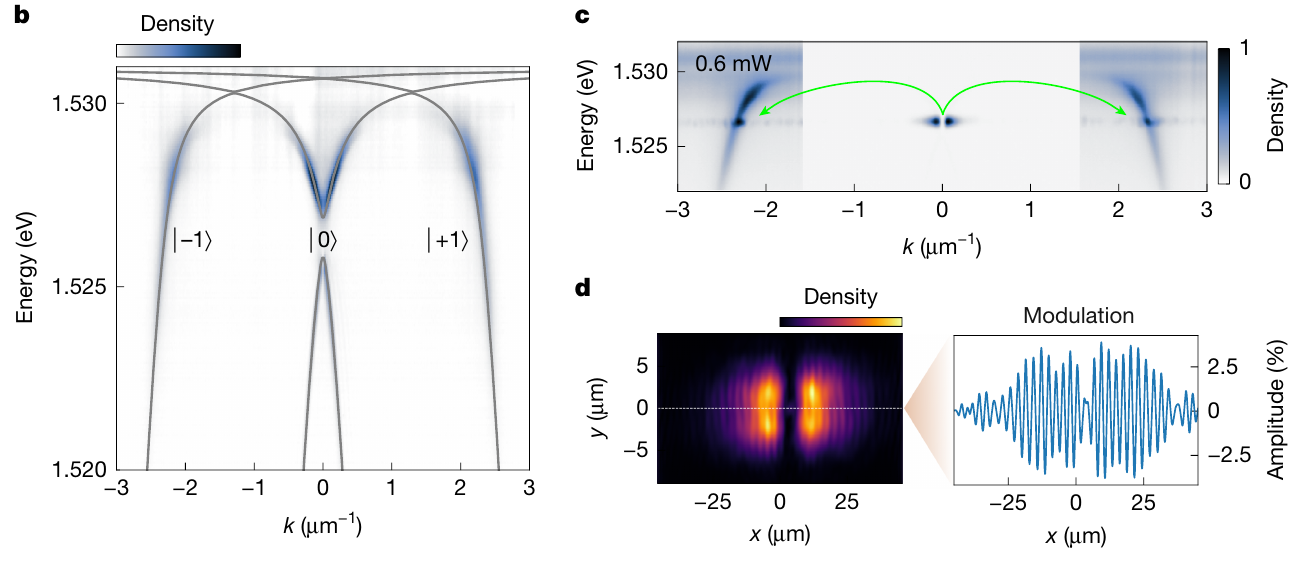}
    \vspace{0.05\textwidth}
 \includegraphics[width=0.3\textwidth]{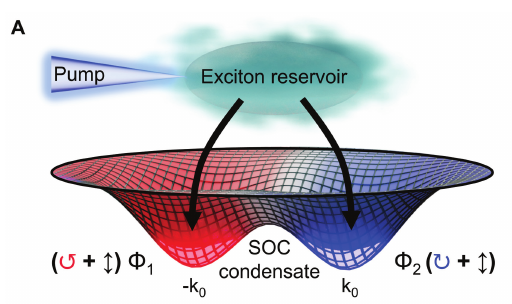}
    \caption{Sketch of the mechanisms for polariton supersolidity. The left and central panels are taken from~\cite{trypogeorgos2025emerging} and illustrate parametric cascade processes: 
    the left (b) panel shows the dispersion of the polariton branches in the linear regime. The $0$ state is at the top of the central branch, while the $\pm 1$ lie on the side branches. The center-up (c) panel the shows parametric process (green arrows) and the wavevector-frequency distribution of light emission in the supersolid state. The center-bottom panel illustrates the spatial modulation in the density profile of the supersolid state. The bound-state-in-continuum nature of the $0$ mode reflects into the bi-lobed shape of its emission in the (c) panel and in the dark stripe in the center of the spatial distribution (d).
    The right panel (A) illustrates the two-mode condensation mechanism of~\cite{Muszynski:arXiv2025}, where condensation concurrently occurs in the two modes ar $\pm k_0$. 
\label{fig:supersolid_scheme}}
\end{figure}

As it is illustrated in Fig.\ref{fig:supersolid_scheme}, a suitably designed photonic crystal waveguide device is used to force condensation to occur first in the $k=0$ state at the top of the central band. Condensation in this specific state is favored by its negative mass, that tends to spatially confine the condensate under the effect of the positive interaction energy~\cite{Tanese:NatComm2013}, as well as by its longer lifetime stemming from its  bound-state-in-continuum~\cite{Hsu:NatMat2016} nature, with an interferentially-suppressed radiative decay rate. 
At a higher pump power above a second threshold, stimulated parametric scattering processes from the $k=0$ condensate into states on the $\pm 1$ sidebands of the photonic crystal waveguide device at finite wavevectors $\pm q$ [green arrows in panel (c)] set in. Interference between these field components is responsible for the spatial density modulation olf the supersolid state that is visible in Fig.\ref{fig:supersolid_scheme}(d).

As a key feature of the supersolid state, the long-range phase coherence was established from the interference between different spatial components of the field. On top of this, the non-rigidity of the supersolid modulation was inferred by the variation of the modulation wavevector with the pump parameters. On the other hand, evidence for a (at least partially) random position of the fringes in the spatial modulation pattern was only indirect in the first experiment~\cite{trypogeorgos2025emerging}. 
A firm proof of the shot-to-shot randomness of the fringe position was reported soon later in~\cite{Meng:NatNano2026} using a device with a similar geometry but based on different materials. Also here, the observation of supersolidity was supported by an interferometric measurement of long-range spatiotemporal coherence. 

A detailed theoretical account of the parametric process underlying supersolidity was reported in~\cite{Nigro:PRL2025}.
In analogy with the optical parametric oscillators discussed in Sec.\ref{sec:expt_OPO}, we can expand the field in the $k=0,\pm q$ modes involved in supersolidity in the $\Psi_{0,\pm}$ components,
\begin{equation}
    \Psi(x,t)= \Psi_{-1} \, e^{-iqx}  + \Psi_0 + \Psi_{+1}\, e^{ i q x} 
\end{equation}
As a main difference from (\ref{eq:OPO_p}-\ref{eq:OPO_i}), in the equation of motion for $\Psi_0$ the coherent pumping is replaced by a incoherent pump one that preserves the $U(1)$ symmetry.
As soon as the external incoherent pump exceeds the threshold for condensation, the $0$ state acquires a macroscopic amplitude $\Psi_0$. When this population grows beyond a second threshold, parametric oscillation into the $\pm 1$ modes sets in and these modes also acquire a macroscopic occupation $\Psi_{\pm 1}$.
Interference between the three components gives rise to the density modulation. 

As the three field components $\Psi_{0,\pm 1}$ are locked by the parametric processes, they oscillate at the same frequency and the overall field pattern is stationary in time.
In the supersolid state, two continuous $U(1)$ symmetries
\begin{eqnarray}
\Psi_{0} \to \Psi_{0} e^{\pm i \varphi} \\
\Psi_{\pm 1} \to \Psi_{\pm 1} e^{i (\varphi\pm \varphi')}\,,
\end{eqnarray}
with arbitrary angles $\varphi,\varphi'$ are spontaneously broken, which respectively correspond to a global rotation of the condensate phase by the angle $\varphi$ and to a shift of the spatial order due to the interference of the $\pm 1$ modes by a distance $\delta x=(\varphi'-\varphi)/q$. In contrast, standard condensates are only characterized by the $\varphi$ symmetry; optical parametric oscillators by the $\varphi'$ one: supersolids enjoy both symmetries.


The threshold towards the supersolid state is associated to a dynamical instability of the initially empty $\pm 1$ modes and is visible in a calculation of the Bogoliubov dispersion around a state with a population restricted to the $0$ mode.
While this instability mechanism is similar to the one at play in atomic supersolids, its onset is controlled by different quantities in the two cases. 
In the photonic case, modes on the $\pm 1$ sidebands resonant with the $0$ mode are always available for parametric scattering; the crossing of the instability threshold is controlled by the increase in the amplitude $\Psi_0$ of the $0$ mode that sets the strength of the parametric coupling. As soon as parametric scattering  exceeds losses, a dynamical instability appears, parametric oscillation sets in and coherent fields $\Psi_\pm$ build up in the modes at $\pm q$. A similar mechanism is at play in the driven-dissipative atomic supersolids of~\cite{Liebster:PRX2025,Liebster:NatPhys2025}.

In standard equilibrium atomic supersolids based, e.g., on dipolar gases~\cite{recati2023supersolidity}, the critical point is instead reached when the roton minima at $\pm q$ in the Bogoliubov dispersion on top of the $0$-only state are pushed to zero energy and the modes becomes correspondingly soft~\cite{Petter:PRL2019}. The frequency of the roton minima are determined by the relative strength of the contact and the long-range dipolar interactions between the atoms; as losses are irrelevant here, the instability of the homogeneous state towards the supersolid modulation develops as soon as the roton minima are close enough to zero-energy.

\paragraph{Concurrent condensation in two modes}

The mechanism for supersolidity in the almost simultaneous experiment reported in~\cite{Muszynski:arXiv2025} is quite different and is closely related to one of the atomic experiments of~\cite{Li:Nature2017}. As it is illustrated in the right panel of Fig.\ref{fig:supersolid_scheme}, supersolidity stems here from the concurrent  condensation into a pair of degenerate minima of the polariton dispersion relation at $\pm k_0$ that concurrently acquire a macroscopic occupation. This peculiar shape of the polariton dispersion with two degenerate minima is due to the coupling of the orbital and polarization degrees of freedom in a suitably engineered liquid crystal layer.

As condensation occurs concurrently but independently in the two degenerate modes, a total of two $U(1)$ symmetries are spontaneously broken in the supersolid.
Interference between the two condensates is then responsible for the overall spatial modulation of the fluid at $2k_0$ and the position of the modulation is fixed in space by the relative phase of the two condensates. This was shown to randomly fluctuate from shot to shot of the experiment. Quite interestingly, evidence was also reported for dislocations in the spatial modulation fringes, that can be associated to quantized vortices in the relative phase of the condensates.


\subsubsection{Open questions}

In the previous Subsection, we have seen how the first generation of experiments~\cite{trypogeorgos2025emerging,Muszynski:arXiv2025,Meng:NatNano2026} has highlighted a number of key properties of the supersolid state of light, in particular the concurrent presence of long-range order in both the phase and the spatial modulation and the shot-to-shot randomness of the modulation position. On the other hand, the collective excitation modes and the mechanical superfluid properties are still a topic of intense debates.

Concerning collective excitations, first theoretical steps are appearing in the literature~\cite{Muszynski:arXiv2025,Grudinina:arxiv2026,Verdier:MSc2026}, but no consensus has yet been reached on the basic features of their dispersion and, in particular, on the impact of the driven-dissipative nature on them. We anticipate that a deep experimental insight on their properties will be achievable via 
the same spectroscopical techniques used in~\cite{Claude:PRL2022,claude2025observation} to measure the collective excitation in spatially uniform fluids. 

So far, all theoretical works on supersolids of light have focused on one-dimensional geometries and very little is known about supersolids with two-dimensional spatial modulations. The generation of such states does not appear to be a straightforward task in the existing platforms where one-dimensionality is hard-wired in the sample structure. Realization of two-dimensional supersolids will anyway be of utmost relevance to complete our understanding of the solidity properties with a measurement of the transverse sound modes. The properties of this mode depend in fact on the shear modulus, a quantity that characterizes the solid nature of the system. Quite interestingly, experimental evidence for this mode is still lacking also in the atomic case~\cite{Senarath:PRA2025}.

The superfluidity properties of photon supersolids are perhaps even more intriguing. Superfluidity played a crucial role in solid-He experiments, where the (eventually retracted) claims of supersolidity originated from the interpretation that the sudden drop of the moment of inertia observed in experiments was a signature of the onset of superfluidity (no evidence was ever reported for a long-range phase coherence of the matter field in these systems).
In atomic supersolids, evidence for superfluidity has been searched in Josephson supercurrents~\cite{Biagioni:Nature2024} and scissor modes~\cite{Tanzi:Science2021,Norcia:PRL2022}. 

As we have discussed in Sec.\ref{sec:superfluidity}, superfluidity is a much more subtle concept in driven-dissipative photonic systems and obtaining a firm evidence of it requires quite some attention. 
A naive application of the Landau criterion is made difficult by the possibly diffusive nature of the Goldstone modes which may hinder a precise identification of a critical speed from a measurement of the friction exerted onto a moving defect. A further difficulty may come from umklapp processes that, in analogy to condensates in periodic potentials~\cite{Price:PRA2023}, may be responsible for scattering processes between the two condensates at $\pm k_0$ even for very small defect speeds.

Measurements of metastability of supercurrents in ring-shaped geometries offer direct evidence of superfluidity and are expected to be applicable to supersolid systems. To this purpose, one needs to stabilize the supersolid state in the ring geometry: the fluid has to uniformly occupy the full ring, 
the fringes of the spatial modulation should be at all points oriented along the radial direction, and the fluid has to be put into a supercurrent state around the ring. 
Interesting first steps in this direction was recently reported in~\cite{Kozhevin:arXiv2025}.

Another interesting strategy is to 
directly measure the superfluid fraction from the response of the fluid to a transverse vector potential according to \eqref{eq:f_n_AT} In the supersolid state, we anticipate that the motion of the normal component will show up as a global motion of the density modulation pattern.  Given the relatively large length scale of supersolid samples, a promising candidate to generate the required synthetic magnetic field may be based on the electro-magneto-static schemes developed in~\cite{lim_electrically_2017}.

\subsection{Strongly interacting fluids}
\label{subsec:strongly_interacting}
All our discussion so far has focussed on weakly interacting photon gases, where the optical nonlinearity per photon is weak and the observation of significant interaction effects requires the presence of a huge number of photons. In this regime, the mean-field theory is accurate and a formulation in purely nonlinear optical terms with no mention of many-body concepts would be formally equivalent, even though it might hide much of the underlying physics. 

The situation is totally different when the optical nonlinearities are so strong that single photons are able to significantly affect the propagation of other photons~\cite{chang2014quantum}. In this regime, the photon gas is a strongly interacting one and sizable quantum correlations develop among the different individual photons forming the quantum fluid of light.
In this final section we will briefly outline how the concept of collective excitation can be fruitfully applied also in this regime and how measurements of their dispersion relation can offer a deep insight into the many-body physics of driven-dissipative, strongly correlated fluids of light.

An active research is presently being devoted to the realization of different states of matter in this context, from quantum Hall fluids~\cite{Roushan:2016NatPhys,clark2020observation,Wang:Science2024} to Mott insulator states~\cite{Ma:Nature2019}. In what follows we will concentrate on this latter example, for which not only a strongly correlated steady-state was observed but first observations of collective excitation dynamics were also reported.

\begin{figure}
    \centering
    \includegraphics[width=0.8\textwidth]{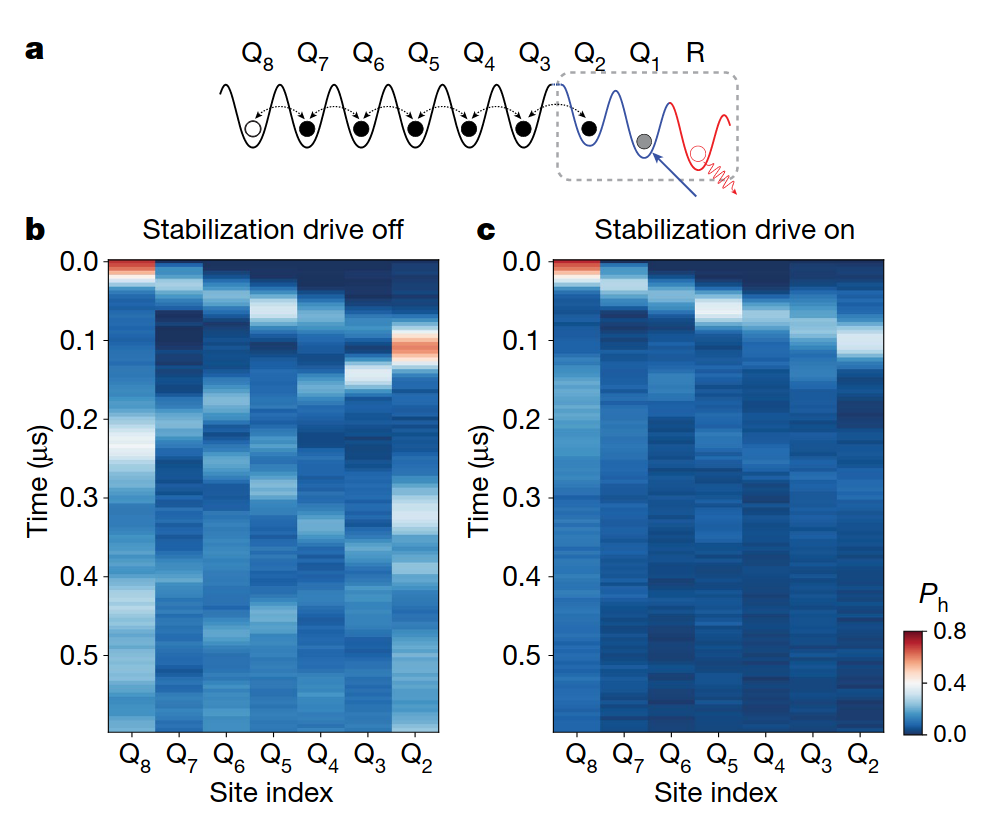}
    \caption{Colorplot of the temporal evolution of the probability of having a hole in each of the $Q_8-Q_2$ sites of a Mott insulator state of photons. The hole is initially prepared in the leftmost site $Q_2$ and left to freely evolve in time. In the left panel (b), no replenishing mechanism is active, so the hole perform a free back-and-forth motion. In the right panel (c), a frequency-dependent incoherent pump is active on the $Q_2$ site [panel (a)] and fills the hole as soon as it arrives there. 
    Figure adapted from~\cite{Ma:Nature2019}.
\label{fig:Mott_dyn}}
\end{figure}

\subsubsection{Mott insulators of photons}

After preliminary steps including the observation of strong photon antibunching and two-photon bound states in a beam of Rydberg polaritons propagating through an atomic gas~\cite{peyronel2012quantum,firstenberg2013attractive},
the first realization of a spatially-extended Mott insulator state of impenetrable photons was made in~\cite{Ma:Nature2019} using a circuit-QED platform operating in the microwave domain. The physical system can be described as a one-dimensional Bose-Hubbard (BH) model of Hamiltonian 
\begin{equation}
    H=-J\sum_{<i,j>} \left[\ahd_i \ah_j + \textrm{H.c.}\right] + \frac{U}{2}\sum_i  \ahd_i \ahd_i \ah_i \ah_i
\end{equation}
where the $i,j$ indices run over the lattice sites, $J$ is the hopping amplitude between nearest-neaighbor sites, $U$ is the on-site interaction energy, and $\ah_i,\ahd_i$ are destruction and creation operators for a photon on the site $i$. The Mott insulator steady-state occurs at large $U/J$ and is characterized by having a precisely integer numebr of photons per lattice site and suppressed density fluctuations~\cite{bloch2008many}.

More in specific, in the experiment~\cite{Ma:Nature2019} each site of the BH lattice consists of a transmon acting as a nonlinear resonator: the anharmonicity is provided by a Josephson junction element and is responsible for strong on-site two-body interactions between photons. Each transmon is then capacitively coupled to its neighbors, which produces nearest-neighbor tunneling.

Stabilization of the Mott insulator state with one photon per site is then obtained via a suitably designed frequency-dependent incoherent pumping scheme~\cite{Carusotto:CRAS2025}: 
photons get quickly injected when a site is empty but injection of a second photon into the same site is blocked by the many-body gap protecting the Mott insulator state. Such a pumping mechanism is therefore able to quickly refill a hole excitation resulting from a photon loss event and,  at the same time, the frequency-dependence of the pumping prevents injection of extra particle on top of the Mott insulator, which would require crossing the many-body gap.
Theoretical studies of this model including the driven-dissipative features~\cite{Kapit:2014PRX,Lebreuilly:CRAS2016,Biella:PRA2017,Lebreuilly:PRA2017} have anticipated a phase diagram quite similar to the one of equilibrium systems of cold atoms~\cite{bloch2008many}. It is interesting to note that related autonomous stabilization schemes are under study also for other kinds of many-body states~\cite{Mi:Science2024}.

\subsubsection{Collective excitations: experiments}

After having assessed the fidelity of the stabilization procedure by measuring the photon number distribution in each site, the same experiment~\cite{Ma:Nature2019} has moved the first steps in the study of the collective dynamics of the Mott insulator.

The left panel of Fig.\ref{fig:Mott_dyn} shows the motion of a hole excitation across the Mott insulator while the pumping mechanism is switched off. The hole is initially prepared on the left-most site $Q_8$ and is then let freely propagate. Its overall back-and-forth motion across the system, from the left-most site $Q_8$ to the right-most one $Q_2$ and then back to $Q_8$ is well visible in the figure. At longer times, the hole wavefunction expands under the effect of the curvature of its dispersion and its motion is no longer recognizable. This rich dynamics is a clear evidence that the Mott insulator state is not just a collection of independent filled states but displays a non-trivial coherent many-body dynamics determined by the interplay of tunneling and strong on-site interactions.

On the right panel of the same figure, the same hole motion is illustrated in the case when the pumping mechanism is active on the right-most site $Q_2$. Here, the hole gets immediately replenished as soon as it reaches the pumped site. This observation illustrates in a clear way the mechanism under which the perfect Mott insulator state gets restored after a photon loss event. 

\begin{figure}
    \centering
    \includegraphics[width=0.8\textwidth]{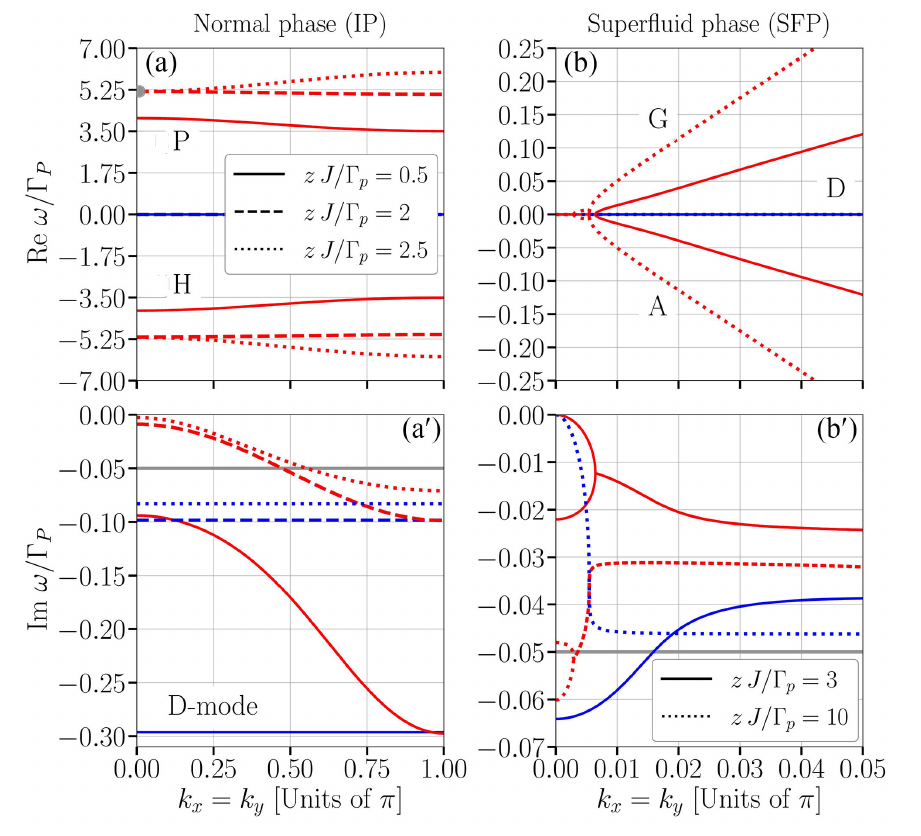}
    \caption{Theoretical calculation of the real (top) and imaginary (bottom) parts of the dispersion of collective excitations in a strongly interacting photon fluid. 
    The left panels refer to operating points in the insulating phase. The right panels refer instead to operating points in the superfluid phase.
    Figure adapted from~\cite{Caleffi:PRL2023}.
\label{fig:Mott_th}}
\end{figure}

\subsubsection{Collective excitations: theory}

This observation of the collective nature  of the excitations of a Mott insulator state of photons called for the development of theoretical models of the dispersion of collective excitations. A first attempt in this direction was presented in~\cite{Caleffi:PRL2023}. 

Here, the oustanding complexity of the driven-dissipative many-body dynamics was unraveled by means of a Gutzwiller approximation where the many-body state is factorized over the different sites. Within this approximation, a phase diagram of the various insulating or superfluid steady-states was determined as a function of the pumping parameters. Analogously to the Bogoliubov calculations for the weakly interacting case, the collective excitation modes were then characterized by linearizing the equations of motion around the stationary-state and extracting their eigenmodes.

The different panels of Fig.\ref{fig:Mott_th} give examples of the dispersion of the collective excitations in the different regimes.
In the insulating phase (IP) illustrated in the left panels, the spectrum includes a pair of gapped particle P and hole H branches with a non-trivial dispersion relation. In particular, the dispersion of the hole branch is related to the motion of holes illustrated in Fig.\ref{fig:Mott_dyn} and its sizable imaginary part corresponds to their quick refilling by the pumping mechanism. 
As the critical point for the transition to the superfluid state is approached (solid to dashed to dotted lines), the gap in the imaginary part closes down (left bottom panel).

The superfluid phase (SFP) illustrated in the right panels is instead characterized by the presence of a soft Goldstone mode whose dispersion tends to zero in both real and imaginary parts in the $k\to 0$ long-wavelength limit. Most remarkably, also in this strongly interacting case, the Goldstone mode features a diffusive rather than propagating behavior at long wavelengths, in close analogy to the weakly interacting case discussed at length in Sec.\ref{sec:incoh_pump}. This suggests the generality of this behavior for any $U(1)$-breaking phase transitions in a driven-dissipative context. 

While insight on the spatial propagation of particle and hole excitations can be extracted by direct imaging of the photon density as shown in Fig.\ref{fig:Mott_dyn}, application of the pump-and-probe techniques reviewed in App.\ref{sec:experimental_techniques_PP} may provide rich information on the dispersion of collective excitations also in the case of a strongly correlated fluid of light.


\section{Conclusions}
\label{sec:conclusions}

In this review article, we have shown how the concept of collective excitations provides crucial insight on the physics of quantum fluids of light and highlights the novel physics stemming from their driven-dissipative nature. 

Depending on the specific pumping configuration used to compensate photon losses, dispersions with very different properties have been theoretically predicted and experimentally observed, culminating in the observation of the diffusive nature of the Goldstone mode of non-equilibrium Bose-Einstein condensates. Thanks to the additional possibilities offered by optical systems, subtle features of the spontaneous symmetry breaking mechanism underlying non-equilibrium condensation could be revealed such as the opening of a gap in the dispersion when the phase of the order parameter is pinned by an external field.
The fundamental understanding of the collective excitations has then offered a comprehensive framework to unravel the rich physics of non-equilibrium superfluids and has helped unraveling the subtle consequences of driving and dissipation on the different exoerimental manifestations of superfluid behavior.

Among the most exciting developments that we may foresee for the coming years in the field of fluids of light, a central place will be occupied by supersolid states combining a long-range phase order with a periodic spatial modulation, and by strongly correlated states of the photon fluid such as Mott insulators and fractional quantum Hall fluids.
We are sure that the concepts of collective excitations reviewed in the article will constitute a powerful workhorse to uncover exciting new physics in a novel context where the driven-dissipative condition interplays with strong interparticle interactions, synthetic gauge fields and complex broken symmetries to realize unprecedented states of strongly correlated matter.  

\section{Acknowledgments}

IC acknowledges financial support by: Provincia Autonoma di Trento (PAT); from Q@TN, the joint lab between University of Trento, FBK-Fondazione Bruno Kessler, INFN-National Institute for Nuclear Physics and CNR-National Research Council; the National Quantum Science and Technology Institute through the PNRR MUR Project under Grant PE0000023-NQSTI, co-funded by the European Union - NextGeneration EU; the Deutsche Forschungsgemeinschaft (DFG, German Research Foundation) via the Research Unit FOR 5688 ``Driven-dissipative many-body systems of ultracold atoms'', project number 521530974.\\ 
AB acknowledges financial support by: the European Union EIC Pathfinder Challenges project ``Quantum Optical Networks based on Exciton-polaritons'' (Q-ONE, Id: 101115575); the French ANR projects HAWQ ("Analogue quantum simulation of the Hawking effect and black hole superradiance"  Grant number: ANR-25-CE47-7323) and FracTrans ("Dissipative phase transitions in fractal and complex nonlinear optical media with in-situ tunability" Grant number: ANR-24-CE30-6983); the National Research Foundation, Singapore (NRF) under its CREATE program via the N-GAP (Nanophotonic GAtes with exciton-Polaritons) project hosted by CNRS@CREATE.
AB is member on the Institut Universitaire de France (IUF)

\appendix


\section{Experimental techniques}
\label{subsec:experimental_techniques}

In this Appendix, we give a few more details on the conceptual ideas underlying the experimental techniques used to measure the dispersion relation of the collective excitation of quantum fluids of light.
A technique is based on measurements of the angle- and frequency-selective spectrum of fluctuations around the stationary state state, the other is based on measurements of the response of the system to small perturbations. While in conservative systems at thermal equilibrium, the two quantities are strictly related by fluctuation-dissipation theorems~\cite{pitaevskii2016bose}, in driven-dissipative systems they are in principle independent and effective equilibrium conditions  dynamically arise only in special cases~\cite{Sieberer:RPP2016}. 

\subsection{Luminescence measurements}
\label{sec:experimental_techniques_luminescence}

Finite temperature as well as quantum fluctuation effects naturally lead to some excitation of the collective modes~\cite{Frerot:PRX2023}. Many theoretical tools are available to estimate the spectrum and the amplitude of these fluctuations, including a Keldysh formalism~\cite{Szymanska:PRL2006,szymanska2007coherence} or quantum optical input-output techniques~\cite{QuantumOptics}.

In what follows, we focus on a formalism based on the stochastic extensions of the field equation \eqref{eq:dd_psi} including noise terms. 
In our weakly-interaction fluid regime, thermal and quantum excitations can be in fact modeled as stochastic fluctuations of the classical field $\psi(\rr,t)$ around the stationary state of \eqref{eq:dd_psi}. 
Such a classical field approach can be derived from the Wigner representation of the state of the in-cavity photon field~\cite{carusotto2013quantum}. 

Under the assumption of weak fluctuations~\footnote{While this assumption may seem innocuous, it is important to keep in mind that it misses the Kardar-Parisi-Zhang physics that stems from nonlinear interactions between fluctuations~\cite{fontaine2022kardar}.}, one can include Gaussian noise terms to the linearized evolution equation \eqref{eq:linearized_coh} or \eqref{eq:linearized_incoh}. This leads to stochastic differential equations in the form
\begin{equation}
i\frac{d}{dt}\left( \begin{array}{cc}
 \delta \psi_\kk \\
  (\delta \psi_{-\kk})^*
\end{array}
\right)= \mathcal{L}(\kk)\, \left( \begin{array}{cc}
 \delta \psi_\kk \\
  (\delta \psi_{-\kk})^*
\end{array}
\right)
+\left( \begin{array}{cc}
 \xi_\kk(t) \\
  (\xi_{-\kk})^*(t)
\end{array}
\right) 
\label{eq:L_stoch}
\end{equation}
for the spatial Fourier components of the field fluctuations around the stationary-state field, that we assume for simplicity to be at rest $\kk=0$.
Here, $\mathcal{L}(\kk)$ is the Bogoliubov matrix describing the linearized deterministic dynamics and $\xi_\kk(t)$ are zero-average, complex-valued and independent noise sources of correlation function
\begin{equation}
\langle \xi^*_\kk(t) \xi_{\kk'}(t') \rangle = (2\pi)^2\,D\,\delta^{(2)}(\kk-\kk')\,\delta(t-t')\, 
\end{equation}
whose magnitude $D$ has to be tuned to reproduce the actual magnitude of the fluctuations in the experiments. Some subtleties have to be considered when evaluating observables in the case of quantum fluctuations~\cite{carusotto2013quantum}, in particular when multi-time correlators are involved~\cite{Berg:PRA2009}.

Moving to frequency-space (the frequency $\delta\omega$ is measured from the steady-state oscillation frequency), the fluctuating fields can be obtained by inverting \eqref{eq:L_stoch} as 
\begin{equation}
\delta\phi_\kk(\delta\omega)=\chi_{11}(\kk,\delta \omega)\,\xi_\kk(\delta\omega)+\chi_{12}(\kk,\delta\omega)\,[(\xi_{-\kk})^*](\delta\omega)
\end{equation}
where $\chi_{11}$ and $\chi_{12}$ are the diagonal and off-diagonal elements of the inverted Bogoliubov matrix encoding the wavevector- and frequency-dependent susceptibility,
\begin{equation}
\chi(\kk,\delta\omega)=\frac{1}{\delta\omega-\mathcal{L}(\kk)}\,.
\label{eq:chiBogo}
\end{equation}
This leads to an analytical expression for the frequency- and wavevector-dependent spectrum $I(\kk,\delta\omega)$ of the field fluctuations,
\begin{multline}
\langle \delta\psi^*_\kk(\delta\omega)\,\delta\psi_\kk(\delta\omega)\rangle =2\pi\,\delta(\delta\omega-\delta\omega')\,I(\kk,\delta\omega)= \\=2\pi\,\delta(\delta\omega-\delta\omega')\,D 
\left[|\chi_{11}(\kk,\delta\omega)|^2+|\chi_{12}(\kk,\delta\omega)|^2 \right]
\end{multline}
which shows that the fluctuations are naturally peaked at the poles of the susceptibility \eqref{eq:chiBogo}, 
that is along the dispersion of the collective excitations. Equivalent conclusions are obtained with the other Keldysh or input-output formalisms mentioned above.

Given the direct proportionality of the amplitude of the emitted light field to the amplitude of the in-cavity field, the field fluctuations are directly accessible from angle- and frequency-resolved measurements of the luminescence $I_{\rm lum}(\kk,\omega)$ that is emitted by the fluid at its stationary state. As this luminescence originates from thermal or quantum fluctuations, it typically has a incoherent character. In planar cavities, the in-plane wavevector $\kk$ of the cavity field is then related to the emission angle $\theta_{\rm em}$ measured from the normal by $k=(\omega/c)\, \sin \theta_{\rm em}$. 

Starting from the pioneering work~\cite{stepanov2019dispersion}, this technique has been used to extract information on the dispersion of collective excitations on top of a stationary fluid of light. A crucial issue of these experiments is to efficiently discard the strong emission coming from the coherent component that typically dominates the signal. Depending on the configuration, this can be done either angularly, by removing the coherent component or spectrally, e.g. by removing the emission in the vicinity of the coherent pump frequency. A spectral selection has the advantage of being able to remove also the contribution due to coherent light scattering on the unavoidable cavity disorder.

Interestingly, these luminescence experiments can be seen as a sort of optical analog of angle-resolved photoemission spectroscopy (ARPES)~\cite{zhang2022angle}, a widely used technique to get energy- and momentum-resolved information on the electronic structure of materials, including the shape of the energy bands and the shape of the Fermi surface. For completeness, it is interesting to mention that analogs of ARPES have been developed also for ultracold atomic systems for both bosonic and fermionic systems~\cite{japha1999coherent,dao2007measuring,vale2021spectroscopic}.

\subsection{Pump-and-probe measurements}
\label{sec:experimental_techniques_PP}

While luminescence experiments do not require additional light sources beyond the coherent and/or incoherent pump needed to generate the stationary fluid of light, pump-and-probe experiments  are based on the measurement of the perturbation induced in the fluid by an additional weak probe field. In typical experiments, the measured quantities are the angle- and frequency-dependent reflectivity $R(\kk,\omega)$ or transmittivity $T(\kk,\omega)$ coefficients for the probe. 

Focusing for concreteness on the coherent pump case, the perturbation induced by a weak probe can be described within the linear response theory by including an additional probe term in the linearized equations of motion \eqref{eq:linearized_coh} or \eqref{eq:linearized_incoh},
\begin{equation}
i\frac{d}{dt}\left( \begin{array}{cc}
 \delta \psi_\kk \\
  (\delta \psi_{-\kk})^*
\end{array}
\right)= \mathcal{L}(\kk)\, \left( \begin{array}{cc}
 \delta \psi_\kk \\
  (\delta \psi_{-\kk})^*
\end{array}
\right)
+\left( \begin{array}{cc}
 E^{\rm pr}_\kk(t) \\
  (E^{\rm pr}_{-\kk})^*(t)
\end{array}
\right)\,.
\label{eq:lin_resp}
\end{equation}
where $E^{\rm pr}_\kk(t)$ is the time-dependent amplitude of the probe. Assuming this to be monochromatic at a frequency $\delta\omega$ (as compared to the steady-state oscillation frequency) and concentrated at wavevector ${\kk}$, the motion equation \eqref{eq:lin_resp} is straightforwardly inverted in frequency space and the field response is quantified by the susceptibility matrix \eqref{eq:chiBogo}.

While the transmittivity is directly proportional to the square modulus of the diagonal matrix element
\begin{equation}
T({\kk},\delta\omega)\propto |\chi_{11}({\kk},\delta\omega)|^2,
\label{eq:T_k_Bogo}
\end{equation}
calculation of the reflectivity requires a slightly more careful use of the input-output theory accounting for the difference between sub-critical, critical, and super-critical coupling between the cavity and the incident radiation~\cite{QuantumOptics,yariv2000universal,xu2000scattering}. Finally, the non-diagonal terms describe four-wave mixing processes where light is emitted at the opposite wavevector $-{\kk}$ and at a specular frequency $\omega_{\rm inc}-\delta \omega$ with an amplitude proportional to the non-diagonal matrix element $\chi_{21}({\kk},\delta\omega)$~\cite{Amelio:PRR2020,Claude:PRB2023}.

It is immediate to see that the susceptibility $\chi(\kk,\delta\omega)$ displays resonance peaks at the eigenfrequencies of $\mathcal{L}(\kk)$, that is on the dispersion of the collective excitations. In our context of the Bogoliubov theory, it is important to remind that for each $\kk$ the two eigenvalues describe the two Bogoliubov branches of opposite norms. Explicit calculation of \eqref{eq:T_k_Bogo} shows that the positive-norm branch is more visible in transmission experiments, but the negative-norm one --the so-called {\em ghost} branch-- is also visible, especially in the small-$\kk$ region. An example is shown in the central-bottom panel of Fig.\ref{fig:coherent_exp}. 
Even though theory~\cite{Wouters:PRB2009,Amelio:PRR2020} predicts that both branches should be equally visible in a four-wave-mixing experiment, measurements performed in~\cite{Claude:PRB2023} have shown a strong asymmetry, whose origin is still unknown. 

As a final remark, it is useful to compare the pump-and-probe approach used for fluids of light to the Bragg spectroscopy techniques that are often used to measure the dispersion of collective excitations of atomic Bose-Einstein condensates~\cite{ozeri2005colloquium,stamper2002spinor}. Even though both methods are based on the linear response of the fluid to an external perturbation at finite wavevector and frequency, the operators involved in the process are different in the two cases. In the photonic one, the probe beam is described as an additional incident coherent probe field term in the field equation, that can be derived from a Hamiltonian term of the form 
\begin{equation}
H_{pr}=\eta E_{\rm pr}(\rr,t)\,\Psihd(\rr) + \textrm{h.c.}\,.
\end{equation}
As such, the process can be formalized in terms of a one-body Green's function of the Bose field. 
In the atomic case, instead, Bragg scattering processes are induced by applying a running-wave-shaped external potential $V(\rr,t)=V_0\cos(\kk\cdot  \rr -\omega t)$ that couples to the atomic density via an operator of the form 
\begin{equation}
H_{\rm Bragg}=    V(\rr,t)\,\Psihd(\rr)\,\Psih(\rr)\,.
\end{equation} 
While the two probes display resonance peaks at the same frequencies, the relative intensity of the various peaks can be very different in the two cases.



\bibliography{biblio}


\begin{thebibliography}{189}
\ifx \bisbn   \undefined \def \bisbn  #1{ISBN #1}\fi
\ifx \binits  \undefined \def \binits#1{#1}\fi
\ifx \bauthor  \undefined \def \bauthor#1{#1}\fi
\ifx \batitle  \undefined \def \batitle#1{#1}\fi
\ifx \bjtitle  \undefined \def \bjtitle#1{#1}\fi
\ifx \bvolume  \undefined \def \bvolume#1{\textbf{#1}}\fi
\ifx \byear  \undefined \def \byear#1{#1}\fi
\ifx \bissue  \undefined \def \bissue#1{#1}\fi
\ifx \bfpage  \undefined \def \bfpage#1{#1}\fi
\ifx \blpage  \undefined \def \blpage #1{#1}\fi
\ifx \burl  \undefined \def \burl#1{\textsf{#1}}\fi
\ifx \doiurl  \undefined \def \doiurl#1{\url{https://doi.org/#1}}\fi
\ifx \betal  \undefined \def \betal{\textit{et al.}}\fi
\ifx \binstitute  \undefined \def \binstitute#1{#1}\fi
\ifx \binstitutionaled  \undefined \def \binstitutionaled#1{#1}\fi
\ifx \bctitle  \undefined \def \bctitle#1{#1}\fi
\ifx \beditor  \undefined \def \beditor#1{#1}\fi
\ifx \bpublisher  \undefined \def \bpublisher#1{#1}\fi
\ifx \bbtitle  \undefined \def \bbtitle#1{#1}\fi
\ifx \bedition  \undefined \def \bedition#1{#1}\fi
\ifx \bseriesno  \undefined \def \bseriesno#1{#1}\fi
\ifx \blocation  \undefined \def \blocation#1{#1}\fi
\ifx \bsertitle  \undefined \def \bsertitle#1{#1}\fi
\ifx \bsnm \undefined \def \bsnm#1{#1}\fi
\ifx \bsuffix \undefined \def \bsuffix#1{#1}\fi
\ifx \bparticle \undefined \def \bparticle#1{#1}\fi
\ifx \barticle \undefined \def \barticle#1{#1}\fi
\bibcommenthead
\ifx \bconfdate \undefined \def \bconfdate #1{#1}\fi
\ifx \botherref \undefined \def \botherref #1{#1}\fi
\ifx \url \undefined \def \url#1{\textsf{#1}}\fi
\ifx \bchapter \undefined \def \bchapter#1{#1}\fi
\ifx \bbook \undefined \def \bbook#1{#1}\fi
\ifx \bcomment \undefined \def \bcomment#1{#1}\fi
\ifx \oauthor \undefined \def \oauthor#1{#1}\fi
\ifx \citeauthoryear \undefined \def \citeauthoryear#1{#1}\fi
\ifx \endbibitem  \undefined \def \endbibitem {}\fi
\ifx \bconflocation  \undefined \def \bconflocation#1{#1}\fi
\ifx \arxivurl  \undefined \def \arxivurl#1{\textsf{#1}}\fi
\csname PreBibitemsHook\endcsname

\bibitem[\protect\citeauthoryear{Nozieres and Pines}{1999}]{nozieres1999theory}
\begin{bbook}
\bauthor{\bsnm{Nozieres}, \binits{P.}},
\bauthor{\bsnm{Pines}, \binits{D.}}:
\bbtitle{Theory Of Quantum Liquids}.
\bsertitle{Advanced Books Classics}.
\bpublisher{Avalon Publishing}, \blocation{???}
(\byear{1999}).
\burl{https://books.google.it/books?id=q3wCwaV-gmUC}
\end{bbook}
\endbibitem

\bibitem[\protect\citeauthoryear{Pitaevski{\u\i} and
  Stringari}{2016}]{pitaevskii2016bose}
\begin{bbook}
\bauthor{\bsnm{Pitaevski{\u\i}}, \binits{L.P.}},
\bauthor{\bsnm{Stringari}, \binits{S.}}:
\bbtitle{Bose-Einstein Condensation and Superfluidity}.
\bsertitle{International series of monographs on physics}.
\bpublisher{Oxford University Press}, \blocation{???}
(\byear{2016}).
\burl{https://books.google.it/books?id=k\_ZGCwAAQBAJ}
\end{bbook}
\endbibitem

\bibitem[\protect\citeauthoryear{Kittel and
  McEuen}{2018}]{kittel2018introduction}
\begin{bbook}
\bauthor{\bsnm{Kittel}, \binits{C.}},
\bauthor{\bsnm{McEuen}, \binits{P.}}:
\bbtitle{Introduction to Solid State Physics}.
\bpublisher{John Wiley \& Sons}, \blocation{???}
(\byear{2018})
\end{bbook}
\endbibitem

\bibitem[\protect\citeauthoryear{Carusotto and
  Ciuti}{2013}]{carusotto2013quantum}
\begin{barticle}
\bauthor{\bsnm{Carusotto}, \binits{I.}},
\bauthor{\bsnm{Ciuti}, \binits{C.}}:
\batitle{Quantum fluids of light}.
\bjtitle{Reviews of Modern Physics}
\bvolume{85}(\bissue{1}),
\bfpage{299}
(\byear{2013})
\end{barticle}
\endbibitem

\bibitem[\protect\citeauthoryear{Bloch et~al.}{2022}]{bloch2022non}
\begin{barticle}
\bauthor{\bsnm{Bloch}, \binits{J.}},
\bauthor{\bsnm{Carusotto}, \binits{I.}},
\bauthor{\bsnm{Wouters}, \binits{M.}}:
\batitle{Non-equilibrium bose--einstein condensation in photonic systems}.
\bjtitle{Nature Reviews Physics}
\bvolume{4}(\bissue{7}),
\bfpage{470}--\blpage{488}
(\byear{2022})
\end{barticle}
\endbibitem

\bibitem[\protect\citeauthoryear{Bogoliubov}{1947}]{bogoliubov1947theory}
\begin{barticle}
\bauthor{\bsnm{Bogoliubov}, \binits{N.}}:
\batitle{On the theory of superfluidity}.
\bjtitle{J. Phys}
\bvolume{11}(\bissue{1}),
\bfpage{23}
(\byear{1947})
\end{barticle}
\endbibitem

\bibitem[\protect\citeauthoryear{Glorieux et~al.}{2025}]{glorieux2025paraxial}
\begin{barticle}
\bauthor{\bsnm{Glorieux}, \binits{Q.}},
\bauthor{\bsnm{Piekarski}, \binits{C.}},
\bauthor{\bsnm{Schibler}, \binits{Q.}},
\bauthor{\bsnm{Aladjidi}, \binits{T.}},
\bauthor{\bsnm{Baker-Rasooli}, \binits{M.}}:
\batitle{Paraxial fluids of light}.
\bjtitle{Advances In Atomic, Molecular, and Optical Physics}
\bvolume{74},
\bfpage{157}--\blpage{241}
(\byear{2025})
\end{barticle}
\endbibitem

\bibitem[\protect\citeauthoryear{Lugiato and Lefever}{1987}]{Lugiato:PRL1987}
\begin{barticle}
\bauthor{\bsnm{Lugiato}, \binits{L.A.}},
\bauthor{\bsnm{Lefever}, \binits{R.}}:
\batitle{Spatial dissipative structures in passive optical systems}.
\bjtitle{Physical review letters}
\bvolume{58}(\bissue{21}),
\bfpage{2209}
(\byear{1987})
\end{barticle}
\endbibitem

\bibitem[\protect\citeauthoryear{Lugiato et~al.}{2025}]{Lugiato:Varenna}
\begin{bchapter}
\bauthor{\bsnm{Lugiato}, \binits{L.A.}},
\bauthor{\bsnm{Prati}, \binits{F.}},
\bauthor{\bsnm{Brambilla}, \binits{E.}},
\bauthor{\bsnm{Gatti}, \binits{A.}}:
\bctitle{The cavity kerr medium model and the surprising history around it}.
In: \bbtitle{Quantum Fluids of Light and Matter},
pp. \bfpage{5}--\blpage{21}.
\bpublisher{IOS Press}, \blocation{???}
(\byear{2025})
\end{bchapter}
\endbibitem

\bibitem[\protect\citeauthoryear{Oppo}{2025}]{oppo2025computational}
\begin{botherref}
\oauthor{\bsnm{Oppo}, \binits{G.-L.}}:
Computational physics applied to photonic devices: Gian-luca oppo.
La Rivista del Nuovo Cimento,
1--74
(2025)
\end{botherref}
\endbibitem

\bibitem[\protect\citeauthoryear{Aranson and Kramer}{2002}]{Aranson:RMP2002}
\begin{barticle}
\bauthor{\bsnm{Aranson}, \binits{I.S.}},
\bauthor{\bsnm{Kramer}, \binits{L.}}:
\batitle{The world of the complex ginzburg-landau equation}.
\bjtitle{Reviews of modern physics}
\bvolume{74}(\bissue{1}),
\bfpage{99}
(\byear{2002})
\end{barticle}
\endbibitem

\bibitem[\protect\citeauthoryear{Cross and Hohenberg}{1993}]{Cross:RMP1993}
\begin{barticle}
\bauthor{\bsnm{Cross}, \binits{M.C.}},
\bauthor{\bsnm{Hohenberg}, \binits{P.C.}}:
\batitle{Pattern formation outside of equilibrium}.
\bjtitle{Reviews of modern physics}
\bvolume{65}(\bissue{3}),
\bfpage{851}
(\byear{1993})
\end{barticle}
\endbibitem

\bibitem[\protect\citeauthoryear{Carusotto}{2025}]{Carusotto:CRAS2025}
\begin{barticle}
\bauthor{\bsnm{Carusotto}, \binits{I.}}:
\batitle{How to exploit driving and dissipation to stabilize and manipulate
  quantum many-body states}.
\bjtitle{Comptes Rendus. Physique}
\bvolume{26},
\bfpage{533}--\blpage{569}
(\byear{2025})
\doiurl{10.5802/crphys.258}
\end{barticle}
\endbibitem

\bibitem[\protect\citeauthoryear{Kavokin
  et~al.}{2011}]{kavokin2011microcavities}
\begin{bbook}
\bauthor{\bsnm{Kavokin}, \binits{A.}},
\bauthor{\bsnm{Baumberg}, \binits{J.J.}},
\bauthor{\bsnm{Malpuech}, \binits{G.}},
\bauthor{\bsnm{Laussy}, \binits{F.P.}}:
\bbtitle{Microcavities}.
\bsertitle{Oxford science publications}.
\bpublisher{OUP Oxford}, \blocation{???}
(\byear{2011}).
\burl{https://books.google.it/books?id=2g7wHcMcaJ0C}
\end{bbook}
\endbibitem

\bibitem[\protect\citeauthoryear{Jackson}{1999}]{jackson1999classical}
\begin{botherref}
\oauthor{\bsnm{Jackson}, \binits{J.D.}}:
Classical electrodynamics.
American Association of Physics Teachers
(1999)
\end{botherref}
\endbibitem

\bibitem[\protect\citeauthoryear{Heisenberg and
  Euler}{1936}]{heisenberg1936folgerungen}
\begin{barticle}
\bauthor{\bsnm{Heisenberg}, \binits{W.}},
\bauthor{\bsnm{Euler}, \binits{H.}}:
\batitle{Folgerungen aus der diracschen theorie des positrons}.
\bjtitle{Zeitschrift f{\"u}r Physik}
\bvolume{98}(\bissue{11}),
\bfpage{714}--\blpage{732}
(\byear{1936})
\end{barticle}
\endbibitem

\bibitem[\protect\citeauthoryear{Butcher and Cotter}{2008}]{Butcher}
\begin{bbook}
\bauthor{\bsnm{Butcher}, \binits{P.N.}},
\bauthor{\bsnm{Cotter}, \binits{D.}}:
\bbtitle{The Elements of Nonlinear Optics}.
\bsertitle{Cambridge Studies in Modern Optics}.
\bpublisher{Cambridge University Press}, \blocation{???}
(\byear{2008})
\end{bbook}
\endbibitem

\bibitem[\protect\citeauthoryear{Boyd}{2008}]{Boyd}
\begin{bbook}
\bauthor{\bsnm{Boyd}, \binits{R.W.}}:
\bbtitle{Nonlinear Optics}.
\bpublisher{Academic Press}, \blocation{???}
(\byear{2008})
\end{bbook}
\endbibitem

\bibitem[\protect\citeauthoryear{Carusotto
  et~al.}{2020}]{carusotto2020photonic}
\begin{barticle}
\bauthor{\bsnm{Carusotto}, \binits{I.}},
\bauthor{\bsnm{Houck}, \binits{A.A.}},
\bauthor{\bsnm{Koll{\'a}r}, \binits{A.J.}},
\bauthor{\bsnm{Roushan}, \binits{P.}},
\bauthor{\bsnm{Schuster}, \binits{D.I.}},
\bauthor{\bsnm{Simon}, \binits{J.}}:
\batitle{Photonic materials in circuit quantum electrodynamics}.
\bjtitle{Nature Physics}
\bvolume{16}(\bissue{3}),
\bfpage{268}--\blpage{279}
(\byear{2020})
\end{barticle}
\endbibitem

\bibitem[\protect\citeauthoryear{Blais et~al.}{2021}]{blais_RMP2021}
\begin{barticle}
\bauthor{\bsnm{Blais}, \binits{A.}},
\bauthor{\bsnm{Grimsmo}, \binits{A.L.}},
\bauthor{\bsnm{Girvin}, \binits{S.M.}},
\bauthor{\bsnm{Wallraff}, \binits{A.}}:
\batitle{Circuit quantum electrodynamics}.
\bjtitle{Rev. Mod. Phys.}
\bvolume{93},
\bfpage{025005}
(\byear{2021})
\doiurl{10.1103/RevModPhys.93.025005}
\end{barticle}
\endbibitem

\bibitem[\protect\citeauthoryear{Peyronel et~al.}{2012}]{peyronel2012quantum}
\begin{barticle}
\bauthor{\bsnm{Peyronel}, \binits{T.}},
\bauthor{\bsnm{Firstenberg}, \binits{O.}},
\bauthor{\bsnm{Liang}, \binits{Q.-Y.}},
\bauthor{\bsnm{Hofferberth}, \binits{S.}},
\bauthor{\bsnm{Gorshkov}, \binits{A.V.}},
\bauthor{\bsnm{Pohl}, \binits{T.}},
\bauthor{\bsnm{Lukin}, \binits{M.D.}},
\bauthor{\bsnm{Vuleti{\'c}}, \binits{V.}}:
\batitle{Quantum nonlinear optics with single photons enabled by strongly
  interacting atoms}.
\bjtitle{Nature}
\bvolume{488}(\bissue{7409}),
\bfpage{57}--\blpage{60}
(\byear{2012})
\end{barticle}
\endbibitem

\bibitem[\protect\citeauthoryear{Firstenberg
  et~al.}{2013}]{firstenberg2013attractive}
\begin{barticle}
\bauthor{\bsnm{Firstenberg}, \binits{O.}},
\bauthor{\bsnm{Peyronel}, \binits{T.}},
\bauthor{\bsnm{Liang}, \binits{Q.-Y.}},
\bauthor{\bsnm{Gorshkov}, \binits{A.V.}},
\bauthor{\bsnm{Lukin}, \binits{M.D.}},
\bauthor{\bsnm{Vuleti{\'c}}, \binits{V.}}:
\batitle{Attractive photons in a quantum nonlinear medium}.
\bjtitle{Nature}
\bvolume{502}(\bissue{7469}),
\bfpage{71}--\blpage{75}
(\byear{2013})
\end{barticle}
\endbibitem

\bibitem[\protect\citeauthoryear{Chang et~al.}{2014}]{chang2014quantum}
\begin{barticle}
\bauthor{\bsnm{Chang}, \binits{D.E.}},
\bauthor{\bsnm{Vuleti{\'c}}, \binits{V.}},
\bauthor{\bsnm{Lukin}, \binits{M.D.}}:
\batitle{Quantum nonlinear optics—photon by photon}.
\bjtitle{Nature Photonics}
\bvolume{8}(\bissue{9}),
\bfpage{685}--\blpage{694}
(\byear{2014})
\end{barticle}
\endbibitem

\bibitem[\protect\citeauthoryear{Carusotto and Ciuti}{2004}]{Carusotto:PRL2004}
\begin{barticle}
\bauthor{\bsnm{Carusotto}, \binits{I.}},
\bauthor{\bsnm{Ciuti}, \binits{C.}}:
\batitle{Probing microcavity polariton superfluidity through resonant rayleigh
  scattering}.
\bjtitle{Phys. Rev. Lett.}
\bvolume{93},
\bfpage{166401}
(\byear{2004})
\doiurl{10.1103/PhysRevLett.93.166401}
\end{barticle}
\endbibitem

\bibitem[\protect\citeauthoryear{Ciuti and Carusotto}{2005}]{Ciuti:PSSB2005}
\begin{barticle}
\bauthor{\bsnm{Ciuti}, \binits{C.}},
\bauthor{\bsnm{Carusotto}, \binits{I.}}:
\batitle{Quantum fluid effects and parametric instabilities in microcavities}.
\bjtitle{physica status solidi (b)}
\bvolume{242}(\bissue{11}),
\bfpage{2224}--\blpage{2245}
(\byear{2005})
\end{barticle}
\endbibitem

\bibitem[\protect\citeauthoryear{Amo et~al.}{2009}]{Amo:NPhys2009}
\begin{barticle}
\bauthor{\bsnm{Amo}, \binits{A.}},
\bauthor{\bsnm{Lefrere}, \binits{J.}},
\bauthor{\bsnm{Pigeon}, \binits{S.}},
\bauthor{\bsnm{Adrados}, \binits{C.}},
\bauthor{\bsnm{Ciuti}, \binits{C.}},
\bauthor{\bsnm{Carusotto}, \binits{I.}},
\bauthor{\bsnm{Houdre}, \binits{R.}},
\bauthor{\bsnm{Giacobino}, \binits{E.}},
\bauthor{\bsnm{Bramati}, \binits{A.}}:
\batitle{Superfluidity of polaritons in semiconductor microcavities}.
\bjtitle{Nature Phys.}
\bvolume{5}(\bissue{11}),
\bfpage{805}--\blpage{810}
(\byear{2009})
\doiurl{10.1038/NPHYS1364}
\end{barticle}
\endbibitem

\bibitem[\protect\citeauthoryear{Claude et~al.}{2022}]{Claude:PRL2022}
\begin{barticle}
\bauthor{\bsnm{Claude}, \binits{F.}},
\bauthor{\bsnm{Jacquet}, \binits{M.J.}},
\bauthor{\bsnm{Usciati}, \binits{R.}},
\bauthor{\bsnm{Carusotto}, \binits{I.}},
\bauthor{\bsnm{Giacobino}, \binits{E.}},
\bauthor{\bsnm{Bramati}, \binits{A.}},
\bauthor{\bsnm{Glorieux}, \binits{Q.}}:
\batitle{High-resolution coherent probe spectroscopy of a polariton quantum
  fluid}.
\bjtitle{Physical Review Letters}
\bvolume{129}(\bissue{10}),
\bfpage{103601}
(\byear{2022})
\end{barticle}
\endbibitem

\bibitem[\protect\citeauthoryear{Claude et~al.}{2023}]{Claude:PRB2023}
\begin{barticle}
\bauthor{\bsnm{Claude}, \binits{F.}},
\bauthor{\bsnm{Jacquet}, \binits{M.J.}},
\bauthor{\bsnm{Carusotto}, \binits{I.}},
\bauthor{\bsnm{Glorieux}, \binits{Q.}},
\bauthor{\bsnm{Giacobino}, \binits{E.}},
\bauthor{\bsnm{Bramati}, \binits{A.}}:
\batitle{Spectrum of collective excitations of a quantum fluid of polaritons}.
\bjtitle{Physical Review B}
\bvolume{107}(\bissue{17}),
\bfpage{174507}
(\byear{2023})
\end{barticle}
\endbibitem

\bibitem[\protect\citeauthoryear{Vogel and Risken}{1988}]{Vogel:PRA1988}
\begin{barticle}
\bauthor{\bsnm{Vogel}, \binits{K.}},
\bauthor{\bsnm{Risken}, \binits{H.}}:
\batitle{Quantum-tunneling rates and stationary solutions in dispersive optical
  bistability}.
\bjtitle{Phys. Rev. A}
\bvolume{38},
\bfpage{2409}--\blpage{2422}
(\byear{1988})
\doiurl{10.1103/PhysRevA.38.2409}
\end{barticle}
\endbibitem

\bibitem[\protect\citeauthoryear{Casteels et~al.}{2017}]{casteels2017critical}
\begin{barticle}
\bauthor{\bsnm{Casteels}, \binits{W.}},
\bauthor{\bsnm{Fazio}, \binits{R.}},
\bauthor{\bsnm{Ciuti}, \binits{C.}}:
\batitle{Critical dynamical properties of a first-order dissipative phase
  transition}.
\bjtitle{Physical Review A}
\bvolume{95}(\bissue{1}),
\bfpage{012128}
(\byear{2017})
\end{barticle}
\endbibitem

\bibitem[\protect\citeauthoryear{Rodriguez et~al.}{2017}]{rodriguez2017probing}
\begin{barticle}
\bauthor{\bsnm{Rodriguez}, \binits{S.}},
\bauthor{\bsnm{Casteels}, \binits{W.}},
\bauthor{\bsnm{Storme}, \binits{F.}},
\bauthor{\bsnm{Carlon~Zambon}, \binits{N.}},
\bauthor{\bsnm{Sagnes}, \binits{I.}},
\bauthor{\bsnm{Le~Gratiet}, \binits{L.}},
\bauthor{\bsnm{Galopin}, \binits{E.}},
\bauthor{\bsnm{Lema{\^\i}tre}, \binits{A.}},
\bauthor{\bsnm{Amo}, \binits{A.}},
\bauthor{\bsnm{Ciuti}, \binits{C.}}, \betal:
\batitle{Probing a dissipative phase transition via dynamical optical
  hysteresis}.
\bjtitle{Physical review letters}
\bvolume{118}(\bissue{24}),
\bfpage{247402}
(\byear{2017})
\end{barticle}
\endbibitem

\bibitem[\protect\citeauthoryear{Selvakumaran
  et~al.}{2026}]{Selvakumaran:arXiv2026}
\begin{botherref}
\oauthor{\bsnm{Selvakumaran}, \binits{A.}},
\oauthor{\bsnm{Minguzzi}, \binits{A.}},
\oauthor{\bsnm{Carusotto}, \binits{I.}},
\oauthor{\bsnm{Richard}, \binits{M.}}:
Photon avalanche triggered by a single photon in a bistable nonlinear optical
  cavity
(2026).
\url{https://arxiv.org/abs/2606.27555}
\end{botherref}
\endbibitem

\bibitem[\protect\citeauthoryear{Stepanov
  et~al.}{2019}]{stepanov2019dispersion}
\begin{barticle}
\bauthor{\bsnm{Stepanov}, \binits{P.}},
\bauthor{\bsnm{Amelio}, \binits{I.}},
\bauthor{\bsnm{Rousset}, \binits{J.-G.}},
\bauthor{\bsnm{Bloch}, \binits{J.}},
\bauthor{\bsnm{Lema{\^\i}tre}, \binits{A.}},
\bauthor{\bsnm{Amo}, \binits{A.}},
\bauthor{\bsnm{Minguzzi}, \binits{A.}},
\bauthor{\bsnm{Carusotto}, \binits{I.}},
\bauthor{\bsnm{Richard}, \binits{M.}}:
\batitle{Dispersion relation of the collective excitations in a resonantly
  driven polariton fluid}.
\bjtitle{Nature communications}
\bvolume{10}(\bissue{1}),
\bfpage{3869}
(\byear{2019})
\end{barticle}
\endbibitem

\bibitem[\protect\citeauthoryear{Falque et~al.}{2025}]{Falque:PRL2025}
\begin{botherref}
\oauthor{\bsnm{Falque}, \binits{K.}},
\oauthor{\bsnm{Delhom}, \binits{A.}},
\oauthor{\bsnm{Glorieux}, \binits{Q.}},
\oauthor{\bsnm{Giacobino}, \binits{E.}},
\oauthor{\bsnm{Bramati}, \binits{A.}},
\oauthor{\bsnm{Jacquet}, \binits{M.J.}}:
Polariton fluids as quantum field theory simulators on tailored curved
  spacetimes.
Phys. Rev. Lett.,
(2025)
\doiurl{10.1103/t5dh-rx6w}
\end{botherref}
\endbibitem

\bibitem[\protect\citeauthoryear{Solnyshkov et~al.}{2011}]{Solnyshkov:PRB2011}
\begin{barticle}
\bauthor{\bsnm{Solnyshkov}, \binits{D.D.}},
\bauthor{\bsnm{Flayac}, \binits{H.}},
\bauthor{\bsnm{Malpuech}, \binits{G.}}:
\batitle{Black holes and wormholes in spinor polariton condensates}.
\bjtitle{Phys. Rev. B}
\bvolume{84},
\bfpage{233405}
(\byear{2011})
\doiurl{10.1103/PhysRevB.84.233405}
\end{barticle}
\endbibitem

\bibitem[\protect\citeauthoryear{Gerace and Carusotto}{2012}]{Gerace:PRB2012}
\begin{barticle}
\bauthor{\bsnm{Gerace}, \binits{D.}},
\bauthor{\bsnm{Carusotto}, \binits{I.}}:
\batitle{Analog hawking radiation from an acoustic black hole in a flowing
  polariton superfluid}.
\bjtitle{Physical Review B—Condensed Matter and Materials Physics}
\bvolume{86}(\bissue{14}),
\bfpage{144505}
(\byear{2012})
\end{barticle}
\endbibitem

\bibitem[\protect\citeauthoryear{Nguyen et~al.}{2015}]{Nguyen:PRL2015}
\begin{barticle}
\bauthor{\bsnm{Nguyen}, \binits{H.S.}},
\bauthor{\bsnm{Gerace}, \binits{D.}},
\bauthor{\bsnm{Carusotto}, \binits{I.}},
\bauthor{\bsnm{Sanvitto}, \binits{D.}},
\bauthor{\bsnm{Galopin}, \binits{E.}},
\bauthor{\bsnm{Lema{\^\i}tre}, \binits{A.}},
\bauthor{\bsnm{Sagnes}, \binits{I.}},
\bauthor{\bsnm{Bloch}, \binits{J.}},
\bauthor{\bsnm{Amo}, \binits{A.}}:
\batitle{Acoustic black hole in a stationary hydrodynamic flow of microcavity
  polaritons}.
\bjtitle{Physical review letters}
\bvolume{114}(\bissue{3}),
\bfpage{036402}
(\byear{2015})
\end{barticle}
\endbibitem

\bibitem[\protect\citeauthoryear{Recati et~al.}{2009}]{recati2009bogoliubov}
\begin{barticle}
\bauthor{\bsnm{Recati}, \binits{A.}},
\bauthor{\bsnm{Pavloff}, \binits{N.}},
\bauthor{\bsnm{Carusotto}, \binits{I.}}:
\batitle{Bogoliubov theory of acoustic hawking radiation in bose-einstein
  condensates}.
\bjtitle{Physical Review A—Atomic, Molecular, and Optical Physics}
\bvolume{80}(\bissue{4}),
\bfpage{043603}
(\byear{2009})
\end{barticle}
\endbibitem

\bibitem[\protect\citeauthoryear{Barcelo et~al.}{2011}]{barcelo2011analogue}
\begin{barticle}
\bauthor{\bsnm{Barcelo}, \binits{C.}},
\bauthor{\bsnm{Liberati}, \binits{S.}},
\bauthor{\bsnm{Visser}, \binits{M.}}:
\batitle{Analogue gravity}.
\bjtitle{Living reviews in relativity}
\bvolume{14},
\bfpage{1}--\blpage{159}
(\byear{2011})
\end{barticle}
\endbibitem

\bibitem[\protect\citeauthoryear{Delhom and
  Giacomelli}{2025}]{delhom2025analogue}
\begin{botherref}
\oauthor{\bsnm{Delhom}, \binits{A.}},
\oauthor{\bsnm{Giacomelli}, \binits{L.}}:
Analogue gravity with bose-einstein condensates.
arXiv preprint arXiv:2512.14209
(2025)
\end{botherref}
\endbibitem

\bibitem[\protect\citeauthoryear{Grisins et~al.}{2016}]{Grisins:PRA2016}
\begin{barticle}
\bauthor{\bsnm{Grisins}, \binits{P.}},
\bauthor{\bsnm{Nguyen}, \binits{H.S.}},
\bauthor{\bsnm{Bloch}, \binits{J.}},
\bauthor{\bsnm{Amo}, \binits{A.}},
\bauthor{\bsnm{Carusotto}, \binits{I.}}:
\batitle{Theoretical study of stimulated and spontaneous hawking effects from
  an acoustic black hole in a hydrodynamically flowing fluid of light}.
\bjtitle{Phys. Rev. B}
\bvolume{94},
\bfpage{144518}
(\byear{2016})
\doiurl{10.1103/PhysRevB.94.144518}
\end{barticle}
\endbibitem

\bibitem[\protect\citeauthoryear{Guerrero et~al.}{2025}]{Guerrero:PRL2025}
\begin{barticle}
\bauthor{\bsnm{Guerrero}, \binits{K.}},
\bauthor{\bsnm{Falque}, \binits{K.}},
\bauthor{\bsnm{Giacobino}, \binits{E.}},
\bauthor{\bsnm{Bramati}, \binits{A.}},
\bauthor{\bsnm{Jacquet}, \binits{M.J.}}:
\batitle{Multiply quantized vortex spectroscopy in a quantum fluid of light}.
\bjtitle{Phys. Rev. Lett.}
\bvolume{135},
\bfpage{243801}
(\byear{2025})
\doiurl{10.1103/whgn-6889}
\end{barticle}
\endbibitem

\bibitem[\protect\citeauthoryear{Delhom et~al.}{2024}]{Delhom:PRD2024}
\begin{barticle}
\bauthor{\bsnm{Delhom}, \binits{A.}},
\bauthor{\bsnm{Guerrero}, \binits{K.}},
\bauthor{\bsnm{Calizaya~Cabrera}, \binits{P.}},
\bauthor{\bsnm{Falque}, \binits{K.}},
\bauthor{\bsnm{Bramati}, \binits{A.}},
\bauthor{\bsnm{Brady}, \binits{A.J.}},
\bauthor{\bsnm{Jacquet}, \binits{M.J.}},
\bauthor{\bsnm{Agullo}, \binits{I.}}:
\batitle{Entanglement from superradiance and rotating quantum fluids of light}.
\bjtitle{Phys. Rev. D}
\bvolume{109},
\bfpage{105024}
(\byear{2024})
\doiurl{10.1103/PhysRevD.109.105024}
\end{barticle}
\endbibitem

\bibitem[\protect\citeauthoryear{Amelio and Carusotto}{2020}]{Amelio:PRB2020}
\begin{barticle}
\bauthor{\bsnm{Amelio}, \binits{I.}},
\bauthor{\bsnm{Carusotto}, \binits{I.}}:
\batitle{Perspectives in superfluidity in resonantly driven polariton fluids}.
\bjtitle{Physical Review B}
\bvolume{101}(\bissue{6}),
\bfpage{064505}
(\byear{2020})
\end{barticle}
\endbibitem

\bibitem[\protect\citeauthoryear{Amelio et~al.}{2020}]{Amelio:PRR2020}
\begin{barticle}
\bauthor{\bsnm{Amelio}, \binits{I.}},
\bauthor{\bsnm{Minguzzi}, \binits{A.}},
\bauthor{\bsnm{Richard}, \binits{M.}},
\bauthor{\bsnm{Carusotto}, \binits{I.}}:
\batitle{Galilean boosts and superfluidity of resonantly driven polariton
  fluids in the presence of an incoherent reservoir}.
\bjtitle{Physical Review Research}
\bvolume{2}(\bissue{2}),
\bfpage{023158}
(\byear{2020})
\end{barticle}
\endbibitem

\bibitem[\protect\citeauthoryear{Richard et~al.}{2026}]{Richard:PRR2026}
\begin{barticle}
\bauthor{\bsnm{Richard}, \binits{M.}},
\bauthor{\bsnm{Fr{\'e}rot}, \binits{I.}},
\bauthor{\bsnm{Ravets}, \binits{S.}},
\bauthor{\bsnm{Bloch}, \binits{J.}},
\bauthor{\bsnm{Anton-Solanas}, \binits{C.}},
\bauthor{\bsnm{Claude}, \binits{F.}},
\bauthor{\bsnm{Zhou}, \binits{Y.}},
\bauthor{\bsnm{Morassi}, \binits{M.}},
\bauthor{\bsnm{Lema{\^\i}tre}, \binits{A.}},
\bauthor{\bsnm{Carusotto}, \binits{I.}}, \betal:
\batitle{Excitonic oscillator-strength saturation dominates polariton-polariton
  interactions}.
\bjtitle{Physical Review Research}
\bvolume{8}(\bissue{1}),
\bfpage{012039}
(\byear{2026})
\end{barticle}
\endbibitem

\bibitem[\protect\citeauthoryear{Wouters and Carusotto}{2007}]{Wouters:PRL2007}
\begin{botherref}
\oauthor{\bsnm{Wouters}, \binits{M.}},
\oauthor{\bsnm{Carusotto}, \binits{I.}}:
Excitations in a nonequilibrium bose-einstein condensate of exciton polaritons.
Phys. Rev. Lett.
\textbf{99}(14)
(2007)
\doiurl{10.1103/PhysRevLett.99.140402}
\end{botherref}
\endbibitem

\bibitem[\protect\citeauthoryear{Ji et~al.}{2015}]{Ji:PRA2015}
\begin{barticle}
\bauthor{\bsnm{Ji}, \binits{K.}},
\bauthor{\bsnm{Gladilin}, \binits{V.N.}},
\bauthor{\bsnm{Wouters}, \binits{M.}}:
\batitle{Temporal coherence of one-dimensional nonequilibrium quantum fluids}.
\bjtitle{Phys. Rev. B}
\bvolume{91},
\bfpage{045301}
(\byear{2015})
\doiurl{10.1103/PhysRevB.91.045301}
\end{barticle}
\endbibitem

\bibitem[\protect\citeauthoryear{Wouters and Carusotto}{2006}]{Wouters:PRB2006}
\begin{barticle}
\bauthor{\bsnm{Wouters}, \binits{M.}},
\bauthor{\bsnm{Carusotto}, \binits{I.}}:
\batitle{Absence of long-range coherence in the parametric emission of photonic
  wires}.
\bjtitle{Phys. Rev. B}
\bvolume{74},
\bfpage{245316}
(\byear{2006})
\doiurl{10.1103/PhysRevB.74.245316}
\end{barticle}
\endbibitem

\bibitem[\protect\citeauthoryear{Szymanska et~al.}{2006}]{Szymanska:PRL2006}
\begin{barticle}
\bauthor{\bsnm{Szymanska}, \binits{M.H.}},
\bauthor{\bsnm{Keeling}, \binits{J.}},
\bauthor{\bsnm{Littlewood}, \binits{P.B.}}:
\batitle{Nonequilibrium quantum condensation in an incoherently pumped
  dissipative system}.
\bjtitle{Phys. Rev. Lett.}
\bvolume{96},
\bfpage{230602}
(\byear{2006})
\doiurl{10.1103/PhysRevLett.96.230602}
\end{barticle}
\endbibitem

\bibitem[\protect\citeauthoryear{Wouters and Carusotto}{2007}]{Wouters:PRA2007}
\begin{barticle}
\bauthor{\bsnm{Wouters}, \binits{M.}},
\bauthor{\bsnm{Carusotto}, \binits{I.}}:
\batitle{Goldstone mode of optical parametric oscillators in planar
  semiconductor microcavities in the strong-coupling regime}.
\bjtitle{Phys. Rev. A}
\bvolume{76},
\bfpage{043807}
(\byear{2007})
\doiurl{10.1103/PhysRevA.76.043807}
\end{barticle}
\endbibitem

\bibitem[\protect\citeauthoryear{Sachdev}{2011}]{SachdevBook}
\begin{bbook}
\bauthor{\bsnm{Sachdev}, \binits{S.}}:
\bbtitle{Quantum Phase Transitions}.
\bpublisher{Cambridge University Press}, \blocation{???}
(\byear{2011}).
\burl{https://books.google.it/books?id=F3IkpxwpqSgC}
\end{bbook}
\endbibitem

\bibitem[\protect\citeauthoryear{Richard et~al.}{2005}]{Richard:PRL2005}
\begin{barticle}
\bauthor{\bsnm{Richard}, \binits{M.}},
\bauthor{\bsnm{Kasprzak}, \binits{J.}},
\bauthor{\bsnm{Romestain}, \binits{R.}},
\bauthor{\bsnm{Andre}, \binits{R.}},
\bauthor{\bsnm{Dang}, \binits{L.}}:
\batitle{Spontaneous coherent phase transition of polaritons in cdte
  microcavities}.
\bjtitle{Phys. Rev. Lett.}
\bvolume{94}(\bissue{18}),
\bfpage{187401}
(\byear{2005})
\doiurl{10.1103/PhysRevLett.94.187401}
\end{barticle}
\endbibitem

\bibitem[\protect\citeauthoryear{Wouters et~al.}{2008}]{Wouters:PRB2008}
\begin{barticle}
\bauthor{\bsnm{Wouters}, \binits{M.}},
\bauthor{\bsnm{Carusotto}, \binits{I.}},
\bauthor{\bsnm{Ciuti}, \binits{C.}}:
\batitle{Spatial and spectral shape of inhomogeneous nonequilibrium
  exciton-polariton condensates}.
\bjtitle{Phys. Rev. B}
\bvolume{77},
\bfpage{115340}
(\byear{2008})
\doiurl{10.1103/PhysRevB.77.115340}
\end{barticle}
\endbibitem

\bibitem[\protect\citeauthoryear{Bahari et~al.}{2017}]{Bahari:Science2017}
\begin{barticle}
\bauthor{\bsnm{Bahari}, \binits{B.}},
\bauthor{\bsnm{Ndao}, \binits{A.}},
\bauthor{\bsnm{Vallini}, \binits{F.}},
\bauthor{\bsnm{El~Amili}, \binits{A.}},
\bauthor{\bsnm{Fainman}, \binits{Y.}},
\bauthor{\bsnm{Kant{\'e}}, \binits{B.}}:
\batitle{Nonreciprocal lasing in topological cavities of arbitrary geometries}.
\bjtitle{Science}
\bvolume{358}(\bissue{6363}),
\bfpage{636}--\blpage{640}
(\byear{2017})
\end{barticle}
\endbibitem

\bibitem[\protect\citeauthoryear{Harari et~al.}{2018}]{Harari:Science2018}
\begin{barticle}
\bauthor{\bsnm{Harari}, \binits{G.}},
\bauthor{\bsnm{Bandres}, \binits{M.A.}},
\bauthor{\bsnm{Lumer}, \binits{Y.}},
\bauthor{\bsnm{Rechtsman}, \binits{M.C.}},
\bauthor{\bsnm{Chong}, \binits{Y.D.}},
\bauthor{\bsnm{Khajavikhan}, \binits{M.}},
\bauthor{\bsnm{Christodoulides}, \binits{D.N.}},
\bauthor{\bsnm{Segev}, \binits{M.}}:
\batitle{Topological insulator laser: Theory}.
\bjtitle{Science}
\bvolume{359}(\bissue{6381}),
\bfpage{4003}
(\byear{2018})
\end{barticle}
\endbibitem

\bibitem[\protect\citeauthoryear{Bandres et~al.}{2018}]{Bandres:Science2018}
\begin{barticle}
\bauthor{\bsnm{Bandres}, \binits{M.A.}},
\bauthor{\bsnm{Wittek}, \binits{S.}},
\bauthor{\bsnm{Harari}, \binits{G.}},
\bauthor{\bsnm{Parto}, \binits{M.}},
\bauthor{\bsnm{Ren}, \binits{J.}},
\bauthor{\bsnm{Segev}, \binits{M.}},
\bauthor{\bsnm{Christodoulides}, \binits{D.N.}},
\bauthor{\bsnm{Khajavikhan}, \binits{M.}}:
\batitle{Topological insulator laser: Experiments}.
\bjtitle{Science}
\bvolume{359}(\bissue{6381}),
\bfpage{4005}
(\byear{2018})
\end{barticle}
\endbibitem

\bibitem[\protect\citeauthoryear{Price et~al.}{2022}]{Price:JPhysPhot2022}
\begin{barticle}
\bauthor{\bsnm{Price}, \binits{H.}},
\bauthor{\bsnm{Chong}, \binits{Y.}},
\bauthor{\bsnm{Khanikaev}, \binits{A.}},
\bauthor{\bsnm{Schomerus}, \binits{H.}},
\bauthor{\bsnm{Maczewsky}, \binits{L.J.}},
\bauthor{\bsnm{Kremer}, \binits{M.}},
\bauthor{\bsnm{Heinrich}, \binits{M.}},
\bauthor{\bsnm{Szameit}, \binits{A.}},
\bauthor{\bsnm{Zilberberg}, \binits{O.}},
\bauthor{\bsnm{Yang}, \binits{Y.}},
\bauthor{\bsnm{Zhang}, \binits{B.}},
\bauthor{\bsnm{Alù}, \binits{A.}},
\bauthor{\bsnm{Thomale}, \binits{R.}},
\bauthor{\bsnm{Carusotto}, \binits{I.}},
\bauthor{\bsnm{St-Jean}, \binits{P.}},
\bauthor{\bsnm{Amo}, \binits{A.}},
\bauthor{\bsnm{Dutt}, \binits{A.}},
\bauthor{\bsnm{Yuan}, \binits{L.}},
\bauthor{\bsnm{Fan}, \binits{S.}},
\bauthor{\bsnm{Yin}, \binits{X.}},
\bauthor{\bsnm{Peng}, \binits{C.}},
\bauthor{\bsnm{Ozawa}, \binits{T.}},
\bauthor{\bsnm{Blanco-Redondo}, \binits{A.}}:
\batitle{Roadmap on topological photonics}.
\bjtitle{Journal of Physics: Photonics}
\bvolume{4}(\bissue{3}),
\bfpage{032501}
(\byear{2022})
\doiurl{10.1088/2515-7647/ac4ee4}
\end{barticle}
\endbibitem

\bibitem[\protect\citeauthoryear{Amelio and Carusotto}{2020}]{amelio:PRX2020}
\begin{barticle}
\bauthor{\bsnm{Amelio}, \binits{I.}},
\bauthor{\bsnm{Carusotto}, \binits{I.}}:
\batitle{Theory of the coherence of topological lasers}.
\bjtitle{Physical Review X}
\bvolume{10}(\bissue{4}),
\bfpage{041060}
(\byear{2020})
\end{barticle}
\endbibitem

\bibitem[\protect\citeauthoryear{Zapletal et~al.}{2020}]{Zapletal:Optica2020}
\begin{barticle}
\bauthor{\bsnm{Zapletal}, \binits{P.}},
\bauthor{\bsnm{Galilo}, \binits{B.}},
\bauthor{\bsnm{Nunnenkamp}, \binits{A.}}:
\batitle{Long-lived elementary excitations and light coherence in topological
  lasers}.
\bjtitle{Optica}
\bvolume{7}(\bissue{9}),
\bfpage{1045}--\blpage{1055}
(\byear{2020})
\end{barticle}
\endbibitem

\bibitem[\protect\citeauthoryear{Loirette-Pelous
  et~al.}{2021}]{Loirette:PRA2021}
\begin{barticle}
\bauthor{\bsnm{Loirette-Pelous}, \binits{A.}},
\bauthor{\bsnm{Amelio}, \binits{I.}},
\bauthor{\bsnm{Secl\`{\i}}, \binits{M.}},
\bauthor{\bsnm{Carusotto}, \binits{I.}}:
\batitle{Linearized theory of the fluctuation dynamics in two-dimensional
  topological lasers}.
\bjtitle{Phys. Rev. A}
\bvolume{104},
\bfpage{053516}
(\byear{2021})
\doiurl{10.1103/PhysRevA.104.053516}
\end{barticle}
\endbibitem

\bibitem[\protect\citeauthoryear{Amelio and Carusotto}{2022}]{Amelio:PRA2022}
\begin{barticle}
\bauthor{\bsnm{Amelio}, \binits{I.}},
\bauthor{\bsnm{Carusotto}, \binits{I.}}:
\batitle{Bogoliubov theory of the laser linewidth and application to polariton
  condensates}.
\bjtitle{Phys. Rev. A}
\bvolume{105},
\bfpage{023527}
(\byear{2022})
\doiurl{10.1103/PhysRevA.105.023527}
\end{barticle}
\endbibitem

\bibitem[\protect\citeauthoryear{Brunelli
  et~al.}{2023}]{Brunelli:SciPostPhys2023}
\begin{barticle}
\bauthor{\bsnm{Brunelli}, \binits{M.}},
\bauthor{\bsnm{Wanjura}, \binits{C.C.}},
\bauthor{\bsnm{Nunnenkamp}, \binits{A.}}:
\batitle{Restoration of the non-hermitian bulk-boundary correspondence via
  topological amplification}.
\bjtitle{SciPost Phys.}
\bvolume{15},
\bfpage{173}
(\byear{2023})
\doiurl{10.21468/SciPostPhys.15.4.173}
\end{barticle}
\endbibitem

\bibitem[\protect\citeauthoryear{Baboux et~al.}{2018}]{Baboux:Optica2018}
\begin{barticle}
\bauthor{\bsnm{Baboux}, \binits{F.}},
\bauthor{\bsnm{De~Bernardis}, \binits{D.}},
\bauthor{\bsnm{Goblot}, \binits{V.}},
\bauthor{\bsnm{Gladilin}, \binits{V.}},
\bauthor{\bsnm{Gomez}, \binits{C.}},
\bauthor{\bsnm{Galopin}, \binits{E.}},
\bauthor{\bsnm{Le~Gratiet}, \binits{L.}},
\bauthor{\bsnm{Lema{\^\i}tre}, \binits{A.}},
\bauthor{\bsnm{Sagnes}, \binits{I.}},
\bauthor{\bsnm{Carusotto}, \binits{I.}}, \betal:
\batitle{Unstable and stable regimes of polariton condensation}.
\bjtitle{Optica}
\bvolume{5}(\bissue{10}),
\bfpage{1163}--\blpage{1170}
(\byear{2018})
\end{barticle}
\endbibitem

\bibitem[\protect\citeauthoryear{Claude et~al.}{2025}]{claude2025observation}
\begin{barticle}
\bauthor{\bsnm{Claude}, \binits{F.}},
\bauthor{\bsnm{Jacquet}, \binits{M.J.}},
\bauthor{\bsnm{Glorieux}, \binits{Q.}},
\bauthor{\bsnm{Wouters}, \binits{M.}},
\bauthor{\bsnm{Giacobino}, \binits{E.}},
\bauthor{\bsnm{Carusotto}, \binits{I.}},
\bauthor{\bsnm{Bramati}, \binits{A.}}:
\batitle{Observation of the diffusive nambu--goldstone mode of a
  non-equilibrium phase transition}.
\bjtitle{Nature Physics}
\bvolume{21}(\bissue{6}),
\bfpage{924}--\blpage{930}
(\byear{2025})
\end{barticle}
\endbibitem

\bibitem[\protect\citeauthoryear{Berceanu et~al.}{2015}]{Berceanu:PRB2015}
\begin{barticle}
\bauthor{\bsnm{Berceanu}, \binits{A.}},
\bauthor{\bsnm{Dominici}, \binits{L.}},
\bauthor{\bsnm{Carusotto}, \binits{I.}},
\bauthor{\bsnm{Ballarini}, \binits{D.}},
\bauthor{\bsnm{Cancellieri}, \binits{E.}},
\bauthor{\bsnm{Gigli}, \binits{G.}},
\bauthor{\bsnm{Szyma{\'n}ska}, \binits{M.}},
\bauthor{\bsnm{Sanvitto}, \binits{D.}},
\bauthor{\bsnm{Marchetti}, \binits{F.M.}}:
\batitle{Multicomponent polariton superfluidity in the optical parametric
  oscillator regime}.
\bjtitle{Physical Review B}
\bvolume{92}(\bissue{3}),
\bfpage{035307}
(\byear{2015})
\end{barticle}
\endbibitem

\bibitem[\protect\citeauthoryear{A{\ss}mann
  et~al.}{2011}]{assmann2011polariton}
\begin{barticle}
\bauthor{\bsnm{A{\ss}mann}, \binits{M.}},
\bauthor{\bsnm{Tempel}, \binits{J.-S.}},
\bauthor{\bsnm{Veit}, \binits{F.}},
\bauthor{\bsnm{Bayer}, \binits{M.}},
\bauthor{\bsnm{Rahimi-Iman}, \binits{A.}},
\bauthor{\bsnm{L{\"o}ffler}, \binits{A.}},
\bauthor{\bsnm{H{\"o}fling}, \binits{S.}},
\bauthor{\bsnm{Reitzenstein}, \binits{S.}},
\bauthor{\bsnm{Worschech}, \binits{L.}},
\bauthor{\bsnm{Forchel}, \binits{A.}}:
\batitle{From polariton condensates to highly photonic quantum degenerate
  states of bosonic matter}.
\bjtitle{Proceedings of the National Academy of Sciences}
\bvolume{108}(\bissue{5}),
\bfpage{1804}--\blpage{1809}
(\byear{2011})
\doiurl{10.1073/pnas.1009847108}
\end{barticle}
\endbibitem

\bibitem[\protect\citeauthoryear{Nakayama and
  Ueda}{2017}]{nakayama2017observation}
\begin{barticle}
\bauthor{\bsnm{Nakayama}, \binits{M.}},
\bauthor{\bsnm{Ueda}, \binits{M.}}:
\batitle{Observation of diffusive and dispersive profiles of the nonequilibrium
  polariton-condensate dispersion relation in a cubr microcavity}.
\bjtitle{Physical Review B}
\bvolume{95}(\bissue{12}),
\bfpage{125315}
(\byear{2017})
\end{barticle}
\endbibitem

\bibitem[\protect\citeauthoryear{Ballarini
  et~al.}{2009}]{ballarini_observation_2009}
\begin{barticle}
\bauthor{\bsnm{Ballarini}, \binits{D.}},
\bauthor{\bsnm{Sanvitto}, \binits{D.}},
\bauthor{\bsnm{Amo}, \binits{A.}},
\bauthor{\bsnm{Viña}, \binits{L.}},
\bauthor{\bsnm{Wouters}, \binits{M.}},
\bauthor{\bsnm{Carusotto}, \binits{I.}},
\bauthor{\bsnm{Lemaitre}, \binits{A.}},
\bauthor{\bsnm{Bloch}, \binits{J.}}:
\batitle{Observation of {Long}-{Lived} {Polariton} {States} in {Semiconductor}
  {Microcavities} across the {Parametric} {Threshold}}.
\bjtitle{Physical Review Letters}
\bvolume{102},
\bfpage{056402}
(\byear{2009})
\doiurl{10.1103/PhysRevLett.102.056402}
\end{barticle}
\endbibitem

\bibitem[\protect\citeauthoryear{Ballarini
  et~al.}{2020}]{ballarini_directional_2020}
\begin{barticle}
\bauthor{\bsnm{Ballarini}, \binits{D.}},
\bauthor{\bsnm{Caputo}, \binits{D.}},
\bauthor{\bsnm{Dagvadorj}, \binits{G.}},
\bauthor{\bsnm{Juggins}, \binits{R.}},
\bauthor{\bsnm{Giorgi}, \binits{M.D.}},
\bauthor{\bsnm{Dominici}, \binits{L.}},
\bauthor{\bsnm{West}, \binits{K.}},
\bauthor{\bsnm{Pfeiffer}, \binits{L.N.}},
\bauthor{\bsnm{Gigli}, \binits{G.}},
\bauthor{\bsnm{Szyma\'nska}, \binits{M.H.}},
\bauthor{\bsnm{Sanvitto}, \binits{D.}}:
\batitle{Directional {Goldstone} waves in polariton condensates close to
  equilibrium}.
\bjtitle{Nature Communications}
\bvolume{11},
\bfpage{217}
(\byear{2020})
\doiurl{10.1038/s41467-019-13733-x}
\end{barticle}
\endbibitem

\bibitem[\protect\citeauthoryear{Adler}{1946}]{Adler1946ASO}
\begin{barticle}
\bauthor{\bsnm{Adler}, \binits{R.}}:
\batitle{A study of locking phenomena in oscillators}.
\bjtitle{Proceedings of the IRE}
\bvolume{34},
\bfpage{351}--\blpage{357}
(\byear{1946})
\end{barticle}
\endbibitem

\bibitem[\protect\citeauthoryear{Liu and Slav\'{i}k}{2020}]{Liu:JLT2020}
\begin{barticle}
\bauthor{\bsnm{Liu}, \binits{Z.}},
\bauthor{\bsnm{Slav\'{i}k}, \binits{R.}}:
\batitle{Optical injection locking: From principle to applications}.
\bjtitle{J. Lightwave Technol.}
\bvolume{38}(\bissue{1}),
\bfpage{43}--\blpage{59}
(\byear{2020})
\end{barticle}
\endbibitem

\bibitem[\protect\citeauthoryear{Markovic et~al.}{2019}]{Huard:PRApp2019}
\begin{barticle}
\bauthor{\bsnm{Markovic}, \binits{D.}},
\bauthor{\bsnm{Pillet}, \binits{J.D.}},
\bauthor{\bsnm{Flurin}, \binits{E.}},
\bauthor{\bsnm{Roch}, \binits{N.}},
\bauthor{\bsnm{Huard}, \binits{B.}}:
\batitle{Injection locking and parametric locking in a superconducting
  circuit}.
\bjtitle{Phys. Rev. Appl.}
\bvolume{12},
\bfpage{024034}
(\byear{2019})
\doiurl{10.1103/PhysRevApplied.12.024034}
\end{barticle}
\endbibitem

\bibitem[\protect\citeauthoryear{Gunton and Buckingham}{1968}]{Gunton:PR1968}
\begin{barticle}
\bauthor{\bsnm{Gunton}, \binits{J.D.}},
\bauthor{\bsnm{Buckingham}, \binits{M.J.}}:
\batitle{Condensation of the ideal bose gas as a cooperative transition}.
\bjtitle{Phys. Rev.}
\bvolume{166},
\bfpage{152}--\blpage{158}
(\byear{1968})
\doiurl{10.1103/PhysRev.166.152}
\end{barticle}
\endbibitem

\bibitem[\protect\citeauthoryear{Huang}{1987}]{Huang}
\begin{bbook}
\bauthor{\bsnm{Huang}, \binits{K.}}:
\bbtitle{Statistical Mechanics}.
\bpublisher{Wiley}, \blocation{???}
(\byear{1987})
\end{bbook}
\endbibitem

\bibitem[\protect\citeauthoryear{Stazzu et~al.}{2026}]{Stazzu:PRA2026}
\begin{barticle}
\bauthor{\bsnm{Stazzu}, \binits{E.}},
\bauthor{\bsnm{Sacchetto}, \binits{G.}},
\bauthor{\bsnm{Carusotto}, \binits{I.}}:
\batitle{Opening a gap in the dispersion of the collective excitations of a
  driven-dissipative condensate subject to an external coherent drive}.
\bjtitle{Physical Review A}
\bvolume{113}(\bissue{5}),
\bfpage{053303}
(\byear{2026})
\end{barticle}
\endbibitem

\bibitem[\protect\citeauthoryear{Kapitsa}{1938}]{Kapitsa:Nature1938}
\begin{barticle}
\bauthor{\bsnm{Kapitsa}, \binits{P.L.}}:
\batitle{Viscosity of liquid helium below the $\lambda$-point}.
\bjtitle{Nature}
\bvolume{141},
\bfpage{74}
(\byear{1938})
\end{barticle}
\endbibitem

\bibitem[\protect\citeauthoryear{Allen and Misener}{1938}]{Allen:Nature1938}
\begin{barticle}
\bauthor{\bsnm{Allen}, \binits{J.F.}},
\bauthor{\bsnm{Misener}, \binits{A.D.}}:
\batitle{Flow of liquid helium ii}.
\bjtitle{Nature}
\bvolume{141},
\bfpage{75}
(\byear{1938})
\end{barticle}
\endbibitem

\bibitem[\protect\citeauthoryear{Leggett}{1999}]{Leggett:RMP1999}
\begin{barticle}
\bauthor{\bsnm{Leggett}, \binits{A.}}:
\batitle{Superfluidity}.
\bjtitle{Rev. Mod. Phys.}
\bvolume{71}(\bissue{2, Sp. Iss. SI}),
\bfpage{318}--\blpage{323}
(\byear{1999})
\end{barticle}
\endbibitem

\bibitem[\protect\citeauthoryear{Carusotto and
  Rousseaux}{2013}]{Carusotto:Ducks2013}
\begin{bbook}
\bauthor{\bsnm{Carusotto}, \binits{I.}},
\bauthor{\bsnm{Rousseaux}, \binits{G.}}:
In: \beditor{\bsnm{Faccio}, \binits{D.}},
\beditor{\bsnm{Belgiorno}, \binits{F.}},
\beditor{\bsnm{Cacciatori}, \binits{S.}},
\beditor{\bsnm{Gorini}, \binits{V.}},
\beditor{\bsnm{Liberati}, \binits{S.}},
\beditor{\bsnm{Moschella}, \binits{U.}} (eds.)
\bbtitle{The Cerenkov Effect Revisited: From Swimming Ducks to Zero Modes in
  Gravitational Analogues},
pp. \bfpage{109}--\blpage{144}.
\bpublisher{Springer},
\blocation{Cham}
(\byear{2013})
\end{bbook}
\endbibitem

\bibitem[\protect\citeauthoryear{Onofrio et~al.}{2000}]{Onofrio:PRL2000}
\begin{barticle}
\bauthor{\bsnm{Onofrio}, \binits{R.}},
\bauthor{\bsnm{Raman}, \binits{C.}},
\bauthor{\bsnm{Vogels}, \binits{J.M.}},
\bauthor{\bsnm{Abo-Shaeer}, \binits{J.R.}},
\bauthor{\bsnm{Chikkatur}, \binits{A.P.}},
\bauthor{\bsnm{Ketterle}, \binits{W.}}:
\batitle{Observation of superfluid flow in a bose-einstein condensed gas}.
\bjtitle{Phys. Rev. Lett.}
\bvolume{85},
\bfpage{2228}--\blpage{2231}
(\byear{2000})
\doiurl{10.1103/PhysRevLett.85.2228}
\end{barticle}
\endbibitem

\bibitem[\protect\citeauthoryear{Frisch et~al.}{1992}]{Frisch:PRL1992}
\begin{barticle}
\bauthor{\bsnm{Frisch}, \binits{T.}},
\bauthor{\bsnm{Pomeau}, \binits{Y.}},
\bauthor{\bsnm{Rica}, \binits{S.}}:
\batitle{Transition to dissipation in a model of superflow}.
\bjtitle{Phys. Rev. Lett.}
\bvolume{69},
\bfpage{1644}--\blpage{1647}
(\byear{1992})
\doiurl{10.1103/PhysRevLett.69.1644}
\end{barticle}
\endbibitem

\bibitem[\protect\citeauthoryear{Hakim}{1997}]{Hakim:PRE1997}
\begin{barticle}
\bauthor{\bsnm{Hakim}, \binits{V.}}:
\batitle{Nonlinear schr\"odinger flow past an obstacle in one dimension}.
\bjtitle{Phys. Rev. E}
\bvolume{55},
\bfpage{2835}--\blpage{2845}
(\byear{1997})
\doiurl{10.1103/PhysRevE.55.2835}
\end{barticle}
\endbibitem

\bibitem[\protect\citeauthoryear{Pavloff}{2002}]{Pavloff:PRA2002}
\begin{barticle}
\bauthor{\bsnm{Pavloff}, \binits{N.}}:
\batitle{Breakdown of superfluidity of an atom laser past an obstacle}.
\bjtitle{Phys. Rev. A}
\bvolume{66},
\bfpage{013610}
(\byear{2002})
\doiurl{10.1103/PhysRevA.66.013610}
\end{barticle}
\endbibitem

\bibitem[\protect\citeauthoryear{Engels and Atherton}{2007}]{Engels:PRL2007}
\begin{barticle}
\bauthor{\bsnm{Engels}, \binits{P.}},
\bauthor{\bsnm{Atherton}, \binits{C.}}:
\batitle{Stationary and nonstationary fluid flow of a bose-einstein condensate
  through a penetrable barrier}.
\bjtitle{Phys. Rev. Lett.}
\bvolume{99},
\bfpage{160405}
(\byear{2007})
\doiurl{10.1103/PhysRevLett.99.160405}
\end{barticle}
\endbibitem

\bibitem[\protect\citeauthoryear{Neely et~al.}{2010}]{Neely:PRL2010}
\begin{barticle}
\bauthor{\bsnm{Neely}, \binits{T.W.}},
\bauthor{\bsnm{Samson}, \binits{E.C.}},
\bauthor{\bsnm{Bradley}, \binits{A.S.}},
\bauthor{\bsnm{Davis}, \binits{M.J.}},
\bauthor{\bsnm{Anderson}, \binits{B.P.}}:
\batitle{Observation of vortex dipoles in an oblate bose-einstein condensate}.
\bjtitle{Phys. Rev. Lett.}
\bvolume{104},
\bfpage{160401}
(\byear{2010})
\doiurl{10.1103/PhysRevLett.104.160401}
\end{barticle}
\endbibitem

\bibitem[\protect\citeauthoryear{Price et~al.}{2023}]{Price:PRA2023}
\begin{barticle}
\bauthor{\bsnm{Price}, \binits{H.M.}},
\bauthor{\bsnm{Wimmer}, \binits{M.}},
\bauthor{\bsnm{Monika}, \binits{M.}},
\bauthor{\bsnm{Peschel}, \binits{U.}},
\bauthor{\bsnm{Carusotto}, \binits{I.}}:
\batitle{Theory of hydrodynamic phenomena in optical mesh lattices}.
\bjtitle{Phys. Rev. A}
\bvolume{108},
\bfpage{063517}
(\byear{2023})
\doiurl{10.1103/PhysRevA.108.063517}
\end{barticle}
\endbibitem

\bibitem[\protect\citeauthoryear{Andronikashvili and
  Mamaladze}{1966}]{Andronikashvili:RMP1966}
\begin{barticle}
\bauthor{\bsnm{Andronikashvili}, \binits{E.L.}},
\bauthor{\bsnm{Mamaladze}, \binits{Y.G.}}:
\batitle{Quantization of macroscopic motions and hydrodynamics of rotating
  helium ii}.
\bjtitle{Rev. Mod. Phys.}
\bvolume{38},
\bfpage{567}--\blpage{625}
(\byear{1966})
\doiurl{10.1103/RevModPhys.38.567}
\end{barticle}
\endbibitem

\bibitem[\protect\citeauthoryear{Carusotto and
  Castin}{2004}]{Carusotto:CRAS2004}
\begin{barticle}
\bauthor{\bsnm{Carusotto}, \binits{I.}},
\bauthor{\bsnm{Castin}, \binits{Y.}}:
\batitle{Superfluidity of the 1d bose gas}.
\bjtitle{Comptes Rendus Physique}
\bvolume{5}(\bissue{1}),
\bfpage{107}--\blpage{127}
(\byear{2004})
\doiurl{10.1016/j.crhy.2004.01.007} .
\bcomment{Bose-Einstein condensates: recent advances in collective effects}
\end{barticle}
\endbibitem

\bibitem[\protect\citeauthoryear{Carusotto and
  Castin}{2007}]{Carusotto:IHP2007}
\begin{botherref}
\oauthor{\bsnm{Carusotto}, \binits{I.}},
\oauthor{\bsnm{Castin}, \binits{Y.}}:
Theoretical studies of thermal vortices in a 2D Bose gas.
Talk at the Conference ``Recent progress in the studies of quantum gases:
  theory and experiments'', Institut Henri Poincar\'e, Paris
(2007)
\end{botherref}
\endbibitem

\bibitem[\protect\citeauthoryear{Hess and Fairbank}{1967}]{Hess:PRL1967}
\begin{barticle}
\bauthor{\bsnm{Hess}, \binits{G.B.}},
\bauthor{\bsnm{Fairbank}, \binits{W.M.}}:
\batitle{Measurements of angular momentum in superfluid helium}.
\bjtitle{Phys. Rev. Lett.}
\bvolume{19},
\bfpage{216}--\blpage{218}
(\byear{1967})
\doiurl{10.1103/PhysRevLett.19.216}
\end{barticle}
\endbibitem

\bibitem[\protect\citeauthoryear{Dalfovo and Stringari}{1992}]{Dalfovo:PRB1992}
\begin{barticle}
\bauthor{\bsnm{Dalfovo}, \binits{F.}},
\bauthor{\bsnm{Stringari}, \binits{S.}}:
\batitle{Static response function for longitudinal and transverse excitations
  in superfluid helium}.
\bjtitle{Phys. Rev. B}
\bvolume{46},
\bfpage{13991}--\blpage{13996}
(\byear{1992})
\doiurl{10.1103/PhysRevB.46.13991}
\end{barticle}
\endbibitem

\bibitem[\protect\citeauthoryear{Carusotto and
  Castin}{2011}]{Carusotto:PRA2011}
\begin{barticle}
\bauthor{\bsnm{Carusotto}, \binits{I.}},
\bauthor{\bsnm{Castin}, \binits{Y.}}:
\batitle{Nonequilibrium and local detection of the normal fraction of a trapped
  two-dimensional bose gas}.
\bjtitle{Phys. Rev. A}
\bvolume{84},
\bfpage{053637}
(\byear{2011})
\doiurl{10.1103/PhysRevA.84.053637}
\end{barticle}
\endbibitem

\bibitem[\protect\citeauthoryear{Cohen-Tannoudji}{2001-2}]{CCT:CdF}
\begin{botherref}
\oauthor{\bsnm{Cohen-Tannoudji}, \binits{C.}}:
Condensats de {B}ose-{E}instein en rotation.
Lectures at Coll\`ege de France
(2001-2)
\end{botherref}
\endbibitem

\bibitem[\protect\citeauthoryear{Cooper and Hadzibabic}{2010}]{Cooper:PRL2010}
\begin{barticle}
\bauthor{\bsnm{Cooper}, \binits{N.R.}},
\bauthor{\bsnm{Hadzibabic}, \binits{Z.}}:
\batitle{Measuring the superfluid fraction of an ultracold atomic gas}.
\bjtitle{Phys. Rev. Lett.}
\bvolume{104},
\bfpage{030401}
(\byear{2010})
\doiurl{10.1103/PhysRevLett.104.030401}
\end{barticle}
\endbibitem

\bibitem[\protect\citeauthoryear{Tinkham}{2004}]{Tinkham}
\begin{bbook}
\bauthor{\bsnm{Tinkham}, \binits{M.}}:
\bbtitle{Introduction to Superconductivity}.
\bpublisher{Dover}, \blocation{???}
(\byear{2004})
\end{bbook}
\endbibitem

\bibitem[\protect\citeauthoryear{Loss et~al.}{1990}]{Loss:PRL1990}
\begin{barticle}
\bauthor{\bsnm{Loss}, \binits{D.}},
\bauthor{\bsnm{Goldbart}, \binits{P.}},
\bauthor{\bsnm{Balatsky}, \binits{A.}}:
\batitle{Berry’s phase and persistent charge and spin currents in textured
  mesoscopic rings}.
\bjtitle{Physical Review Letters}
\bvolume{65}(\bissue{13}),
\bfpage{1655}
(\byear{1990})
\end{barticle}
\endbibitem

\bibitem[\protect\citeauthoryear{Fetter and
  Svidzinsky}{2001}]{Fetter:JPhys2001}
\begin{barticle}
\bauthor{\bsnm{Fetter}, \binits{A.L.}},
\bauthor{\bsnm{Svidzinsky}, \binits{A.A.}}:
\batitle{Vortices in a trapped dilute bose-einstein condensate}.
\bjtitle{Journal of Physics: condensed matter}
\bvolume{13}(\bissue{12}),
\bfpage{135}--\blpage{194}
(\byear{2001})
\end{barticle}
\endbibitem

\bibitem[\protect\citeauthoryear{Abo-Shaeer et~al.}{2001}]{Abo:Science2001}
\begin{barticle}
\bauthor{\bsnm{Abo-Shaeer}, \binits{J.R.}},
\bauthor{\bsnm{Raman}, \binits{C.}},
\bauthor{\bsnm{Vogels}, \binits{J.M.}},
\bauthor{\bsnm{Ketterle}, \binits{W.}}:
\batitle{Observation of vortex lattices in bose-einstein condensates}.
\bjtitle{Science}
\bvolume{292}(\bissue{5516}),
\bfpage{476}--\blpage{479}
(\byear{2001})
\end{barticle}
\endbibitem

\bibitem[\protect\citeauthoryear{Peretti et~al.}{2023}]{Peretti:SciAdv2023}
\begin{barticle}
\bauthor{\bsnm{Peretti}, \binits{C.}},
\bauthor{\bsnm{Vessaire}, \binits{J.}},
\bauthor{\bsnm{Durozoy}, \binits{{\'E}.}},
\bauthor{\bsnm{Gibert}, \binits{M.}}:
\batitle{Direct visualization of the quantum vortex lattice structure,
  oscillations, and destabilization in rotating 4he}.
\bjtitle{Science advances}
\bvolume{9}(\bissue{30}),
\bfpage{2899}
(\byear{2023})
\end{barticle}
\endbibitem

\bibitem[\protect\citeauthoryear{Hall}{1957}]{Hall:ProcRoySoc1957}
\begin{barticle}
\bauthor{\bsnm{Hall}, \binits{H.E.}}:
\batitle{The angular acceleration of liquid helium. ii}.
\bjtitle{Philosophical Transactions of the Royal Society of London. Series A,
  Mathematical and Physical Sciences}
\bvolume{250}(\bissue{980}),
\bfpage{359}--\blpage{385}
(\byear{1957}).
Accessed 2026-09-30
\end{barticle}
\endbibitem

\bibitem[\protect\citeauthoryear{Vinen}{1958}]{Vinen:Nature1958}
\begin{barticle}
\bauthor{\bsnm{Vinen}, \binits{W.}}:
\batitle{Detection of single quanta of circulation in rotating helium ii}.
\bjtitle{Nature}
\bvolume{181}(\bissue{4622}),
\bfpage{1524}--\blpage{1525}
(\byear{1958})
\end{barticle}
\endbibitem

\bibitem[\protect\citeauthoryear{Davis et~al.}{1991}]{Davis:PRL1991}
\begin{barticle}
\bauthor{\bsnm{Davis}, \binits{J.C.}},
\bauthor{\bsnm{Close}, \binits{J.D.}},
\bauthor{\bsnm{Zieve}, \binits{R.}},
\bauthor{\bsnm{Packard}, \binits{R.E.}}:
\batitle{Observation of quantized circulation in superfluid $^{3}\mathit{B}$}.
\bjtitle{Phys. Rev. Lett.}
\bvolume{66},
\bfpage{329}--\blpage{332}
(\byear{1991})
\doiurl{10.1103/PhysRevLett.66.329}
\end{barticle}
\endbibitem

\bibitem[\protect\citeauthoryear{Ryu et~al.}{2007}]{Ryu:PRL2007}
\begin{barticle}
\bauthor{\bsnm{Ryu}, \binits{C.}},
\bauthor{\bsnm{Andersen}, \binits{M.F.}},
\bauthor{\bsnm{Clad{\'e}}, \binits{P.}},
\bauthor{\bsnm{Natarajan}, \binits{V.}},
\bauthor{\bsnm{Helmerson}, \binits{K.}},
\bauthor{\bsnm{Phillips}, \binits{W.D.}}:
\batitle{Observation of persistent flow of a bose-einstein condensate in a
  toroidal trap}.
\bjtitle{Phys. Rev. Lett.}
\bvolume{99},
\bfpage{260401}
(\byear{2007})
\doiurl{10.1103/PhysRevLett.99.260401}
\end{barticle}
\endbibitem

\bibitem[\protect\citeauthoryear{Ramanathan et~al.}{2011}]{Ramanathan:PRL2011}
\begin{barticle}
\bauthor{\bsnm{Ramanathan}, \binits{A.}},
\bauthor{\bsnm{Wright}, \binits{K.C.}},
\bauthor{\bsnm{Muniz}, \binits{S.R.}},
\bauthor{\bsnm{Zelan}, \binits{M.}},
\bauthor{\bsnm{Hill}, \binits{W.T.}},
\bauthor{\bsnm{Lobb}, \binits{C.J.}},
\bauthor{\bsnm{Helmerson}, \binits{K.}},
\bauthor{\bsnm{Phillips}, \binits{W.D.}},
\bauthor{\bsnm{Campbell}, \binits{G.K.}}:
\batitle{Superflow in a toroidal bose-einstein condensate: An atom circuit with
  a tunable weak link}.
\bjtitle{Phys. Rev. Lett.}
\bvolume{106},
\bfpage{130401}
(\byear{2011})
\doiurl{10.1103/PhysRevLett.106.130401}
\end{barticle}
\endbibitem

\bibitem[\protect\citeauthoryear{Moulder et~al.}{2012}]{Moulder:PRA2012}
\begin{barticle}
\bauthor{\bsnm{Moulder}, \binits{S.}},
\bauthor{\bsnm{Beattie}, \binits{S.}},
\bauthor{\bsnm{Smith}, \binits{R.P.}},
\bauthor{\bsnm{Tammuz}, \binits{N.}},
\bauthor{\bsnm{Hadzibabic}, \binits{Z.}}:
\batitle{Quantized supercurrent decay in an annular bose-einstein condensate}.
\bjtitle{Phys. Rev. A}
\bvolume{86},
\bfpage{013629}
(\byear{2012})
\doiurl{10.1103/PhysRevA.86.013629}
\end{barticle}
\endbibitem

\bibitem[\protect\citeauthoryear{Packard and Vitale}{1992}]{Packard:PRB1992}
\begin{barticle}
\bauthor{\bsnm{Packard}, \binits{R.E.}},
\bauthor{\bsnm{Vitale}, \binits{S.}}:
\batitle{Principles of superfluid-helium gyroscopes}.
\bjtitle{Phys. Rev. B}
\bvolume{46},
\bfpage{3540}--\blpage{3549}
(\byear{1992})
\doiurl{10.1103/PhysRevB.46.3540}
\end{barticle}
\endbibitem

\bibitem[\protect\citeauthoryear{Avenel et~al.}{1997}]{Avenel:PRL1997}
\begin{barticle}
\bauthor{\bsnm{Avenel}, \binits{O.}},
\bauthor{\bsnm{Hakonen}, \binits{P.}},
\bauthor{\bsnm{Varoquaux}, \binits{E.}}:
\batitle{Detection of the rotation of the earth with a superfluid gyrometer}.
\bjtitle{Phys. Rev. Lett.}
\bvolume{78},
\bfpage{3602}--\blpage{3605}
(\byear{1997})
\doiurl{10.1103/PhysRevLett.78.3602}
\end{barticle}
\endbibitem

\bibitem[\protect\citeauthoryear{Avenel and Varoquaux}{1985}]{Avenel:PRL1985}
\begin{barticle}
\bauthor{\bsnm{Avenel}, \binits{O.}},
\bauthor{\bsnm{Varoquaux}, \binits{E.}}:
\batitle{Observation of singly quantized dissipation events obeying the
  josephson frequency relation in the critical flow of superfluid
  $^{4}\mathrm{He}$ through an aperture}.
\bjtitle{Phys. Rev. Lett.}
\bvolume{55},
\bfpage{2704}--\blpage{2707}
(\byear{1985})
\doiurl{10.1103/PhysRevLett.55.2704}
\end{barticle}
\endbibitem

\bibitem[\protect\citeauthoryear{Amo et~al.}{2009}]{Amo:Nature2009}
\begin{barticle}
\bauthor{\bsnm{Amo}, \binits{A.}},
\bauthor{\bsnm{Sanvitto}, \binits{D.}},
\bauthor{\bsnm{Laussy}, \binits{F.}},
\bauthor{\bsnm{Ballarini}, \binits{D.}},
\bauthor{\bsnm{Valle}, \binits{E.d.}},
\bauthor{\bsnm{Martin}, \binits{M.}},
\bauthor{\bsnm{Lemaitre}, \binits{A.}},
\bauthor{\bsnm{Bloch}, \binits{J.}},
\bauthor{\bsnm{Krizhanovskii}, \binits{D.}},
\bauthor{\bsnm{Skolnick}, \binits{M.}}, \betal:
\batitle{Collective fluid dynamics of a polariton condensate in a semiconductor
  microcavity}.
\bjtitle{Nature}
\bvolume{457}(\bissue{7227}),
\bfpage{291}--\blpage{295}
(\byear{2009})
\end{barticle}
\endbibitem

\bibitem[\protect\citeauthoryear{Amo et~al.}{2011}]{Amo:2011Science}
\begin{barticle}
\bauthor{\bsnm{Amo}, \binits{A.}},
\bauthor{\bsnm{Pigeon}, \binits{S.}},
\bauthor{\bsnm{Sanvitto}, \binits{D.}},
\bauthor{\bsnm{Sala}, \binits{V.G.}},
\bauthor{\bsnm{Hivet}, \binits{R.}},
\bauthor{\bsnm{Carusotto}, \binits{I.}},
\bauthor{\bsnm{Pisanello}, \binits{F.}},
\bauthor{\bsnm{Leménager}, \binits{G.}},
\bauthor{\bsnm{Houdré}, \binits{R.}},
\bauthor{\bsnm{Giacobino}, \binits{E.}},
\bauthor{\bsnm{Ciuti}, \binits{C.}},
\bauthor{\bsnm{Bramati}, \binits{A.}}:
\batitle{Polariton superfluids reveal quantum hydrodynamic solitons}.
\bjtitle{Science}
\bvolume{332}(\bissue{6034}),
\bfpage{1167}--\blpage{1170}
(\byear{2011})
\doiurl{10.1126/science.1202307}
{\href{https://arxiv.org/abs/http://www.sciencemag.org/content/332/6034/1167.full.pdf}{{http://www.sciencemag.org/content/332/6034/1167.full.pdf}}}
\end{barticle}
\endbibitem

\bibitem[\protect\citeauthoryear{Nardin et~al.}{2011}]{Nardin:2011NatPhys}
\begin{barticle}
\bauthor{\bsnm{Nardin}, \binits{G.}},
\bauthor{\bsnm{Grosso}, \binits{G.}},
\bauthor{\bsnm{Leger}, \binits{Y.}},
\bauthor{\bsnm{Pietka}, \binits{B.}},
\bauthor{\bsnm{Morier-Genoud}, \binits{F.}},
\bauthor{\bsnm{Deveaud-Pledran}, \binits{B.}}:
\batitle{Hydrodynamic nucleation of quantized vortex pairs in a polariton
  quantum fluid}.
\bjtitle{Nat. Phys.}
\bvolume{7}(\bissue{8}),
\bfpage{635}--\blpage{641}
(\byear{2011})
\doiurl{10.1038/NPHYS1959}
\end{barticle}
\endbibitem

\bibitem[\protect\citeauthoryear{Sanvitto et~al.}{2011}]{Sanvitto:NPhot2011}
\begin{barticle}
\bauthor{\bsnm{Sanvitto}, \binits{D.}},
\bauthor{\bsnm{Pigeon}, \binits{S.}},
\bauthor{\bsnm{Amo}, \binits{A.}},
\bauthor{\bsnm{Ballarini}, \binits{D.}},
\bauthor{\bsnm{Giorgi}, \binits{M.D.}},
\bauthor{\bsnm{Carusotto}, \binits{I.}},
\bauthor{\bsnm{Hivet}, \binits{R.}},
\bauthor{\bsnm{Pisanello}, \binits{F.}},
\bauthor{\bsnm{Sala}, \binits{V.G.}},
\bauthor{\bsnm{Guimaraes}, \binits{P.S.S.}},
\bauthor{\bsnm{Houdr\'e}, \binits{R.}},
\bauthor{\bsnm{Giacobino}, \binits{E.}},
\bauthor{\bsnm{Ciuti}, \binits{C.}},
\bauthor{\bsnm{Bramati}, \binits{A.}},
\bauthor{\bsnm{Gigli}, \binits{G.}}:
\batitle{All-optical control of the quantum flow of a polariton condensate}.
\bjtitle{Nat. Phot.}
\bvolume{5},
\bfpage{610}--\blpage{614}
(\byear{2011})
\end{barticle}
\endbibitem

\bibitem[\protect\citeauthoryear{Wouters and Carusotto}{2010}]{Wouters:PRL2010}
\begin{barticle}
\bauthor{\bsnm{Wouters}, \binits{M.}},
\bauthor{\bsnm{Carusotto}, \binits{I.}}:
\batitle{Superfluidity and critical velocities in nonequilibrium bose-einstein
  condensates}.
\bjtitle{Phys. Rev. Lett.}
\bvolume{105}(\bissue{2}),
\bfpage{020602}
(\byear{2010})
\doiurl{10.1103/PhysRevLett.105.020602}
\end{barticle}
\endbibitem

\bibitem[\protect\citeauthoryear{Astrakharchik and
  Pitaevskii}{2004}]{Astrakharchik:PRA2004}
\begin{barticle}
\bauthor{\bsnm{Astrakharchik}, \binits{G.}},
\bauthor{\bsnm{Pitaevskii}, \binits{L.}}:
\batitle{Motion of a heavy impurity through a bose-einstein condensate}.
\bjtitle{Physical Review A—Atomic, Molecular, and Optical Physics}
\bvolume{70}(\bissue{1}),
\bfpage{013608}
(\byear{2004})
\end{barticle}
\endbibitem

\bibitem[\protect\citeauthoryear{Gnusov et~al.}{2023}]{Gnusov:SciAdv2023}
\begin{barticle}
\bauthor{\bsnm{Gnusov}, \binits{I.}},
\bauthor{\bsnm{Harrison}, \binits{S.}},
\bauthor{\bsnm{Alyatkin}, \binits{S.}},
\bauthor{\bsnm{Sitnik}, \binits{K.}},
\bauthor{\bsnm{T{\"o}pfer}, \binits{J.}},
\bauthor{\bsnm{Sigurdsson}, \binits{H.}},
\bauthor{\bsnm{Lagoudakis}, \binits{P.}}:
\batitle{Quantum vortex formation in the “rotating bucket” experiment with
  polariton condensates}.
\bjtitle{Science advances}
\bvolume{9}(\bissue{4}),
\bfpage{1299}
(\byear{2023})
\end{barticle}
\endbibitem

\bibitem[\protect\citeauthoryear{del Valle-Inclan~Redondo
  et~al.}{2023}]{DelValle:NanoLett2023}
\begin{barticle}
\bauthor{\bsnm{Valle-Inclan~Redondo}, \binits{Y.}},
\bauthor{\bsnm{Schneider}, \binits{C.}},
\bauthor{\bsnm{Klembt}, \binits{S.}},
\bauthor{\bsnm{H\"{o}fling}, \binits{S.}},
\bauthor{\bsnm{Tarucha}, \binits{S.}},
\bauthor{\bsnm{Fraser}, \binits{M.D.}}:
\batitle{Optically driven rotation of exciton--polariton condensates}.
\bjtitle{Nano Letters}
\bvolume{23}(\bissue{10}),
\bfpage{4564}--\blpage{4571}
(\byear{2023})
\end{barticle}
\endbibitem

\bibitem[\protect\citeauthoryear{Ozawa et~al.}{2019}]{ozawaRMP2019topological}
\begin{barticle}
\bauthor{\bsnm{Ozawa}, \binits{T.}},
\bauthor{\bsnm{Price}, \binits{H.M.}},
\bauthor{\bsnm{Amo}, \binits{A.}},
\bauthor{\bsnm{Goldman}, \binits{N.}},
\bauthor{\bsnm{Hafezi}, \binits{M.}},
\bauthor{\bsnm{Lu}, \binits{L.}},
\bauthor{\bsnm{Rechtsman}, \binits{M.C.}},
\bauthor{\bsnm{Schuster}, \binits{D.}},
\bauthor{\bsnm{Simon}, \binits{J.}},
\bauthor{\bsnm{Zilberberg}, \binits{O.}}, \betal:
\batitle{Topological photonics}.
\bjtitle{Reviews of Modern Physics}
\bvolume{91}(\bissue{1}),
\bfpage{015006}
(\byear{2019})
\end{barticle}
\endbibitem

\bibitem[\protect\citeauthoryear{Keeling}{2011}]{Keeling:PRL2011}
\begin{barticle}
\bauthor{\bsnm{Keeling}, \binits{J.}}:
\batitle{Superfluid density of an open dissipative condensate}.
\bjtitle{Phys. Rev. Lett.}
\bvolume{107},
\bfpage{080402}
(\byear{2011})
\doiurl{10.1103/PhysRevLett.107.080402}
\end{barticle}
\endbibitem

\bibitem[\protect\citeauthoryear{Juggins et~al.}{2018}]{Juggins:NatComm2018}
\begin{barticle}
\bauthor{\bsnm{Juggins}, \binits{R.}},
\bauthor{\bsnm{Keeling}, \binits{J.}},
\bauthor{\bsnm{Szyma{\'n}ska}, \binits{M.}}:
\batitle{Coherently driven microcavity-polaritons and the question of
  superfluidity}.
\bjtitle{Nature communications}
\bvolume{9}(\bissue{1}),
\bfpage{4062}
(\byear{2018})
\end{barticle}
\endbibitem

\bibitem[\protect\citeauthoryear{Lagoudakis
  et~al.}{2008}]{Lagoudakis:NatPhys2008}
\begin{barticle}
\bauthor{\bsnm{Lagoudakis}, \binits{K.G.}},
\bauthor{\bsnm{Wouters}, \binits{M.}},
\bauthor{\bsnm{Richard}, \binits{M.}},
\bauthor{\bsnm{Baas}, \binits{A.}},
\bauthor{\bsnm{Carusotto}, \binits{I.}},
\bauthor{\bsnm{Andre}, \binits{R.}},
\bauthor{\bsnm{Dang}, \binits{L.E.S.I.}},
\bauthor{\bsnm{Deveaud-Pledran}, \binits{B.}}:
\batitle{Quantized vortices in an exciton-polariton condensate}.
\bjtitle{Nature Phys.}
\bvolume{4}(\bissue{9}),
\bfpage{706}--\blpage{710}
(\byear{2008})
\doiurl{10.1038/nphys1051}
\end{barticle}
\endbibitem

\bibitem[\protect\citeauthoryear{Sanvitto et~al.}{2010}]{Sanvitto:NatPhys2010}
\begin{barticle}
\bauthor{\bsnm{Sanvitto}, \binits{D.}},
\bauthor{\bsnm{Marchetti}, \binits{F.M.}},
\bauthor{\bsnm{Szymanska}, \binits{M.H.}},
\bauthor{\bsnm{Tosi}, \binits{G.}},
\bauthor{\bsnm{Baudisch}, \binits{M.}},
\bauthor{\bsnm{Laussy}, \binits{F.P.}},
\bauthor{\bsnm{Krizhanovskii}, \binits{D.N.}},
\bauthor{\bsnm{Skolnick}, \binits{M.S.}},
\bauthor{\bsnm{Marrucci}, \binits{L.}},
\bauthor{\bsnm{Lemaitre}, \binits{A.}},
\bauthor{\bsnm{Bloch}, \binits{J.}},
\bauthor{\bsnm{Tejedor}, \binits{C.}},
\bauthor{\bsnm{Vina}, \binits{L.}}:
\batitle{Persistent currents and quantized vortices in a polariton superfluid}.
\bjtitle{Nature Phys.}
\bvolume{6}(\bissue{7}),
\bfpage{527}--\blpage{533}
(\byear{2010})
\doiurl{10.1038/NPHYS1668}
\end{barticle}
\endbibitem

\bibitem[\protect\citeauthoryear{Yao et~al.}{2025}]{Yao:Optica2025}
\begin{barticle}
\bauthor{\bsnm{Yao}, \binits{Q.}},
\bauthor{\bsnm{Comaron}, \binits{P.}},
\bauthor{\bsnm{Alnatah}, \binits{H.}},
\bauthor{\bsnm{Beaumariage}, \binits{J.}},
\bauthor{\bsnm{Mukherjee}, \binits{S.}},
\bauthor{\bsnm{West}, \binits{K.}},
\bauthor{\bsnm{Pfeiffer}, \binits{L.}},
\bauthor{\bsnm{Baldwin}, \binits{K.}},
\bauthor{\bsnm{Szyma{\'n}ska}, \binits{M.H.}},
\bauthor{\bsnm{Snoke}, \binits{D.}}:
\batitle{Persistent, controllable circulation of a polariton ring condensate}.
\bjtitle{Optica}
\bvolume{12}(\bissue{7}),
\bfpage{991}--\blpage{996}
(\byear{2025})
\end{barticle}
\endbibitem

\bibitem[\protect\citeauthoryear{Macek and Davis}{1963}]{Macek:APL1963}
\begin{barticle}
\bauthor{\bsnm{Macek}, \binits{W.M.}},
\bauthor{\bsnm{Davis}, \binits{J.} \bsuffix{D.~T.~M.}}:
\batitle{Rotation rate sensing with traveling‐wave ring lasers}.
\bjtitle{Applied Physics Letters}
\bvolume{2}(\bissue{3}),
\bfpage{67}--\blpage{68}
(\byear{1963})
\end{barticle}
\endbibitem

\bibitem[\protect\citeauthoryear{Giovinetti et~al.}{2024}]{Giovinetti:FQST2024}
\begin{barticle}
\bauthor{\bsnm{Giovinetti}, \binits{F.}},
\bauthor{\bsnm{Altucci}, \binits{C.}},
\bauthor{\bsnm{Bajardi}, \binits{F.}},
\bauthor{\bsnm{Basti}, \binits{A.}},
\bauthor{\bsnm{Beverini}, \binits{N.}},
\bauthor{\bsnm{Capozziello}, \binits{S.}},
\bauthor{\bsnm{Carelli}, \binits{G.}},
\bauthor{\bsnm{Castellano}, \binits{S.}},
\bauthor{\bsnm{Ciampini}, \binits{D.}},
\bauthor{\bsnm{Di~Somma}, \binits{G.}}, \betal:
\batitle{Gingerino: a high sensitivity ring laser gyroscope for fundamental and
  quantum physics investigation}.
\bjtitle{Frontiers in Quantum Science and Technology}
\bvolume{3},
\bfpage{1363409}
(\byear{2024})
\end{barticle}
\endbibitem

\bibitem[\protect\citeauthoryear{{Andreev} and
  {Lifshitz}}{1969}]{Andreev:JETP1969}
\begin{barticle}
\bauthor{\bsnm{{Andreev}}, \binits{A.F.}},
\bauthor{\bsnm{{Lifshitz}}, \binits{I.M.}}:
\batitle{{Quantum Theory of Defects in Crystals}}.
\bjtitle{Soviet Journal of Experimental and Theoretical Physics}
\bvolume{29},
\bfpage{1107}
(\byear{1969})
\end{barticle}
\endbibitem

\bibitem[\protect\citeauthoryear{Chester}{1970}]{Chester:PRA1970}
\begin{barticle}
\bauthor{\bsnm{Chester}, \binits{G.V.}}:
\batitle{Speculations on {Bose}-{Einstein} {Condensation} and {Quantum}
  {Crystals}}.
\bjtitle{Physical Review A}
\bvolume{2}(\bissue{1}),
\bfpage{256}--\blpage{258}
(\byear{1970})
\doiurl{10.1103/PhysRevA.2.256} .
Accessed 2026-01-11
\end{barticle}
\endbibitem

\bibitem[\protect\citeauthoryear{Leggett}{1970}]{Leggett:PRL1970}
\begin{barticle}
\bauthor{\bsnm{Leggett}, \binits{A.J.}}:
\batitle{Can a {Solid} {Be} "{Superfluid}"?}
\bjtitle{Physical Review Letters}
\bvolume{25}(\bissue{22}),
\bfpage{1543}--\blpage{1546}
(\byear{1970})
\doiurl{10.1103/PhysRevLett.25.1543} .
Accessed 2026-01-11
\end{barticle}
\endbibitem

\bibitem[\protect\citeauthoryear{Kim and Chan}{2004a}]{Kim:Science2004}
\begin{barticle}
\bauthor{\bsnm{Kim}, \binits{E.}},
\bauthor{\bsnm{Chan}, \binits{M.H.W.}}:
\batitle{Observation of {Superflow} in {Solid} {Helium}}.
\bjtitle{Science}
\bvolume{305}(\bissue{5692}),
\bfpage{1941}--\blpage{1944}
(\byear{2004})
\doiurl{10.1126/science.1101501} .
Accessed 2026-01-17
\end{barticle}
\endbibitem

\bibitem[\protect\citeauthoryear{Kim and Chan}{2004b}]{Kim:Nature2004}
\begin{barticle}
\bauthor{\bsnm{Kim}, \binits{E.}},
\bauthor{\bsnm{Chan}, \binits{M.H.W.}}:
\batitle{Probable observation of a supersolid helium phase}.
\bjtitle{Nature}
\bvolume{427}(\bissue{6971}),
\bfpage{225}--\blpage{227}
(\byear{2004})
\doiurl{10.1038/nature02220} .
Accessed 2026-01-13
\end{barticle}
\endbibitem

\bibitem[\protect\citeauthoryear{Day and Beamish}{2007}]{Day:Nature2007}
\begin{barticle}
\bauthor{\bsnm{Day}, \binits{J.}},
\bauthor{\bsnm{Beamish}, \binits{J.}}:
\batitle{Low-temperature shear modulus changes in solid 4he and connection to
  supersolidity}.
\bjtitle{Nature}
\bvolume{450}(\bissue{7171}),
\bfpage{853}--\blpage{856}
(\byear{2007})
\end{barticle}
\endbibitem

\bibitem[\protect\citeauthoryear{Syshchenko et~al.}{2010}]{Syshchenko:PRL2010}
\begin{barticle}
\bauthor{\bsnm{Syshchenko}, \binits{O.}},
\bauthor{\bsnm{Day}, \binits{J.}},
\bauthor{\bsnm{Beamish}, \binits{J.}}:
\batitle{Frequency {Dependence} and {Dissipation} in the {Dynamics} of {Solid}
  {Helium}}.
\bjtitle{Physical Review Letters}
\bvolume{104}(\bissue{19}),
\bfpage{195301}
(\byear{2010})
\doiurl{10.1103/PhysRevLett.104.195301} .
Accessed 2026-01-14
\end{barticle}
\endbibitem

\bibitem[\protect\citeauthoryear{Balibar}{2010}]{Balibar:Nature2010}
\begin{barticle}
\bauthor{\bsnm{Balibar}, \binits{S.}}:
\batitle{The enigma of supersolidity}.
\bjtitle{Nature}
\bvolume{464}(\bissue{7286}),
\bfpage{176}--\blpage{182}
(\byear{2010})
\doiurl{10.1038/nature08913} .
Accessed 2026-01-09
\end{barticle}
\endbibitem

\bibitem[\protect\citeauthoryear{Kim and Chan}{2012}]{Kim:PRL2012}
\begin{barticle}
\bauthor{\bsnm{Kim}, \binits{D.Y.}},
\bauthor{\bsnm{Chan}, \binits{M.H.W.}}:
\batitle{Absence of {Supersolidity} in {Solid} {Helium} in {Porous} {Vycor}
  {Glass}}.
\bjtitle{Physical Review Letters}
\bvolume{109}(\bissue{15}),
\bfpage{155301}
(\byear{2012})
\doiurl{10.1103/PhysRevLett.109.155301} .
Accessed 2026-01-11
\end{barticle}
\endbibitem

\bibitem[\protect\citeauthoryear{Hallock}{2015}]{Hallock:PhysToday2015}
\begin{barticle}
\bauthor{\bsnm{Hallock}, \binits{R.}}:
\batitle{Is solid helium a supersolid?}
\bjtitle{Physics Today}
\bvolume{68}(\bissue{5}),
\bfpage{30}--\blpage{35}
(\byear{2015})
\doiurl{10.1063/PT.3.2782} .
Accessed 2026-01-10
\end{barticle}
\endbibitem

\bibitem[\protect\citeauthoryear{Boninsegni and
  Prokof’ev}{2012}]{Boninsegni:RMP2012}
\begin{barticle}
\bauthor{\bsnm{Boninsegni}, \binits{M.}},
\bauthor{\bsnm{Prokof’ev}, \binits{N.V.}}:
\batitle{Colloquium: {Supersolids}: {What} and where are they?}
\bjtitle{Reviews of Modern Physics}
\bvolume{84}(\bissue{2}),
\bfpage{759}--\blpage{776}
(\byear{2012})
\doiurl{10.1103/RevModPhys.84.759} .
Accessed 2026-01-14
\end{barticle}
\endbibitem

\bibitem[\protect\citeauthoryear{Recati and
  Stringari}{2023}]{Recati:NatRevPhys2023}
\begin{barticle}
\bauthor{\bsnm{Recati}, \binits{A.}},
\bauthor{\bsnm{Stringari}, \binits{S.}}:
\batitle{Supersolidity in ultracold dipolar gases}.
\bjtitle{Nature Reviews Physics}
\bvolume{5}(\bissue{12}),
\bfpage{735}--\blpage{743}
(\byear{2023})
\end{barticle}
\endbibitem

\bibitem[\protect\citeauthoryear{Li et~al.}{2017}]{Li:Nature2017}
\begin{barticle}
\bauthor{\bsnm{Li}, \binits{J.-R.}},
\bauthor{\bsnm{Lee}, \binits{J.}},
\bauthor{\bsnm{Huang}, \binits{W.}},
\bauthor{\bsnm{Burchesky}, \binits{S.}},
\bauthor{\bsnm{Shteynas}, \binits{B.}},
\bauthor{\bsnm{Top}, \binits{F.C.}},
\bauthor{\bsnm{Jamison}, \binits{A.O.}},
\bauthor{\bsnm{Ketterle}, \binits{W.}}:
\batitle{A stripe phase with supersolid properties in spin–orbit-coupled
  bose–einstein condensates}.
\bjtitle{Nature (London)}
\bvolume{543},
\bfpage{91}
(\byear{2017})
\end{barticle}
\endbibitem

\bibitem[\protect\citeauthoryear{L{\'e}onard
  et~al.}{2017}]{leonard2017supersolid}
\begin{barticle}
\bauthor{\bsnm{L{\'e}onard}, \binits{J.}},
\bauthor{\bsnm{Morales}, \binits{A.}},
\bauthor{\bsnm{Zupancic}, \binits{P.}},
\bauthor{\bsnm{Esslinger}, \binits{T.}},
\bauthor{\bsnm{Donner}, \binits{T.}}:
\batitle{Supersolid formation in a quantum gas breaking a continuous
  translational symmetry}.
\bjtitle{Nature}
\bvolume{543}(\bissue{7643}),
\bfpage{87}--\blpage{90}
(\byear{2017})
\end{barticle}
\endbibitem

\bibitem[\protect\citeauthoryear{Biagioni et~al.}{2024}]{Biagioni:Nature2024}
\begin{barticle}
\bauthor{\bsnm{Biagioni}, \binits{G.}},
\bauthor{\bsnm{Antolini}, \binits{N.}},
\bauthor{\bsnm{Donelli}, \binits{B.}},
\bauthor{\bsnm{Pezz{\`e}}, \binits{L.}},
\bauthor{\bsnm{Smerzi}, \binits{A.}},
\bauthor{\bsnm{Fattori}, \binits{M.}},
\bauthor{\bsnm{Fioretti}, \binits{A.}},
\bauthor{\bsnm{Gabbanini}, \binits{C.}},
\bauthor{\bsnm{Inguscio}, \binits{M.}},
\bauthor{\bsnm{Tanzi}, \binits{L.}}, \betal:
\batitle{Measurement of the superfluid fraction of a supersolid by josephson
  effect}.
\bjtitle{Nature}
\bvolume{629}(\bissue{8013}),
\bfpage{773}--\blpage{777}
(\byear{2024})
\end{barticle}
\endbibitem

\bibitem[\protect\citeauthoryear{Casotti et~al.}{2024}]{Casotti:Nature2024}
\begin{barticle}
\bauthor{\bsnm{Casotti}, \binits{E.}},
\bauthor{\bsnm{Poli}, \binits{E.}},
\bauthor{\bsnm{Klaus}, \binits{L.}},
\bauthor{\bsnm{Litvinov}, \binits{A.}},
\bauthor{\bsnm{Ulm}, \binits{C.}},
\bauthor{\bsnm{Politi}, \binits{C.}},
\bauthor{\bsnm{Mark}, \binits{M.J.}},
\bauthor{\bsnm{Bland}, \binits{T.}},
\bauthor{\bsnm{Ferlaino}, \binits{F.}}:
\batitle{Observation of vortices in a dipolar supersolid}.
\bjtitle{Nature}
\bvolume{635}(\bissue{8038}),
\bfpage{327}--\blpage{331}
(\byear{2024})
\end{barticle}
\endbibitem

\bibitem[\protect\citeauthoryear{Guo et~al.}{2019}]{Guo:Nature2019}
\begin{barticle}
\bauthor{\bsnm{Guo}, \binits{M.}},
\bauthor{\bsnm{B{\"o}ttcher}, \binits{F.}},
\bauthor{\bsnm{Hertkorn}, \binits{J.}},
\bauthor{\bsnm{Schmidt}, \binits{J.-N.}},
\bauthor{\bsnm{Wenzel}, \binits{M.}},
\bauthor{\bsnm{B{\"u}chler}, \binits{H.P.}},
\bauthor{\bsnm{Langen}, \binits{T.}},
\bauthor{\bsnm{Pfau}, \binits{T.}}:
\batitle{The low-energy goldstone mode in a trapped dipolar supersolid}.
\bjtitle{Nature}
\bvolume{574}(\bissue{7778}),
\bfpage{386}--\blpage{389}
(\byear{2019})
\end{barticle}
\endbibitem

\bibitem[\protect\citeauthoryear{Natale et~al.}{2019}]{Natale:PRL2019}
\begin{barticle}
\bauthor{\bsnm{Natale}, \binits{G.}},
\bauthor{\bsnm{Van~Bijnen}, \binits{R.}},
\bauthor{\bsnm{Patscheider}, \binits{A.}},
\bauthor{\bsnm{Petter}, \binits{D.}},
\bauthor{\bsnm{Mark}, \binits{M.J.}},
\bauthor{\bsnm{Chomaz}, \binits{L.}},
\bauthor{\bsnm{Ferlaino}, \binits{F.}}:
\batitle{Excitation spectrum of a trapped dipolar supersolid and its
  experimental evidence}.
\bjtitle{Physical review letters}
\bvolume{123}(\bissue{5}),
\bfpage{050402}
(\byear{2019})
\end{barticle}
\endbibitem

\bibitem[\protect\citeauthoryear{Tanzi et~al.}{2019}]{Tanzi:Nature2019}
\begin{barticle}
\bauthor{\bsnm{Tanzi}, \binits{L.}},
\bauthor{\bsnm{Roccuzzo}, \binits{S.}},
\bauthor{\bsnm{Lucioni}, \binits{E.}},
\bauthor{\bsnm{Fam{\`a}}, \binits{F.}},
\bauthor{\bsnm{Fioretti}, \binits{A.}},
\bauthor{\bsnm{Gabbanini}, \binits{C.}},
\bauthor{\bsnm{Modugno}, \binits{G.}},
\bauthor{\bsnm{Recati}, \binits{A.}},
\bauthor{\bsnm{Stringari}, \binits{S.}}:
\batitle{Supersolid symmetry breaking from compressional oscillations in a
  dipolar quantum gas}.
\bjtitle{Nature}
\bvolume{574}(\bissue{7778}),
\bfpage{382}--\blpage{385}
(\byear{2019})
\end{barticle}
\endbibitem

\bibitem[\protect\citeauthoryear{Chisholm et~al.}{2026}]{Chisholm:Science2026}
\begin{barticle}
\bauthor{\bsnm{Chisholm}, \binits{C.}},
\bauthor{\bsnm{Hirthe}, \binits{S.}},
\bauthor{\bsnm{Makhalov}, \binits{V.}},
\bauthor{\bsnm{Ramos}, \binits{R.}},
\bauthor{\bsnm{Vatr{\'e}}, \binits{R.}},
\bauthor{\bsnm{Cabedo}, \binits{J.}},
\bauthor{\bsnm{Celi}, \binits{A.}},
\bauthor{\bsnm{Tarruell}, \binits{L.}}:
\batitle{Probing supersolidity through excitations in a spin-orbit--coupled
  bose-einstein condensate}.
\bjtitle{Science}
\bvolume{391}(\bissue{6784}),
\bfpage{480}--\blpage{484}
(\byear{2026})
\end{barticle}
\endbibitem

\bibitem[\protect\citeauthoryear{Tanzi et~al.}{2021}]{Tanzi:Science2021}
\begin{barticle}
\bauthor{\bsnm{Tanzi}, \binits{L.}},
\bauthor{\bsnm{Maloberti}, \binits{J.}},
\bauthor{\bsnm{Biagioni}, \binits{G.}},
\bauthor{\bsnm{Fioretti}, \binits{A.}},
\bauthor{\bsnm{Gabbanini}, \binits{C.}},
\bauthor{\bsnm{Modugno}, \binits{G.}}:
\batitle{Evidence of superfluidity in a dipolar supersolid from nonclassical
  rotational inertia}.
\bjtitle{Science}
\bvolume{371}(\bissue{6534}),
\bfpage{1162}--\blpage{1165}
(\byear{2021})
\end{barticle}
\endbibitem

\bibitem[\protect\citeauthoryear{Norcia et~al.}{2022}]{Norcia:PRL2022}
\begin{barticle}
\bauthor{\bsnm{Norcia}, \binits{M.A.}},
\bauthor{\bsnm{Poli}, \binits{E.}},
\bauthor{\bsnm{Politi}, \binits{C.}},
\bauthor{\bsnm{Klaus}, \binits{L.}},
\bauthor{\bsnm{Bland}, \binits{T.}},
\bauthor{\bsnm{Mark}, \binits{M.J.}},
\bauthor{\bsnm{Santos}, \binits{L.}},
\bauthor{\bsnm{Bisset}, \binits{R.N.}},
\bauthor{\bsnm{Ferlaino}, \binits{F.}}:
\batitle{Can angular oscillations probe superfluidity in dipolar supersolids?}
\bjtitle{Phys. Rev. Lett.}
\bvolume{129},
\bfpage{040403}
(\byear{2022})
\doiurl{10.1103/PhysRevLett.129.040403}
\end{barticle}
\endbibitem

\bibitem[\protect\citeauthoryear{Senarath~Yapa and
  Bland}{2025}]{Senarath:PRA2025}
\begin{barticle}
\bauthor{\bsnm{Senarath~Yapa}, \binits{P.}},
\bauthor{\bsnm{Bland}, \binits{T.}}:
\batitle{Anomalous dispersion of shear waves in dipolar supersolids}.
\bjtitle{Phys. Rev. A}
\bvolume{112},
\bfpage{021303}
(\byear{2025})
\doiurl{10.1103/xz57-52ft}
\end{barticle}
\endbibitem

\bibitem[\protect\citeauthoryear{Liebster et~al.}{2025a}]{Liebster:PRX2025}
\begin{barticle}
\bauthor{\bsnm{Liebster}, \binits{N.}},
\bauthor{\bsnm{Sparn}, \binits{M.}},
\bauthor{\bsnm{Kath}, \binits{E.}},
\bauthor{\bsnm{Duchene}, \binits{J.}},
\bauthor{\bsnm{Fujii}, \binits{K.}},
\bauthor{\bsnm{G\"orlitz}, \binits{S.L.}},
\bauthor{\bsnm{Enss}, \binits{T.}},
\bauthor{\bsnm{Strobel}, \binits{H.}},
\bauthor{\bsnm{Oberthaler}, \binits{M.K.}}:
\batitle{Observation of pattern stabilization in a driven superfluid}.
\bjtitle{Phys. Rev. X}
\bvolume{15},
\bfpage{011026}
(\byear{2025})
\doiurl{10.1103/PhysRevX.15.011026}
\end{barticle}
\endbibitem

\bibitem[\protect\citeauthoryear{Liebster et~al.}{2025b}]{Liebster:NatPhys2025}
\begin{barticle}
\bauthor{\bsnm{Liebster}, \binits{N.}},
\bauthor{\bsnm{Sparn}, \binits{M.}},
\bauthor{\bsnm{Kath}, \binits{E.}},
\bauthor{\bsnm{Duchene}, \binits{J.}},
\bauthor{\bsnm{Strobel}, \binits{H.}},
\bauthor{\bsnm{Oberthaler}, \binits{M.K.}}:
\batitle{Supersolid-like sound modes in a driven quantum gas}.
\bjtitle{Nature Physics}
\bvolume{21}(\bissue{7}),
\bfpage{1064}--\blpage{1070}
(\byear{2025})
\end{barticle}
\endbibitem

\bibitem[\protect\citeauthoryear{Bao et~al.}{2020}]{Bao:PRR2020}
\begin{barticle}
\bauthor{\bsnm{Bao}, \binits{H.}},
\bauthor{\bsnm{Olivieri}, \binits{L.}},
\bauthor{\bsnm{Rowley}, \binits{M.}},
\bauthor{\bsnm{Chu}, \binits{S.T.}},
\bauthor{\bsnm{Little}, \binits{B.E.}},
\bauthor{\bsnm{Morandotti}, \binits{R.}},
\bauthor{\bsnm{Moss}, \binits{D.J.}},
\bauthor{\bsnm{Totero~Gongora}, \binits{J.S.}},
\bauthor{\bsnm{Peccianti}, \binits{M.}},
\bauthor{\bsnm{Pasquazi}, \binits{A.}}:
\batitle{Turing patterns in a fiber laser with a nested microresonator:
  {Robust} and controllable microcomb generation}.
\bjtitle{Physical Review Research}
\bvolume{2}(\bissue{2}),
\bfpage{023395}
(\byear{2020})
\doiurl{10.1103/PhysRevResearch.2.023395} .
Accessed 2026-01-15
\end{barticle}
\endbibitem

\bibitem[\protect\citeauthoryear{Trypogeorgos
  et~al.}{2025}]{trypogeorgos2025emerging}
\begin{botherref}
\oauthor{\bsnm{Trypogeorgos}, \binits{D.}},
\oauthor{\bsnm{Gianfrate}, \binits{A.}},
\oauthor{\bsnm{Landini}, \binits{M.}},
\oauthor{\bsnm{Nigro}, \binits{D.}},
\oauthor{\bsnm{Gerace}, \binits{D.}},
\oauthor{\bsnm{Carusotto}, \binits{I.}},
\oauthor{\bsnm{Riminucci}, \binits{F.}},
\oauthor{\bsnm{Baldwin}, \binits{K.W.}},
\oauthor{\bsnm{Pfeiffer}, \binits{L.N.}},
\oauthor{\bsnm{Martone}, \binits{G.I.}}, et al.:
Emerging supersolidity in photonic-crystal polariton condensates.
Nature,
1--5
(2025)
\end{botherref}
\endbibitem

\bibitem[\protect\citeauthoryear{Muszyński et~al.}{2025}]{Muszynski:arXiv2025}
\begin{botherref}
\oauthor{\bsnm{Muszyński}, \binits{M.}},
\oauthor{\bsnm{Kokhanchik}, \binits{P.}},
\oauthor{\bsnm{Mirek}, \binits{R.}},
\oauthor{\bsnm{Urbonas}, \binits{D.}},
\oauthor{\bsnm{Tassan}, \binits{P.}},
\oauthor{\bsnm{Kapuściński}, \binits{P.}},
\oauthor{\bsnm{Oliwa}, \binits{P.}},
\oauthor{\bsnm{Georgakilas}, \binits{I.}},
\oauthor{\bsnm{Stöferle}, \binits{T.}},
\oauthor{\bsnm{Mahrt}, \binits{R.F.}},
\oauthor{\bsnm{Forster}, \binits{M.}},
\oauthor{\bsnm{Scherf}, \binits{U.}},
\oauthor{\bsnm{Dovzhenko}, \binits{D.}},
\oauthor{\bsnm{Mazur}, \binits{R.}},
\oauthor{\bsnm{Morawiak}, \binits{P.}},
\oauthor{\bsnm{Piecek}, \binits{W.}},
\oauthor{\bsnm{Kula}, \binits{P.}},
\oauthor{\bsnm{Pietka}, \binits{B.}},
\oauthor{\bsnm{Solnyshkov}, \binits{D.}},
\oauthor{\bsnm{Malpuech}, \binits{G.}},
\oauthor{\bsnm{Szczytko}, \binits{J.}}:
Observation of a supersolid phase in a spin-orbit coupled exciton-polariton
  {Bose}-{Einstein} condensate at room temperature.
arXiv.
arXiv:2407.02406 [cond-mat]
(2025).
\doiurl{10.48550/arXiv.2407.02406} .
\url{http://arxiv.org/abs/2407.02406}
Accessed 2026-02-14
\end{botherref}
\endbibitem

\bibitem[\protect\citeauthoryear{Tanese et~al.}{2013}]{Tanese:NatComm2013}
\begin{barticle}
\bauthor{\bsnm{Tanese}, \binits{D.}},
\bauthor{\bsnm{Flayac}, \binits{H.}},
\bauthor{\bsnm{Solnyshkov}, \binits{D.}},
\bauthor{\bsnm{Amo}, \binits{A.}},
\bauthor{\bsnm{Lema{\^\i}tre}, \binits{A.}},
\bauthor{\bsnm{Galopin}, \binits{E.}},
\bauthor{\bsnm{Braive}, \binits{R.}},
\bauthor{\bsnm{Senellart}, \binits{P.}},
\bauthor{\bsnm{Sagnes}, \binits{I.}},
\bauthor{\bsnm{Malpuech}, \binits{G.}}, \betal:
\batitle{Polariton condensation in solitonic gap states in a one-dimensional
  periodic potential}.
\bjtitle{Nature communications}
\bvolume{4}(\bissue{1}),
\bfpage{1749}
(\byear{2013})
\end{barticle}
\endbibitem

\bibitem[\protect\citeauthoryear{Hsu et~al.}{2016}]{Hsu:NatMat2016}
\begin{barticle}
\bauthor{\bsnm{Hsu}, \binits{C.W.}},
\bauthor{\bsnm{Zhen}, \binits{B.}},
\bauthor{\bsnm{Stone}, \binits{A.D.}},
\bauthor{\bsnm{Joannopoulos}, \binits{J.D.}},
\bauthor{\bsnm{Solja{\v{c}}i{\'c}}, \binits{M.}}:
\batitle{Bound states in the continuum}.
\bjtitle{Nature Reviews Materials}
\bvolume{1}(\bissue{9}),
\bfpage{16048}
(\byear{2016})
\end{barticle}
\endbibitem

\bibitem[\protect\citeauthoryear{Meng et~al.}{2026}]{Meng:NatNano2026}
\begin{botherref}
\oauthor{\bsnm{Meng}, \binits{Y.}},
\oauthor{\bsnm{Li}, \binits{W.}},
\oauthor{\bsnm{Peng}, \binits{K.}},
\oauthor{\bsnm{Ti}, \binits{C.}},
\oauthor{\bsnm{Dang}, \binits{J.}},
\oauthor{\bsnm{Wu}, \binits{X.}},
\oauthor{\bsnm{Han}, \binits{X.}},
\oauthor{\bsnm{Bao}, \binits{W.}}:
Hybrid perovskite--nanograting photonic architecture enables supersolidity at
  room temperature.
Nature Nanotechnology,
1--9
(2026)
\end{botherref}
\endbibitem

\bibitem[\protect\citeauthoryear{Nigro et~al.}{2025}]{Nigro:PRL2025}
\begin{barticle}
\bauthor{\bsnm{Nigro}, \binits{D.}},
\bauthor{\bsnm{Trypogeorgos}, \binits{D.}},
\bauthor{\bsnm{Gianfrate}, \binits{A.}},
\bauthor{\bsnm{Sanvitto}, \binits{D.}},
\bauthor{\bsnm{Carusotto}, \binits{I.}},
\bauthor{\bsnm{Gerace}, \binits{D.}}:
\batitle{Supersolidity of polariton condensates in photonic crystal
  waveguides}.
\bjtitle{Physical Review Letters}
\bvolume{134}(\bissue{5}),
\bfpage{056002}
(\byear{2025})
\end{barticle}
\endbibitem

\bibitem[\protect\citeauthoryear{Recati and
  Stringari}{2023}]{recati2023supersolidity}
\begin{barticle}
\bauthor{\bsnm{Recati}, \binits{A.}},
\bauthor{\bsnm{Stringari}, \binits{S.}}:
\batitle{Supersolidity in ultracold dipolar gases}.
\bjtitle{Nature Reviews Physics}
\bvolume{5}(\bissue{12}),
\bfpage{735}--\blpage{743}
(\byear{2023})
\end{barticle}
\endbibitem

\bibitem[\protect\citeauthoryear{Petter et~al.}{2019}]{Petter:PRL2019}
\begin{barticle}
\bauthor{\bsnm{Petter}, \binits{D.}},
\bauthor{\bsnm{Natale}, \binits{G.}},
\bauthor{\bsnm{Bijnen}, \binits{R.M.W.}},
\bauthor{\bsnm{Patscheider}, \binits{A.}},
\bauthor{\bsnm{Mark}, \binits{M.J.}},
\bauthor{\bsnm{Chomaz}, \binits{L.}},
\bauthor{\bsnm{Ferlaino}, \binits{F.}}:
\batitle{Probing the roton excitation spectrum of a stable dipolar bose gas}.
\bjtitle{Phys. Rev. Lett.}
\bvolume{122},
\bfpage{183401}
(\byear{2019})
\doiurl{10.1103/PhysRevLett.122.183401}
\end{barticle}
\endbibitem

\bibitem[\protect\citeauthoryear{Grudinina et~al.}{2026}]{Grudinina:arxiv2026}
\begin{botherref}
\oauthor{\bsnm{Grudinina}, \binits{A.}},
\oauthor{\bsnm{Cao}, \binits{J.}},
\oauthor{\bsnm{Kavokin}, \binits{A.}},
\oauthor{\bsnm{Voronova}, \binits{N.}},
\oauthor{\bsnm{Nalitov}, \binits{A.}}:
Collective excitations and stability of nonequilibrium polariton supersolids.
arXiv preprint arXiv:2604.21353
(2026)
\end{botherref}
\endbibitem

\bibitem[\protect\citeauthoryear{Verdier}{2026}]{Verdier:MSc2026}
\begin{botherref}
\oauthor{\bsnm{Verdier}, \binits{S.e.}}:
Collective excitation modes of a supersolid state of a quantum fluid of light.
Master's thesis,
\'Ecole Polytechnique
(2026)
\end{botherref}
\endbibitem

\bibitem[\protect\citeauthoryear{Kozhevin et~al.}{2025}]{Kozhevin:arXiv2025}
\begin{botherref}
\oauthor{\bsnm{Kozhevin}, \binits{P.}},
\oauthor{\bsnm{Liubomirov}, \binits{A.}},
\oauthor{\bsnm{Cherbunin}, \binits{R.}},
\oauthor{\bsnm{Chukeev}, \binits{M.}},
\oauthor{\bsnm{Chestnov}, \binits{I.Y.}},
\oauthor{\bsnm{Kavokin}, \binits{A.}},
\oauthor{\bsnm{Nalitov}, \binits{A.}}:
Supersolidity in optically trapped polariton condensates.
arXiv preprint arXiv:2507.14585
(2025)
\end{botherref}
\endbibitem

\bibitem[\protect\citeauthoryear{Lim et~al.}{2017}]{lim_electrically_2017}
\begin{barticle}
\bauthor{\bsnm{Lim}, \binits{H.-T.}},
\bauthor{\bsnm{Togan}, \binits{E.}},
\bauthor{\bsnm{Kroner}, \binits{M.}},
\bauthor{\bsnm{Miguel-Sanchez}, \binits{J.}},
\bauthor{\bsnm{Imamoğlu}, \binits{A.}}:
\batitle{Electrically tunable artificial gauge potential for polaritons}.
\bjtitle{Nature Communications}
\bvolume{8}(\bissue{1}),
\bfpage{14540}
(\byear{2017})
\doiurl{10.1038/ncomms14540} .
\bcomment{Number: 1}.
Accessed 2023-03-30
\end{barticle}
\endbibitem

\bibitem[\protect\citeauthoryear{Roushan et~al.}{2017}]{Roushan:2016NatPhys}
\begin{barticle}
\bauthor{\bsnm{Roushan}, \binits{P.}},
\bauthor{\bsnm{Neill}, \binits{C.}},
\bauthor{\bsnm{Megrant}, \binits{A.}},
\bauthor{\bsnm{Chen}, \binits{Y.}},
\bauthor{\bsnm{Babbush}, \binits{R.}},
\bauthor{\bsnm{Barends}, \binits{R.}},
\bauthor{\bsnm{Campbell}, \binits{B.}},
\bauthor{\bsnm{Chen}, \binits{Z.}},
\bauthor{\bsnm{Chiaro}, \binits{B.}},
\bauthor{\bsnm{Dunsworth}, \binits{A.}}, \betal:
\batitle{Chiral ground-state currents of interacting photons in a synthetic
  magnetic field}.
\bjtitle{Nature Physics}
\bvolume{13}(\bissue{2}),
\bfpage{146}
(\byear{2017})
\end{barticle}
\endbibitem

\bibitem[\protect\citeauthoryear{Clark et~al.}{2020}]{clark2020observation}
\begin{barticle}
\bauthor{\bsnm{Clark}, \binits{L.W.}},
\bauthor{\bsnm{Schine}, \binits{N.}},
\bauthor{\bsnm{Baum}, \binits{C.}},
\bauthor{\bsnm{Jia}, \binits{N.}},
\bauthor{\bsnm{Simon}, \binits{J.}}:
\batitle{Observation of laughlin states made of light}.
\bjtitle{Nature}
\bvolume{582}(\bissue{7810}),
\bfpage{41}--\blpage{45}
(\byear{2020})
\end{barticle}
\endbibitem

\bibitem[\protect\citeauthoryear{Wang et~al.}{2024}]{Wang:Science2024}
\begin{barticle}
\bauthor{\bsnm{Wang}, \binits{C.}},
\bauthor{\bsnm{Liu}, \binits{F.-M.}},
\bauthor{\bsnm{Chen}, \binits{M.-C.}},
\bauthor{\bsnm{Chen}, \binits{H.}},
\bauthor{\bsnm{Zhao}, \binits{X.-H.}},
\bauthor{\bsnm{Ying}, \binits{C.}},
\bauthor{\bsnm{Shang}, \binits{Z.-X.}},
\bauthor{\bsnm{Wang}, \binits{J.-W.}},
\bauthor{\bsnm{Huo}, \binits{Y.-H.}},
\bauthor{\bsnm{Peng}, \binits{C.-Z.}}, \betal:
\batitle{Realization of fractional quantum hall state with interacting
  photons}.
\bjtitle{Science}
\bvolume{384}(\bissue{6695}),
\bfpage{579}--\blpage{584}
(\byear{2024})
\end{barticle}
\endbibitem

\bibitem[\protect\citeauthoryear{Ma et~al.}{2019}]{Ma:Nature2019}
\begin{barticle}
\bauthor{\bsnm{Ma}, \binits{R.}},
\bauthor{\bsnm{Saxberg}, \binits{B.}},
\bauthor{\bsnm{Owens}, \binits{C.}},
\bauthor{\bsnm{Leung}, \binits{N.}},
\bauthor{\bsnm{Lu}, \binits{Y.}},
\bauthor{\bsnm{Simon}, \binits{J.}},
\bauthor{\bsnm{Schuster}, \binits{D.I.}}:
\batitle{A dissipatively stabilized mott insulator of photons}.
\bjtitle{Nature}
\bvolume{566}(\bissue{7742}),
\bfpage{51}--\blpage{57}
(\byear{2019})
\end{barticle}
\endbibitem

\bibitem[\protect\citeauthoryear{Bloch et~al.}{2008}]{bloch2008many}
\begin{barticle}
\bauthor{\bsnm{Bloch}, \binits{I.}},
\bauthor{\bsnm{Dalibard}, \binits{J.}},
\bauthor{\bsnm{Zwerger}, \binits{W.}}:
\batitle{Many-body physics with ultracold gases}.
\bjtitle{Reviews of modern physics}
\bvolume{80}(\bissue{3}),
\bfpage{885}
(\byear{2008})
\end{barticle}
\endbibitem

\bibitem[\protect\citeauthoryear{Kapit et~al.}{2014}]{Kapit:2014PRX}
\begin{barticle}
\bauthor{\bsnm{Kapit}, \binits{E.}},
\bauthor{\bsnm{Hafezi}, \binits{M.}},
\bauthor{\bsnm{Simon}, \binits{S.H.}}:
\batitle{Induced self-stabilization in fractional quantum hall states of
  light}.
\bjtitle{Phys. Rev. X}
\bvolume{4},
\bfpage{031039}
(\byear{2014})
\end{barticle}
\endbibitem

\bibitem[\protect\citeauthoryear{Lebreuilly et~al.}{2016}]{Lebreuilly:CRAS2016}
\begin{barticle}
\bauthor{\bsnm{Lebreuilly}, \binits{J.}},
\bauthor{\bsnm{Wouters}, \binits{M.}},
\bauthor{\bsnm{Carusotto}, \binits{I.}}:
\batitle{Towards strongly correlated photons in arrays of dissipative nonlinear
  cavities under a frequency-dependent incoherent pumping}.
\bjtitle{Comptes Rendus Physique}
\bvolume{17}(\bissue{8}),
\bfpage{836}--\blpage{860}
(\byear{2016})
\end{barticle}
\endbibitem

\bibitem[\protect\citeauthoryear{Biella et~al.}{2017}]{Biella:PRA2017}
\begin{barticle}
\bauthor{\bsnm{Biella}, \binits{A.}},
\bauthor{\bsnm{Storme}, \binits{F.}},
\bauthor{\bsnm{Lebreuilly}, \binits{J.}},
\bauthor{\bsnm{Rossini}, \binits{D.}},
\bauthor{\bsnm{Fazio}, \binits{R.}},
\bauthor{\bsnm{Carusotto}, \binits{I.}},
\bauthor{\bsnm{Ciuti}, \binits{C.}}:
\batitle{Phase diagram of incoherently driven strongly correlated photonic
  lattices}.
\bjtitle{Physical Review A}
\bvolume{96}(\bissue{2}),
\bfpage{023839}
(\byear{2017})
\end{barticle}
\endbibitem

\bibitem[\protect\citeauthoryear{Lebreuilly et~al.}{2017}]{Lebreuilly:PRA2017}
\begin{barticle}
\bauthor{\bsnm{Lebreuilly}, \binits{J.}},
\bauthor{\bsnm{Biella}, \binits{A.}},
\bauthor{\bsnm{Storme}, \binits{F.}},
\bauthor{\bsnm{Rossini}, \binits{D.}},
\bauthor{\bsnm{Fazio}, \binits{R.}},
\bauthor{\bsnm{Ciuti}, \binits{C.}},
\bauthor{\bsnm{Carusotto}, \binits{I.}}:
\batitle{Stabilizing strongly correlated photon fluids with non-markovian
  reservoirs}.
\bjtitle{Physical Review A}
\bvolume{96}(\bissue{3}),
\bfpage{033828}
(\byear{2017})
\end{barticle}
\endbibitem

\bibitem[\protect\citeauthoryear{Mi et~al.}{2024}]{Mi:Science2024}
\begin{barticle}
\bauthor{\bsnm{Mi}, \binits{X.}},
\bauthor{\bsnm{Michailidis}, \binits{A.A.}},
\bauthor{\bsnm{Shabani}, \binits{S.}},
\bauthor{\bsnm{Miao}, \binits{K.C.}},
\bauthor{\bsnm{Klimov}, \binits{P.V.}},
\bauthor{\bsnm{Lloyd}, \binits{J.}},
\bauthor{\bsnm{Rosenberg}, \binits{E.}},
\bauthor{\bsnm{Acharya}, \binits{R.}},
\bauthor{\bsnm{Aleiner}, \binits{I.}},
\bauthor{\bsnm{Andersen}, \binits{T.I.}},
\bauthor{\bsnm{Ansmann}, \binits{M.}},
\bauthor{\bsnm{Arute}, \binits{F.}},
\bauthor{\bsnm{Arya}, \binits{K.}},
\bauthor{\bsnm{Asfaw}, \binits{A.}},
\bauthor{\bsnm{Atalaya}, \binits{J.}},
\bauthor{\bsnm{Bardin}, \binits{J.C.}},
\bauthor{\bsnm{Bengtsson}, \binits{A.}},
\bauthor{\bsnm{Bortoli}, \binits{G.}},
\bauthor{\bsnm{Bourassa}, \binits{A.}},
\bauthor{\bsnm{Bovaird}, \binits{J.}},
\bauthor{\bsnm{Brill}, \binits{L.}},
\bauthor{\bsnm{Broughton}, \binits{M.}},
\bauthor{\bsnm{Buckley}, \binits{B.B.}},
\bauthor{\bsnm{Buell}, \binits{D.A.}},
\bauthor{\bsnm{Burger}, \binits{T.}},
\bauthor{\bsnm{Burkett}, \binits{B.}},
\bauthor{\bsnm{Bushnell}, \binits{N.}},
\bauthor{\bsnm{Chen}, \binits{Z.}},
\bauthor{\bsnm{Chiaro}, \binits{B.}},
\bauthor{\bsnm{Chik}, \binits{D.}},
\bauthor{\bsnm{Chou}, \binits{C.}},
\bauthor{\bsnm{Cogan}, \binits{J.}},
\bauthor{\bsnm{Collins}, \binits{R.}},
\bauthor{\bsnm{Conner}, \binits{P.}},
\bauthor{\bsnm{Courtney}, \binits{W.}},
\bauthor{\bsnm{Crook}, \binits{A.L.}},
\bauthor{\bsnm{Curtin}, \binits{B.}},
\bauthor{\bsnm{Dau}, \binits{A.G.}},
\bauthor{\bsnm{Debroy}, \binits{D.M.}},
\bauthor{\bsnm{Barba}, \binits{A.D.T.}},
\bauthor{\bsnm{Demura}, \binits{S.}},
\bauthor{\bsnm{Paolo}, \binits{A.D.}},
\bauthor{\bsnm{Drozdov}, \binits{I.K.}},
\bauthor{\bsnm{Dunsworth}, \binits{A.}},
\bauthor{\bsnm{Erickson}, \binits{C.}},
\bauthor{\bsnm{Faoro}, \binits{L.}},
\bauthor{\bsnm{Farhi}, \binits{E.}},
\bauthor{\bsnm{Fatemi}, \binits{R.}},
\bauthor{\bsnm{Ferreira}, \binits{V.S.}},
\bauthor{\bsnm{Burgos}, \binits{L.F.}},
\bauthor{\bsnm{Forati}, \binits{E.}},
\bauthor{\bsnm{Fowler}, \binits{A.G.}},
\bauthor{\bsnm{Foxen}, \binits{B.}},
\bauthor{\bsnm{Genois}, \binits{E.}},
\bauthor{\bsnm{Giang}, \binits{W.}},
\bauthor{\bsnm{Gidney}, \binits{C.}},
\bauthor{\bsnm{Gilboa}, \binits{D.}},
\bauthor{\bsnm{Giustina}, \binits{M.}},
\bauthor{\bsnm{Gosula}, \binits{R.}},
\bauthor{\bsnm{Gross}, \binits{J.A.}},
\bauthor{\bsnm{Habegger}, \binits{S.}},
\bauthor{\bsnm{Hamilton}, \binits{M.C.}},
\bauthor{\bsnm{Hansen}, \binits{M.}},
\bauthor{\bsnm{Harrigan}, \binits{M.P.}},
\bauthor{\bsnm{Harrington}, \binits{S.D.}},
\bauthor{\bsnm{Heu}, \binits{P.}},
\bauthor{\bsnm{Hoffmann}, \binits{M.R.}},
\bauthor{\bsnm{Hong}, \binits{S.}},
\bauthor{\bsnm{Huang}, \binits{T.}},
\bauthor{\bsnm{Huff}, \binits{A.}},
\bauthor{\bsnm{Huggins}, \binits{W.J.}},
\bauthor{\bsnm{Ioffe}, \binits{L.B.}},
\bauthor{\bsnm{Isakov}, \binits{S.V.}},
\bauthor{\bsnm{Iveland}, \binits{J.}},
\bauthor{\bsnm{Jeffrey}, \binits{E.}},
\bauthor{\bsnm{Jiang}, \binits{Z.}},
\bauthor{\bsnm{Jones}, \binits{C.}},
\bauthor{\bsnm{Juhas}, \binits{P.}},
\bauthor{\bsnm{Kafri}, \binits{D.}},
\bauthor{\bsnm{Kechedzhi}, \binits{K.}},
\bauthor{\bsnm{Khattar}, \binits{T.}},
\bauthor{\bsnm{Khezri}, \binits{M.}},
\bauthor{\bsnm{Kieferová}, \binits{M.}},
\bauthor{\bsnm{Kim}, \binits{S.}},
\bauthor{\bsnm{Kitaev}, \binits{A.}},
\bauthor{\bsnm{Klots}, \binits{A.R.}},
\bauthor{\bsnm{Korotkov}, \binits{A.N.}},
\bauthor{\bsnm{Kostritsa}, \binits{F.}},
\bauthor{\bsnm{Kreikebaum}, \binits{J.M.}},
\bauthor{\bsnm{Landhuis}, \binits{D.}},
\bauthor{\bsnm{Laptev}, \binits{P.}},
\bauthor{\bsnm{Lau}, \binits{K.-M.}},
\bauthor{\bsnm{Laws}, \binits{L.}},
\bauthor{\bsnm{Lee}, \binits{J.}},
\bauthor{\bsnm{Lee}, \binits{K.W.}},
\bauthor{\bsnm{Lensky}, \binits{Y.D.}},
\bauthor{\bsnm{Lester}, \binits{B.J.}},
\bauthor{\bsnm{Lill}, \binits{A.T.}},
\bauthor{\bsnm{Liu}, \binits{W.}},
\bauthor{\bsnm{Locharla}, \binits{A.}},
\bauthor{\bsnm{Malone}, \binits{F.D.}},
\bauthor{\bsnm{Martin}, \binits{O.}},
\bauthor{\bsnm{McClean}, \binits{J.R.}},
\bauthor{\bsnm{McEwen}, \binits{M.}},
\bauthor{\bsnm{Mieszala}, \binits{A.}},
\bauthor{\bsnm{Montazeri}, \binits{S.}},
\bauthor{\bsnm{Morvan}, \binits{A.}},
\bauthor{\bsnm{Movassagh}, \binits{R.}},
\bauthor{\bsnm{Mruczkiewicz}, \binits{W.}},
\bauthor{\bsnm{Neeley}, \binits{M.}},
\bauthor{\bsnm{Neill}, \binits{C.}},
\bauthor{\bsnm{Nersisyan}, \binits{A.}},
\bauthor{\bsnm{Newman}, \binits{M.}},
\bauthor{\bsnm{Ng}, \binits{J.H.}},
\bauthor{\bsnm{Nguyen}, \binits{A.}},
\bauthor{\bsnm{Nguyen}, \binits{M.}},
\bauthor{\bsnm{Niu}, \binits{M.Y.}},
\bauthor{\bsnm{O’Brien}, \binits{T.E.}},
\bauthor{\bsnm{Opremcak}, \binits{A.}},
\bauthor{\bsnm{Petukhov}, \binits{A.}},
\bauthor{\bsnm{Potter}, \binits{R.}},
\bauthor{\bsnm{Pryadko}, \binits{L.P.}},
\bauthor{\bsnm{Quintana}, \binits{C.}},
\bauthor{\bsnm{Rocque}, \binits{C.}},
\bauthor{\bsnm{Rubin}, \binits{N.C.}},
\bauthor{\bsnm{Saei}, \binits{N.}},
\bauthor{\bsnm{Sank}, \binits{D.}},
\bauthor{\bsnm{Sankaragomathi}, \binits{K.}},
\bauthor{\bsnm{Satzinger}, \binits{K.J.}},
\bauthor{\bsnm{Schurkus}, \binits{H.F.}},
\bauthor{\bsnm{Schuster}, \binits{C.}},
\bauthor{\bsnm{Shearn}, \binits{M.J.}},
\bauthor{\bsnm{Shorter}, \binits{A.}},
\bauthor{\bsnm{Shutty}, \binits{N.}},
\bauthor{\bsnm{Shvarts}, \binits{V.}},
\bauthor{\bsnm{Skruzny}, \binits{J.}},
\bauthor{\bsnm{Smith}, \binits{W.C.}},
\bauthor{\bsnm{Somma}, \binits{R.}},
\bauthor{\bsnm{Sterling}, \binits{G.}},
\bauthor{\bsnm{Strain}, \binits{D.}},
\bauthor{\bsnm{Szalay}, \binits{M.}},
\bauthor{\bsnm{Torres}, \binits{A.}},
\bauthor{\bsnm{Vidal}, \binits{G.}},
\bauthor{\bsnm{Villalonga}, \binits{B.}},
\bauthor{\bsnm{Heidweiller}, \binits{C.V.}},
\bauthor{\bsnm{White}, \binits{T.}},
\bauthor{\bsnm{Woo}, \binits{B.W.K.}},
\bauthor{\bsnm{Xing}, \binits{C.}},
\bauthor{\bsnm{Yao}, \binits{Z.J.}},
\bauthor{\bsnm{Yeh}, \binits{P.}},
\bauthor{\bsnm{Yoo}, \binits{J.}},
\bauthor{\bsnm{Young}, \binits{G.}},
\bauthor{\bsnm{Zalcman}, \binits{A.}},
\bauthor{\bsnm{Zhang}, \binits{Y.}},
\bauthor{\bsnm{Zhu}, \binits{N.}},
\bauthor{\bsnm{Zobrist}, \binits{N.}},
\bauthor{\bsnm{Neven}, \binits{H.}},
\bauthor{\bsnm{Babbush}, \binits{R.}},
\bauthor{\bsnm{Bacon}, \binits{D.}},
\bauthor{\bsnm{Boixo}, \binits{S.}},
\bauthor{\bsnm{Hilton}, \binits{J.}},
\bauthor{\bsnm{Lucero}, \binits{E.}},
\bauthor{\bsnm{Megrant}, \binits{A.}},
\bauthor{\bsnm{Kelly}, \binits{J.}},
\bauthor{\bsnm{Chen}, \binits{Y.}},
\bauthor{\bsnm{Roushan}, \binits{P.}},
\bauthor{\bsnm{Smelyanskiy}, \binits{V.}},
\bauthor{\bsnm{Abanin}, \binits{D.A.}}:
\batitle{Stable quantum-correlated many-body states through engineered
  dissipation}.
\bjtitle{Science}
\bvolume{383}(\bissue{6689}),
\bfpage{1332}--\blpage{1337}
(\byear{2024})
\doiurl{10.1126/science.adh9932}
{\href{https://arxiv.org/abs/https://www.science.org/doi/pdf/10.1126/science.adh9932}{{https://www.science.org/doi/pdf/10.1126/science.adh9932}}}
\end{barticle}
\endbibitem

\bibitem[\protect\citeauthoryear{Caleffi et~al.}{2023}]{Caleffi:PRL2023}
\begin{barticle}
\bauthor{\bsnm{Caleffi}, \binits{F.}},
\bauthor{\bsnm{Capone}, \binits{M.}},
\bauthor{\bsnm{Carusotto}, \binits{I.}}:
\batitle{Collective excitations of a strongly correlated nonequilibrium photon
  fluid across the insulator-superfluid phase transition}.
\bjtitle{Physical Review Letters}
\bvolume{131}(\bissue{19}),
\bfpage{193604}
(\byear{2023})
\end{barticle}
\endbibitem

\bibitem[\protect\citeauthoryear{Sieberer et~al.}{2016}]{Sieberer:RPP2016}
\begin{barticle}
\bauthor{\bsnm{Sieberer}, \binits{L.M.}},
\bauthor{\bsnm{Buchhold}, \binits{M.}},
\bauthor{\bsnm{Diehl}, \binits{S.}}:
\batitle{Keldysh field theory for driven open quantum systems}.
\bjtitle{Reports on Progress in Physics}
\bvolume{79}(\bissue{9}),
\bfpage{096001}
(\byear{2016})
\end{barticle}
\endbibitem

\bibitem[\protect\citeauthoryear{Fr{\'e}rot et~al.}{2023}]{Frerot:PRX2023}
\begin{barticle}
\bauthor{\bsnm{Fr{\'e}rot}, \binits{I.}},
\bauthor{\bsnm{Vashisht}, \binits{A.}},
\bauthor{\bsnm{Morassi}, \binits{M.}},
\bauthor{\bsnm{Lema{\^\i}tre}, \binits{A.}},
\bauthor{\bsnm{Ravets}, \binits{S.}},
\bauthor{\bsnm{Bloch}, \binits{J.}},
\bauthor{\bsnm{Minguzzi}, \binits{A.}},
\bauthor{\bsnm{Richard}, \binits{M.}}:
\batitle{Bogoliubov excitations driven by thermal lattice phonons in a quantum
  fluid of light}.
\bjtitle{Physical Review X}
\bvolume{13}(\bissue{4}),
\bfpage{041058}
(\byear{2023})
\end{barticle}
\endbibitem

\bibitem[\protect\citeauthoryear{Szyma{\'n}ska
  et~al.}{2007}]{szymanska2007coherence}
\begin{barticle}
\bauthor{\bsnm{Szyma{\'n}ska}, \binits{M.}},
\bauthor{\bsnm{Marchetti}, \binits{F.}},
\bauthor{\bsnm{Keeling}, \binits{J.}},
\bauthor{\bsnm{Littlewood}, \binits{P.}}:
\batitle{Coherence properties and luminescence spectra of condensed polaritons
  in cdte microcavities}.
\bjtitle{Solid state communications}
\bvolume{144}(\bissue{9}),
\bfpage{364}--\blpage{370}
(\byear{2007})
\end{barticle}
\endbibitem

\bibitem[\protect\citeauthoryear{Walls and Milburn}{2006}]{QuantumOptics}
\begin{bbook}
\bauthor{\bsnm{Walls}, \binits{D.F.}},
\bauthor{\bsnm{Milburn}, \binits{G.}}:
\bbtitle{Quantum Optics}.
\bpublisher{Springer}, \blocation{???}
(\byear{2006})
\end{bbook}
\endbibitem

\bibitem[\protect\citeauthoryear{Fontaine et~al.}{2022}]{fontaine2022kardar}
\begin{barticle}
\bauthor{\bsnm{Fontaine}, \binits{Q.}},
\bauthor{\bsnm{Squizzato}, \binits{D.}},
\bauthor{\bsnm{Baboux}, \binits{F.}},
\bauthor{\bsnm{Amelio}, \binits{I.}},
\bauthor{\bsnm{Lema{\^\i}tre}, \binits{A.}},
\bauthor{\bsnm{Morassi}, \binits{M.}},
\bauthor{\bsnm{Sagnes}, \binits{I.}},
\bauthor{\bsnm{Le~Gratiet}, \binits{L.}},
\bauthor{\bsnm{Harouri}, \binits{A.}},
\bauthor{\bsnm{Wouters}, \binits{M.}}, \betal:
\batitle{Kardar--parisi--zhang universality in a one-dimensional polariton
  condensate}.
\bjtitle{Nature}
\bvolume{608}(\bissue{7924}),
\bfpage{687}--\blpage{691}
(\byear{2022})
\end{barticle}
\endbibitem

\bibitem[\protect\citeauthoryear{Berg et~al.}{2009}]{Berg:PRA2009}
\begin{barticle}
\bauthor{\bsnm{Berg}, \binits{B.}},
\bauthor{\bsnm{Plimak}, \binits{L.I.}},
\bauthor{\bsnm{Polkovnikov}, \binits{A.}},
\bauthor{\bsnm{Olsen}, \binits{M.K.}},
\bauthor{\bsnm{Fleischhauer}, \binits{M.}},
\bauthor{\bsnm{Schleich}, \binits{W.P.}}:
\batitle{Commuting heisenberg operators as the quantum response problem:
  Time-normal averages in the truncated wigner representation}.
\bjtitle{Phys. Rev. A}
\bvolume{80},
\bfpage{033624}
(\byear{2009})
\doiurl{10.1103/PhysRevA.80.033624}
\end{barticle}
\endbibitem

\bibitem[\protect\citeauthoryear{Zhang et~al.}{2022}]{zhang2022angle}
\begin{barticle}
\bauthor{\bsnm{Zhang}, \binits{H.}},
\bauthor{\bsnm{Pincelli}, \binits{T.}},
\bauthor{\bsnm{Jozwiak}, \binits{C.}},
\bauthor{\bsnm{Kondo}, \binits{T.}},
\bauthor{\bsnm{Ernstorfer}, \binits{R.}},
\bauthor{\bsnm{Sato}, \binits{T.}},
\bauthor{\bsnm{Zhou}, \binits{S.}}:
\batitle{Angle-resolved photoemission spectroscopy}.
\bjtitle{Nature Reviews Methods Primers}
\bvolume{2}(\bissue{1}),
\bfpage{54}
(\byear{2022})
\end{barticle}
\endbibitem

\bibitem[\protect\citeauthoryear{Japha et~al.}{1999}]{japha1999coherent}
\begin{barticle}
\bauthor{\bsnm{Japha}, \binits{Y.}},
\bauthor{\bsnm{Choi}, \binits{S.}},
\bauthor{\bsnm{Burnett}, \binits{K.}},
\bauthor{\bsnm{Band}, \binits{Y.}}:
\batitle{Coherent output, stimulated quantum evaporation, and pair breaking in
  a trapped atomic bose gas}.
\bjtitle{Physical review letters}
\bvolume{82}(\bissue{6}),
\bfpage{1079}
(\byear{1999})
\end{barticle}
\endbibitem

\bibitem[\protect\citeauthoryear{Dao et~al.}{2007}]{dao2007measuring}
\begin{barticle}
\bauthor{\bsnm{Dao}, \binits{T.-L.}},
\bauthor{\bsnm{Georges}, \binits{A.}},
\bauthor{\bsnm{Dalibard}, \binits{J.}},
\bauthor{\bsnm{Salomon}, \binits{C.}},
\bauthor{\bsnm{Carusotto}, \binits{I.}}:
\batitle{Measuring the one-particle excitations of ultracold fermionic atoms<?
  format?> by stimulated raman spectroscopy}.
\bjtitle{Physical review letters}
\bvolume{98}(\bissue{24}),
\bfpage{240402}
(\byear{2007})
\end{barticle}
\endbibitem

\bibitem[\protect\citeauthoryear{Vale and
  Zwierlein}{2021}]{vale2021spectroscopic}
\begin{barticle}
\bauthor{\bsnm{Vale}, \binits{C.J.}},
\bauthor{\bsnm{Zwierlein}, \binits{M.}}:
\batitle{Spectroscopic probes of quantum gases}.
\bjtitle{Nature Physics}
\bvolume{17}(\bissue{12}),
\bfpage{1305}--\blpage{1315}
(\byear{2021})
\end{barticle}
\endbibitem

\bibitem[\protect\citeauthoryear{Yariv}{2000}]{yariv2000universal}
\begin{barticle}
\bauthor{\bsnm{Yariv}, \binits{A.}}:
\batitle{Universal relations for coupling of optical power between
  microresonators and dielectric waveguides}.
\bjtitle{Electronics letters}
\bvolume{36}(\bissue{4}),
\bfpage{321}--\blpage{322}
(\byear{2000})
\end{barticle}
\endbibitem

\bibitem[\protect\citeauthoryear{Xu et~al.}{2000}]{xu2000scattering}
\begin{barticle}
\bauthor{\bsnm{Xu}, \binits{Y.}},
\bauthor{\bsnm{Li}, \binits{Y.}},
\bauthor{\bsnm{Lee}, \binits{R.K.}},
\bauthor{\bsnm{Yariv}, \binits{A.}}:
\batitle{Scattering-theory analysis of waveguide-resonator coupling}.
\bjtitle{Physical Review E}
\bvolume{62}(\bissue{5}),
\bfpage{7389}
(\byear{2000})
\end{barticle}
\endbibitem

\bibitem[\protect\citeauthoryear{Wouters and Savona}{2009}]{Wouters:PRB2009}
\begin{barticle}
\bauthor{\bsnm{Wouters}, \binits{M.}},
\bauthor{\bsnm{Savona}, \binits{V.}}:
\batitle{Stochastic classical field model for polariton condensates}.
\bjtitle{Phys. Rev. B}
\bvolume{79}(\bissue{16}),
\bfpage{165302}
(\byear{2009})
\doiurl{10.1103/PhysRevB.79.165302}
\end{barticle}
\endbibitem

\bibitem[\protect\citeauthoryear{Ozeri et~al.}{2005}]{ozeri2005colloquium}
\begin{barticle}
\bauthor{\bsnm{Ozeri}, \binits{R.}},
\bauthor{\bsnm{Katz}, \binits{N.}},
\bauthor{\bsnm{Steinhauer}, \binits{J.}},
\bauthor{\bsnm{Davidson}, \binits{N.}}:
\batitle{Colloquium: bulk bogoliubov excitations in a bose-einstein
  condensate}.
\bjtitle{Reviews of Modern Physics}
\bvolume{77}(\bissue{1}),
\bfpage{187}--\blpage{205}
(\byear{2005})
\end{barticle}
\endbibitem

\bibitem[\protect\citeauthoryear{Stamper-Kurn and
  Ketterle}{2002}]{stamper2002spinor}
\begin{bchapter}
\bauthor{\bsnm{Stamper-Kurn}, \binits{D.M.}},
\bauthor{\bsnm{Ketterle}, \binits{W.}}:
\bctitle{Spinor condensates and light scattering from bose-einstein
  condensates}.
In: \bbtitle{Coherent Atomic Matter Waves: 27 July--27 August 1999},
pp. \bfpage{139}--\blpage{217}.
\bpublisher{Springer}, \blocation{???}
(\byear{2002})
\end{bchapter}
\endbibitem

\end{thebibliography}

\end{document}